\documentclass[preprint]{aastex}

\newcommand{\dm}{\Delta m_{15}(B)}
\newcommand{\om}{\Omega_M}
\newcommand{\ol}{\Omega_\Lambda}

\shorttitle{CSP SNe~Ia}
\shortauthors{Folatelli et al.}

\begin{document}

\title{The Carnegie Supernova Project: Analysis of the First
  Sample of Low-Redshift Type-Ia Supernovae\altaffilmark{1}}

\author{Gast\'on~Folatelli\altaffilmark{2,3},
M.~M.~Phillips\altaffilmark{2},
Christopher~R.~Burns\altaffilmark{4},
Carlos~Contreras\altaffilmark{2},
Mario~Hamuy\altaffilmark{3},
W.~L.~Freedman\altaffilmark{4},
S.~E.~Persson\altaffilmark{4},
Maximilian~Stritzinger\altaffilmark{2,9},
Nicholas~B.~Suntzeff\altaffilmark{5,6},
Kevin~Krisciunas\altaffilmark{5,6},
Luis~Boldt\altaffilmark{2},
Sergio~Gonz\'alez\altaffilmark{2},
Wojtek~Krzeminski\altaffilmark{2},
Nidia~Morrell\altaffilmark{2}, 
Miguel~Roth\altaffilmark{2},
Francisco~Salgado\altaffilmark{2,3},
Barry~F.~Madore\altaffilmark{4,7},
David~Murphy\altaffilmark{4},
Pamela~Wyatt\altaffilmark{7},
Weidong~Li\altaffilmark{8},
Alexei~V.~Filippenko\altaffilmark{8}, and
Nicole~Miller\altaffilmark{3}
}

\altaffiltext{1}{This paper includes data gathered with the 
6.5~m Magellan Telescopes located at Las Campanas Observatory, Chile.}
\altaffiltext{2}{Las Campanas Observatory, Carnegie Observatories,
  Casilla 601, La Serena, Chile.}
\altaffiltext{3}{Universidad de Chile, Departamento de Astronom\'{\i}a,
  Casilla 36-D, Santiago, Chile.}
\altaffiltext{4}{Observatories of the Carnegie Institution of
  Washington, 813 Santa Barbara St., Pasadena, CA 91101.}
\altaffiltext{5}{Texas A\&M University, Physics Department, College
  Station, TX 77843-4242.}
\altaffiltext{6}{Mitchell Institute for Fundamental Physics and Astronomy.}
\altaffiltext{7}{Infrared Processing and Analysis Center, Caltech/Jet
  Propulsion Laboratory, Pasadena, CA 91125.}
\altaffiltext{8}{Department of Astronomy, University of California,
  Berkeley, CA 94720-3411.}
\altaffiltext{9}{Dark Cosmology Centre, Niels Bohr Institute, University
of Copenhagen, Juliane Maries Vej 30, 2100 Copenhagen \O, Denmark.}


\begin{abstract}
\noindent
An analysis of the first set of low-redshift ($z$$<$$0.08$) Type~Ia 
supernovae monitored by the Carnegie Supernova Project between 2004 and 2006
is presented. The data consist of well-sampled, high-precision optical 
($ugriBV$) and near-infrared (NIR; $YJHK_s$) light curves in a well-understood 
photometric system.  Methods are described for deriving light-curve 
parameters, and for building template light curves which are used to fit 
Type~Ia supernova data in the $ugriBVYJH$ bands.  The intrinsic colors at 
maximum light are calibrated using a subsample of supernovae assumed to have 
suffered little or no reddening, enabling color excesses to be estimated for 
the full sample. The optical--NIR color excesses allow the properties of the 
reddening law in the host galaxies to be studied.  A low average value of 
the total-to-selective absorption coefficient, 
$R_V \approx 1.7$, is derived when 
using the entire sample of supernovae.  However, when the two highly reddened 
supernovae (SN~2005A and SN~2006X) in the sample are excluded, a value 
$R_V \approx 3.2$ is obtained, similar to the standard value for the Galaxy.  
The red colors of these two events are well matched by a model where multiple 
scattering of photons by circumstellar dust steepens the effective extinction 
law.  The absolute peak magnitudes of the supernovae are studied in all bands 
using a two-parameter linear fit to the decline rates and the colors at 
maximum light, or alternatively, the color excesses.  In both cases, similar 
results are obtained with dispersions in absolute magnitude of 0.12--0.16~mag, 
depending on the specific filter-color combination.  
In contrast to the results obtained from the comparison of the 
color excesses, these fits of absolute magnitude give $R_V \approx$ 1--2 when the 
dispersion is minimized, even when the two highly reddened supernovae are 
excluded.  This discrepancy suggests that, beyond the ``normal'' interstellar
reddening produced in the host galaxies, there is an intrinsic dispersion in 
the colors of Type~Ia supernovae which is correlated with luminosity but 
independent of the decline rate.  Finally, a Hubble diagram for the 
best-observed subsample of supernovae is produced by combining the results 
of the fits of absolute magnitude versus decline rate and color excess for 
each filter.  The resulting scatter of 0.12~mag appears to be limited by the
peculiar velocities of the host galaxies as evidenced by the strong
correlation between the distance-modulus residuals observed in the individual
filters.  The implication is that the actual precision of Type~Ia supernovae 
distances is 3--4\%.
\end{abstract}


\section{INTRODUCTION}
\label{sec:intro}

The discovery in the late-1990s from measurements of distant Type~Ia
supernovae (SNe~Ia) that the expansion of the Universe is currently 
accelerating \citep{riess98,perlmutter99} has given rise to an exciting new 
era in cosmology.  The observations appear to require a 
previously unrecognized form of energy density (``dark energy'') which 
today is the dominant constituent of the Universe.
In recent years, observations of SNe~Ia to increasingly higher redshifts 
have confirmed this finding \citep[e.g.,][]{riess07,astier06,wood-vasey07}.  
Measurements of the angular power spectrum of the cosmic microwave background 
radiation provide independent confirmation, favoring a flat geometry for 
the Universe with a dominant 73\% of the energy density in the form of dark 
energy \citep[e.g., see][]{spergel07}.  Baryon acoustic oscillations
\citep{eisenstein05} and measurements of the ratio of X-ray-emitting gas to 
total mass in galaxy clusters \citep{allen04,allen07} have also provided
convincing confirmation of this picture.

Effort is currently focused on determining the nature of dark energy
by measuring an equation-of-state parameter of the form $w = P/(\rho c^2$), and 
its time derivative $\dot{w}$. Several ongoing and future projects aim
at filling the Hubble diagram with high-redshift ($z>0.1$) SNe~Ia in
order to measure $w$ and $\dot{w}$ with sufficient precision to
distinguish among the various proposed models of dark energy. However, 
these measurements require an improved reference set of SN~Ia observations
at low redshift. Surprisingly, the currently available low-redshift
datasets are neither sufficiently numerous nor homogeneous to adequately
complement the high-redshift data.

The Carnegie Supernova Project (CSP) is one of several ongoing efforts 
to improve the quality of the low-redshift data.\footnote[9]{See
\citet{contreras09} for a summary of other groups that are producing large
databases of low-redshift SN~Ia light curves.}  Over the five-year period 
ending in May~2009, the CSP has obtained densely sampled
optical {\em and}  
near-infrared (NIR) light curves of $\sim$100 SNe~Ia in a well-understood, 
homogeneous photometric system \citep[][hereafter H06]{hamuy06}. 
The NIR was included specifically to address the unknown intrinsic colors
and interstellar reddening to SNe~Ia.  This work presents an analysis of 
the first release of SNe~Ia by the CSP.
 
In \S~\ref{sec:lcs}, we present an analysis of the light curves, both optical 
and NIR, and introduce the methods used to derive parameters which serve to 
determine distances. In \S~\ref{sec:redd}, the non-trivial issue of measuring 
extinction is considered, and the nature of the reddening law in the host 
galaxies is studied. In \S~\ref{sec:absmag}, the properties of SNe~Ia as 
{\em standardizable\,} candles is examined by using Hubble-flow distances to 
fit the relationship between absolute peak magnitudes, decline rates, and 
colors or reddenings for different bands.  Finally, in \S~\ref{sec:concl}, 
the implications of our findings are discussed.

This paper is the second of three companion papers presenting first
results from CSP observations of low- and high-redshift SNe~Ia.  The 
low-redshift data used in the present analysis are described in detail 
in the first paper by
\citet[][hereafter C09]{contreras09}.  In the third paper, NIR light 
curves of 35 high-redshift SNe~Ia observed with the Magellan Baade 6.5-m 
telescope are combined with the low-redshift data in order to construct the 
first rest-frame $i$-band Hubble diagram of SNe~Ia \citep{freedman09}.

\section{PROPERTIES OF THE LIGHT CURVES}
\label{sec:lcs}

The CSP carries out its follow-up observations with the Las Campanas 
Observatory (LCO) Swope 1-m telescope using optical $ugriBV$ bands, 
and NIR $YJH$ bands. Additionally, $YJHK_s$ photometry is obtained with 
the Wide-field Infrared Camera (WIRC) at the du~Pont 2.5~m telescope of LCO. 
As mentioned in H06, the NIR monitoring of the CSP during the first year 
of observations was generally not as thorough as that for the optical due 
to the fact that RetroCam, the NIR camera at the Swope telescope, was not 
available until September 2005.  This situation was much improved during
the final four campaigns of the project.

The photometry of each SN is performed relative to nearby field stars which
are, in turn, calibrated from observations of standard stars in the 
\cite{landolt92} ($BV$), \cite{smith02} ($ugri$), and \cite{persson04} 
($YJHK_s$) systems during several photometric nights. Point-spread function 
(PSF) photometry of the SN is measured after the underlying light 
from the host galaxy has been subtracted from every follow-up image.  
Details of the measurement procedures and photometric system are 
given by H06.  Magnitudes are expressed in the natural photometric 
system of the Swope+CSP bands, as explained in C09. In accordance with 
that paper, in the remainder of this work we will refer to our optical 
natural photometry as $ugriBV$. The final photometry for the SNe is
published by C09.

In the following, the methods employed for deriving
light-curve parameters for the SN sample are presented, including the 
construction of template light curves for the $ugriBVYJH$ bands.

\subsection{Direct Measurements}
\label{sec:direc}

The SNe~Ia included in this analysis are listed in Table~\ref{tab:sne}.  
Twenty-six of these SNe have optical observations
beginning before $B$-band maximum light and with a cadence of a few
days or less, which allows direct functional fits to their data. 
A subsample of nine SNe in this group also have enough $YJH$ coverage 
to allow the same type of fits around maximum light. These fits provide 
direct measurements of light-curve parameters such as peak magnitude
$m^{\mathrm{max}}$, time of maximum $t^{\mathrm{max}}$, and decline 
rate $\dm$, the latter measured from the $B$-band light curves
\citep{phillips93}.\footnote[10]{Throughout this paper we use the 
superscript or subscript ``max'' to denote the quantity measured at the 
time of maximum light {\em in the corresponding band}.}
We therefore decided to directly measure the 
light-curve shape parameter, $\dm$, whenever possible rather than derive it 
from template 
fits in order to tie our luminosity corrections to a purely empirical 
parameter.  These resulting parameters are given in
Table~\ref{tab:lcpar}, labeled as ``Spline'' in column 7. The fits
also serve to build a set of template $ugriBVYJH$ light curves 
which may be used to fit more poorly
observed SNe, as described in \S~\ref{sec:tmplfit}.  

Spline functions are preferred for these direct fits since they are more 
readily adaptable to the shapes occurring in SN~Ia light curves than are 
other functions such as polynomials. We use the FITPACK library 
\citep{dierckx93}, which employs a variable number of knots 
depending on the sampling and precision of the data.  These spline functions 
allow the whole time span of the observations to be fit without introducing
undesirable oscillations where the coverage is less dense.  The high quality 
of the photometry and the good time sampling of the observations ensures the
accuracy of the fits.  Errors in the parameters derived from the spline
fits are modeled via a Monte Carlo calculation performed by randomly 
perturbing each data point according to its quoted photometric error 
and then measuring the scatter of the output measurements.

As mentioned by \citet{contreras09}, one can expect a certain
  degree of correlation among light-curve points introduced by the
  host-galaxy subtraction process because the same template image is
  used to subtract all of the follow-up images. Not taking the correlation
  into account may
  lead to underestimating the derived uncertainties of the light-curve
  parameters. The effect is expected to be more noticeable when the
  host-galaxy subtraction is more uncertain, and the photometric
  precision and coverage are larger. 
  We have estimated this effect by computing the covariance matrix
  for the data points 
  and including this in the Monte Carlo calculations.
  The covariance matrix is estimated by examining the photometry of artificial
  stars which are added to the subtracted images at fixed locations
  around the SN and which are set to have the same flux as the SN at
  each epoch. In general, these experiments have yielded results in
  agreement with the case in which the correlations are
  neglected. Only in a few cases have the uncertainties in the peak
  magnitudes increased by more than a few thousandths of a
  magnitude. Therefore, we consider this effect to be negligible for the
  current analysis.

K-corrections are applied to the observed magnitudes to convert these
to the rest frame.  For this
purpose, we utilize a library of SN~Ia template spectra provided by
\citet{hsiao07} in the optical, and \citet{hsiao09} in the
NIR. The K-corrections are computed simultaneously with the
light-curve fits.  The template spectra corresponding to the epochs of the
observations are first warped by a smooth function of wavelength in order to 
make the synthetic colors match the observed ones for all available bands. 
Since the template spectra cover a limited range of epochs from 
$-15$ to $+85$ days relative to $B$-band maximum, observations outside
these limits are ignored.

The accuracy of the K-corrections can be tested by comparing them with 
values computed from the 194 follow-up spectra obtained by the CSP of the 
sample SNe. The spectra 
typically cover the optical $griBV$ bands and a range of epochs from $-12$ 
to over $+100$ days with respect to $B$-band maximum light. In most cases, 
the agreement with the K-corrections derived from the template spectra is 
better than $\pm0.02$~mag.  Unfortunately, the lack of spectral coverage in 
the $u$ and NIR bands prevents this test from being performed in those bands.

\begin{figure}
\epsscale{1.0}
\plotone{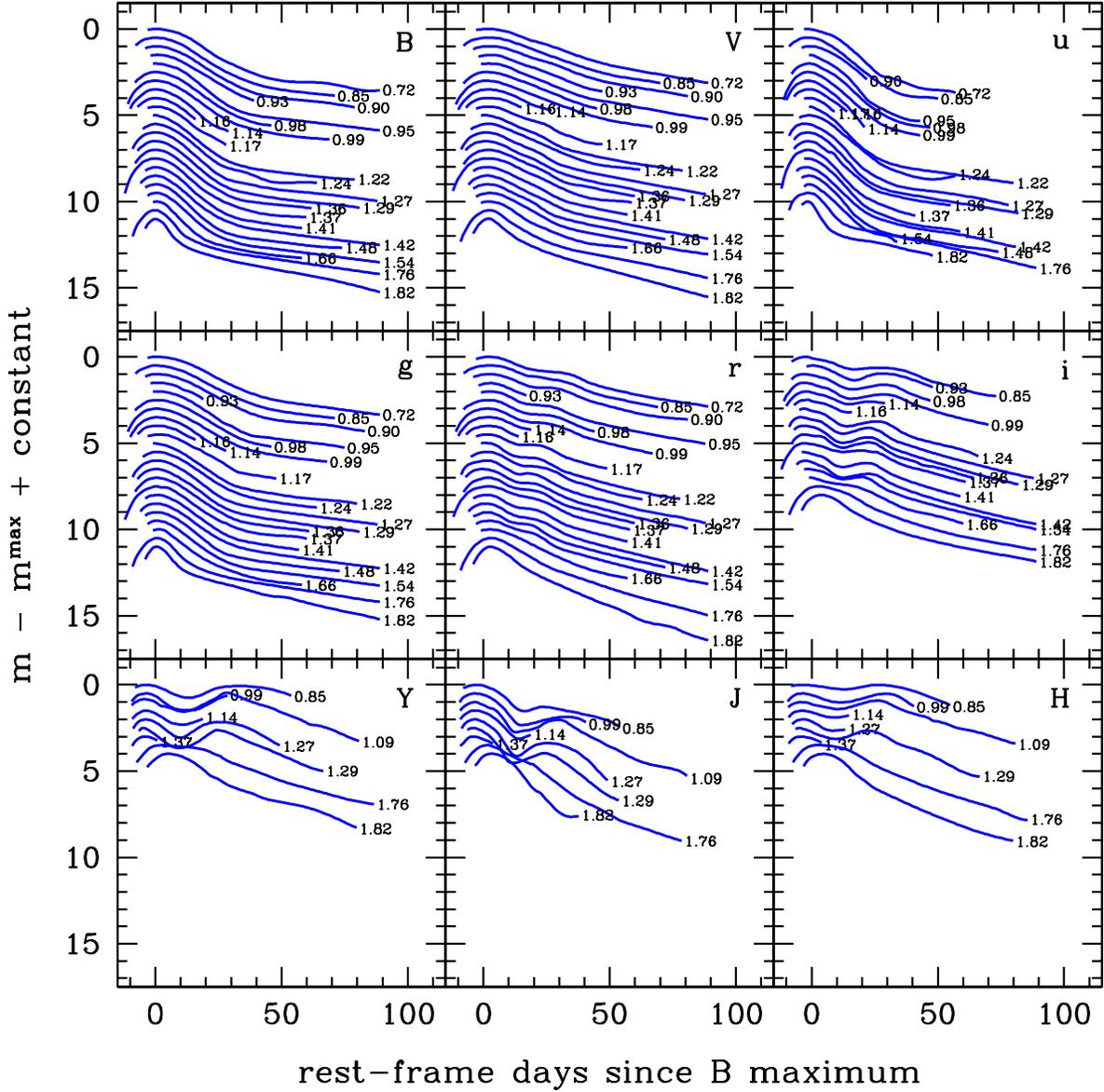}
\caption{Spline fits to the $ugriBVYJH$ light curves of the SNe
    used as templates. The light curves are sorted by $\dm$ and are
    plotted relative to the magnitude at maximum light in each band, 
    shifted by $0.5$~mag with respect to each other. The values
    of $\dm$ label each curve at the right end.\label{fig:templcs}}
\end{figure}

Figure~\ref{fig:templcs} shows the spline fits for all
SNe with pre-maximum data in each band. It should be noted that since
normal SNe~Ia peak earlier in $u$ and $i$ than they peak in $B$ (see
\S~\ref{sec:lcpar} and Figure~\ref{fig:tmaxd}), some SNe were not
observed before maximum light in these two bands, reducing the sample
from 26 in $B$, to 21 and 19 (respectively) in $u$ and $i$. 

The value of $\dm$ obtained for each SN is indicated next to the
corresponding curves in Figure~\ref{fig:templcs}. The range of decline
rates covered by the sample is $0.7 \,<\, \dm \,<\, 1.9$ for the
$BVgr$ bands, and $0.8 \,<\, \dm \,<\, 1.9$ for the $uiYJH$ bands,
which corresponds well to the full range of observed values of 
$\dm$ for SNe~Ia.

The optical data for these 26 SNe start between $-12$ and
$-1$ days relative to $B$-band maximum light, and end between $17$ and
over $100$ days past maximum, typically extending more than one month
after maximum light. The $YJH$ data for the nine SNe with pre-maximum
coverage start between $-10$ and $-1$ days, and end between $5$ and $90$
days past maximum.

\begin{figure}
\includegraphics[angle=-90,width=16cm]{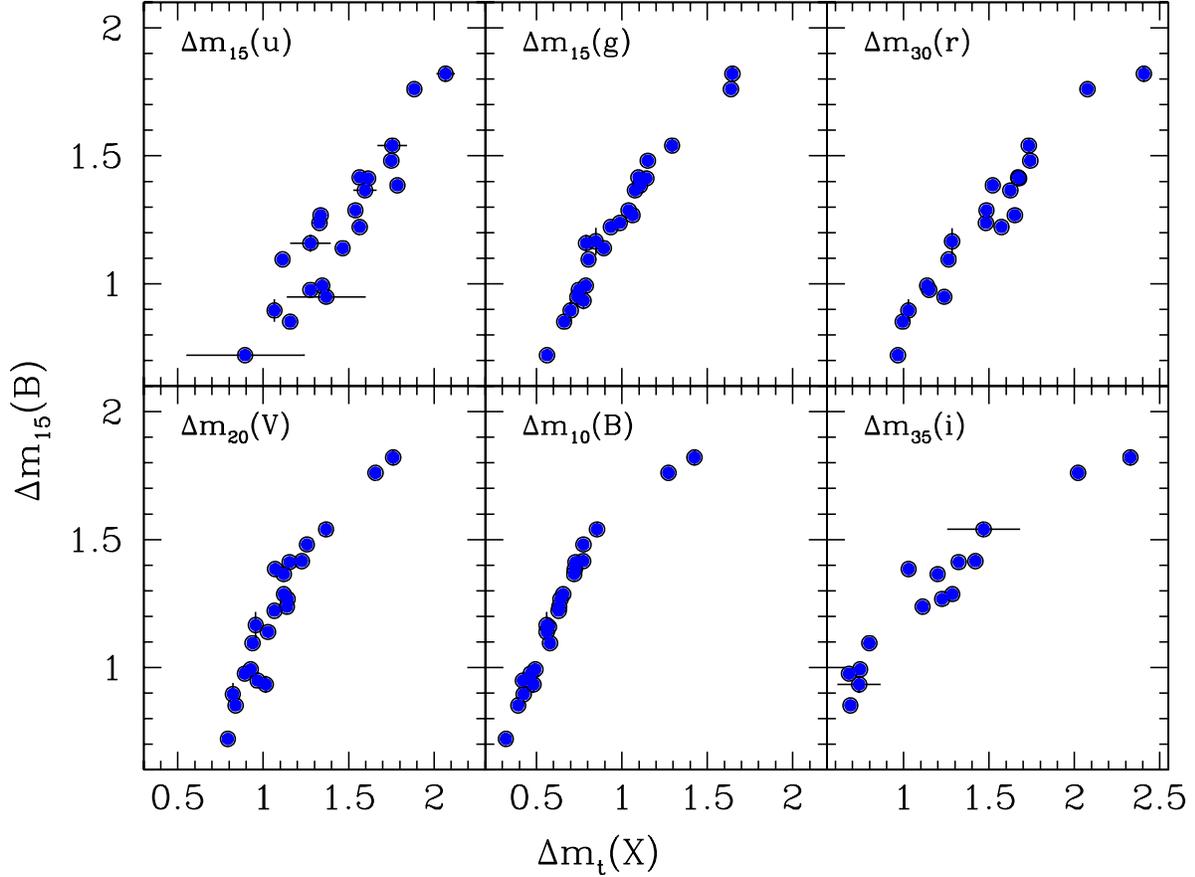}
\caption{Comparison of decline rates $\dm$ with similar quantities
  obtained from other optical bands and epochs since maximum light. 
  Decline rates $\Delta m_t(X)$ are
  measured from the light curve in $X$ as the difference in magnitudes
  between $t$ rest-frame days since maximum light (in that band) and 
  the peak magnitude. Only results of spline-function fits to SNe with
  pre-maximum coverage are shown for each band.  Except where plotted,
  error bars are smaller than the points.\label{fig:declro}}
\end{figure}

\begin{figure}
\epsscale{0.75}
\plotone{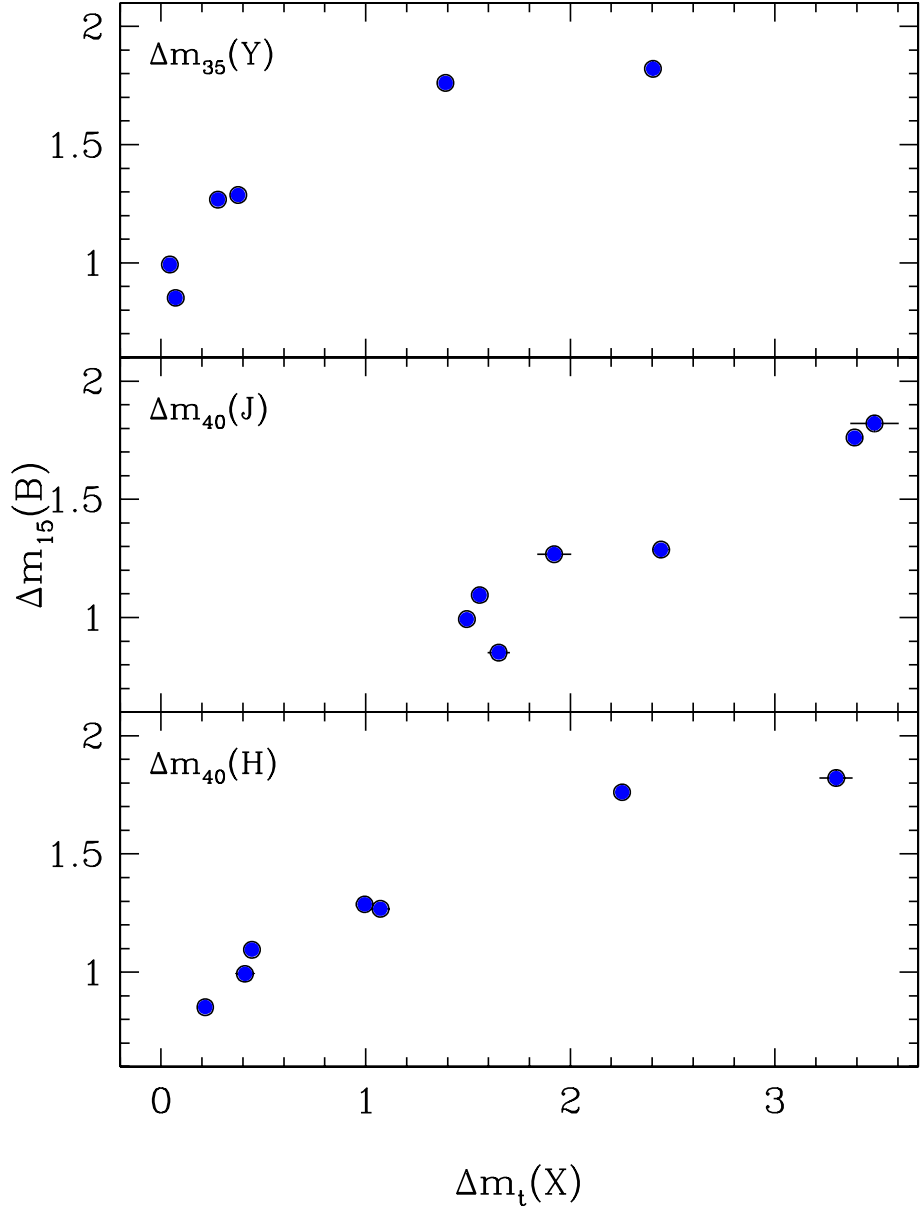}
\caption{Comparison of decline rates $\dm$ with similar quantities
  obtained from $YJH$ bands. Decline rates $\Delta m_t(X)$ are
  measured from the light curve in $X$ as the difference in magnitudes
  between $t$ rest-frame days since maximum light (in that band) and
  the peak magnitude. Only results of spline-function fits to SNe with
  pre-maximum coverage are shown for each band.  Except where plotted,
  error bars are smaller than the points.\label{fig:declri}}
\end{figure}

\begin{figure}
\epsscale{1.0}
\plotone{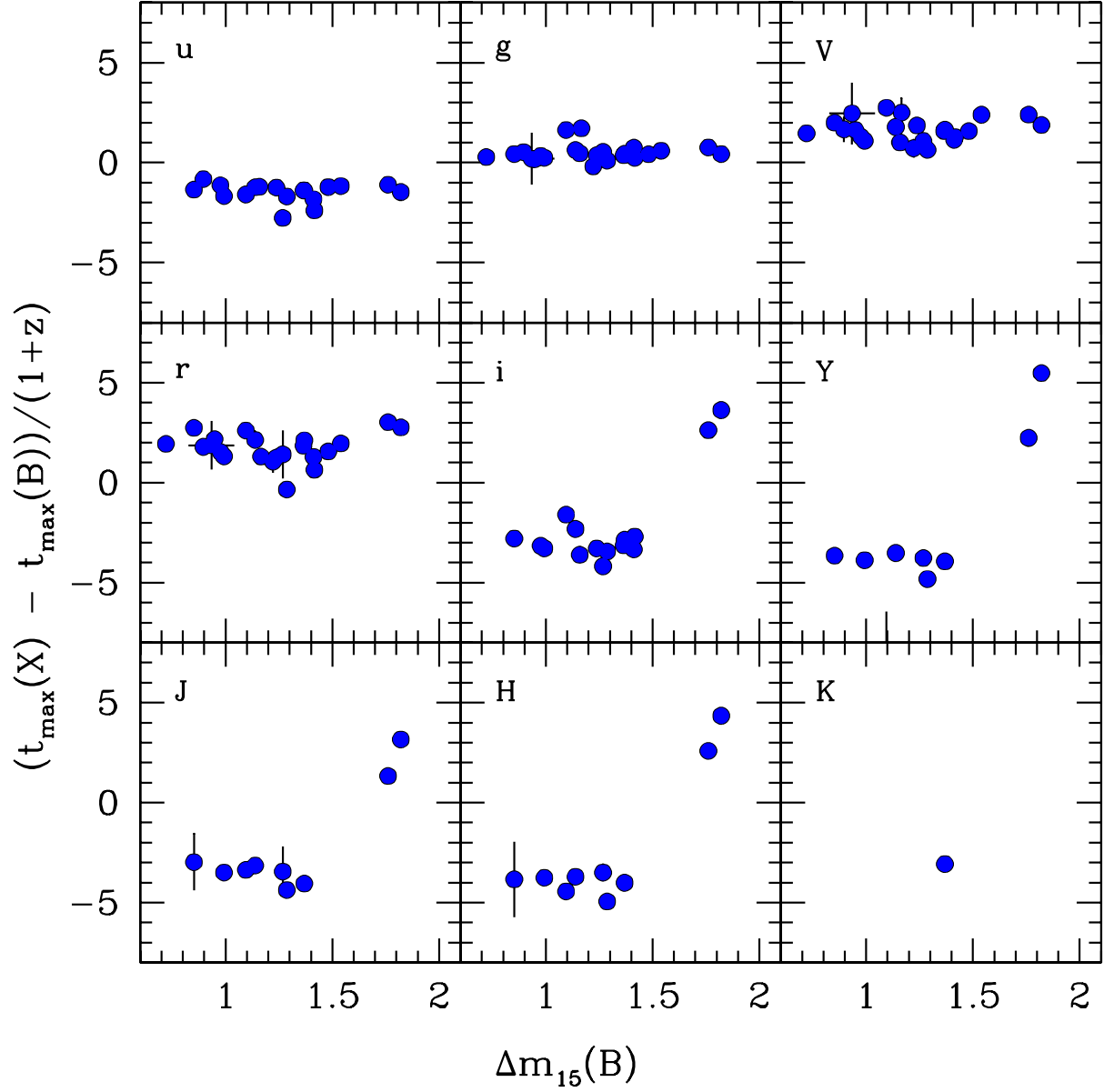}
\caption{Differences in rest-frame days between the time of maximum in
  $ugriVYJHK_s$ with respect to $B$. Only results of spline-function
  fits to SNe with pre-maximum coverage are shown for each
  band.\label{fig:tmaxd}}
\end{figure}

Following the work of \citet{hamuy96b} and \citet{elias-rosa08}, 
we searched for alternative 
characterizations of the SN~Ia decline rates by comparing $\dm$ with 
values of $\Delta m_{t}(X)$ for different bands $X$, and times $t$, 
since maximum. Figure~\ref{fig:declro} shows
the cases for $ugriBV$ which provide the strongest
correlations with $\dm$; Figure~\ref{fig:declri} shows the same
results for the NIR $YJH$ bands.  In general, at longer wavelengths, the
best correlations with $\dm$ are found
at later epochs. Even in the case of the NIR $iYJH$ bands, where the
double-peaked shape of the light curves complicates matters, reasonable 
correlations with $\dm$ are obtained at $t \sim 35$--$40$ days --- i.e., 
during the radiative decline.

The behavior of the timing of maximum light for different bands with 
respect to $B$ band was also studied. This is 
shown in Figure~\ref{fig:tmaxd}. For $ugVr$,
the delay (or advance for $u$) is roughly constant for SNe of all
decline rates. In the $iYJH$ bands, there is a nearly constant advance of
2--4 days for SNe with $\dm < 1.5$, whereas there is a delay of several 
days for SNe with $\dm \approx 1.8$.  Note that the scatter in 
Figure~\ref{fig:tmaxd} is generally greater than the error bars, implying 
that there is an intrinsic component.

\subsection{Template Light-Curve Fits}
\label{sec:tmplfit}

The availability of such a complete set of $ugriBVYJH$ light
curves with pre-maximum coverage allows template light curves to be 
built which can be used to derive light-curve
parameters for SNe with inadequate coverage around maximum
brightness.  For this purpose, we used the SNOOPy (SuperNovae in Object-Oriented
Python) package \citep{burns09}, which is derived from the the multi-filter
light-curve fitting method developed by \citet{prieto06}, with a few modifications.  
The technique for generating templates is briefly summarized below, and in 
more detail by \citet{burns09}. 

The SNe that were fit with spline functions (see \S~\ref{sec:direc})
form the training dataset. The time
of $B$-band maximum light $t^{\mathrm{max}}(B)$,
the decline rate measured as $\dm$, and the peak magnitude
$m_X^{\mathrm{max}}$ in each band $X$ are used to
place the photometric data points of the training set
in a three-dimensional (3-D) space for each band defined by the following
variables: 
\begin{enumerate}
\item The rest-frame epoch with respect to $B$-band maximum, $\Delta t =
[t-t^{\mathrm{max}}(B)]/(1+z)$, where $t$ is the time of the
observation and $z$ is the heliocentric redshift of the SN;
\item the decline-rate parameter, $\dm$, considered as a single parameter to
characterize all bands; and
\item the K-corrected magnitude relative to maximum light, $\Delta m = m_X(t) -
m_X^{\mathrm{max}}- K_X(t)$, for each observed magnitude $m_X(t)$.
\end{enumerate}

SN~2006X was removed from the training dataset in $ugriBV$ because of its 
peculiar behavior, especially in the $B$ band, which showed an abnormally 
flat decline rate beginning one month after maximum light. This morphology 
has recently been explained as a light echo produced by dust surrounding 
this heavily reddened SN \citep{wangx08b}.  However, since there is no
evidence that the NIR light curves of SN~2006X suffered such contamination,
the data for this event {\em were} maintained
in the training set for building $YJH$ templates.

The data points of the training set 
define a surface in the 3-D space
($\Delta t$, $\dm$, $\Delta m$) for each band.  To produce a smooth
interpolation to any point in the 3-D space, we use a 2-D variation of the 
``gloess'' algorithm, which is a Gaussian-windowed and error-weighted 
extension of data-smoothing methods outlined by \citet{cleveland79} that was
first implemented by \citet{persson04} for fitting Cepheid light curves.
This is the main difference between the method of \citet{prieto06} and 
that of the SNOOPy package \citep{burns09}.  While the former method 
uses the fit spline functions as templates and interpolates among them, 
SNOOPy performs an interpolation among
the {\em data points\,} of the SNe. The gloess algorithm ensures a
smooth interpolation even with heterogeneously sampled data. 

The SNOOPy package generates templates for a given value of
$\dm$ by making a slice through the 
($\Delta t$, $\dm$, $\Delta m$) surface, interpolating along a constant $\dm$ 
line.  The best-fit time of maximum light $t^{\mathrm{max}}$, decline rate
$\dm$, and peak magnitudes $m^{\mathrm{max}}$ are then determined by fitting 
these templates to the $ugriBVYJH$ data of each SN via $\chi^2$
minimization, with K-corrections computed in an iterative fashion, as 
described in \S~\ref{sec:direc}. Uncertainties in the fit
  parameters are taken from the diagonals of the covariance matrix output 
from the Levenbert-Marquardt least-square routine used to fit the
templates \citep{burns09}. Additionally, we have tested the possible
  effect of correlated errors in the photometry due to the use of a
  single host-galaxy image in the subtraction process
  \citep[see][]{contreras09}. Similarly to the results of the tests
  for spline fits that are given in \S~\ref{sec:direc}, in the case of
  template fits we find that the effect is very small in all but a
  few examples. We have thus decided to use the uncertainties that are
  derived from uncorrelated errors.

This way, it is possible to derive the light-curve 
parameters for each SN, including the objects in our sample which were 
not observed until after maximum light. Table~\ref{tab:lcpar} lists the 
results of these template fits for the cases in which a spline function could
not be fit; these are labeled as ``Templ.'' in column 7. 
The template fits themselves are shown by C09 together with 
the observed light curves for all SNe in the sample.

\begin{figure}
\epsscale{1.1}
\plottwo{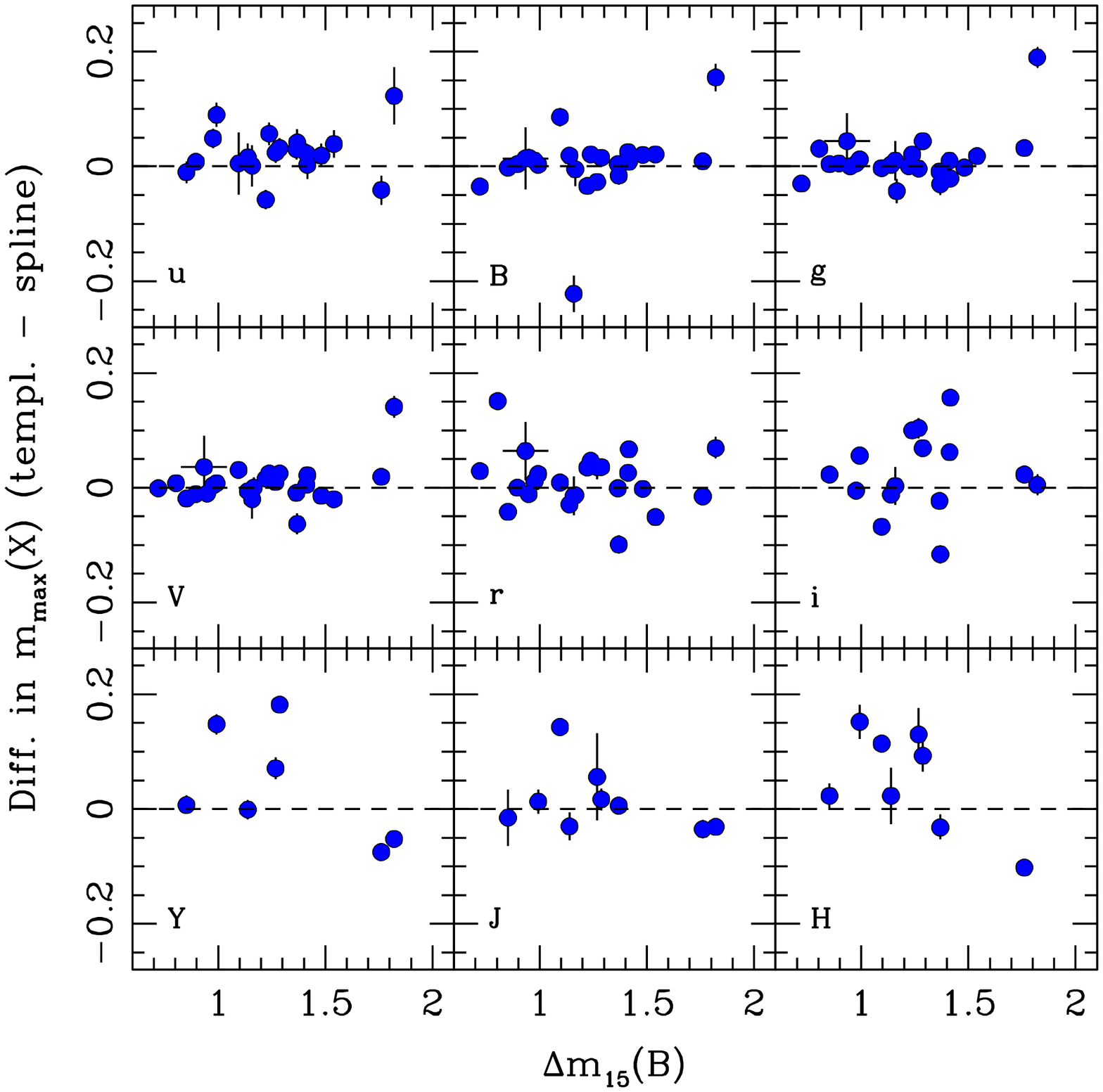}{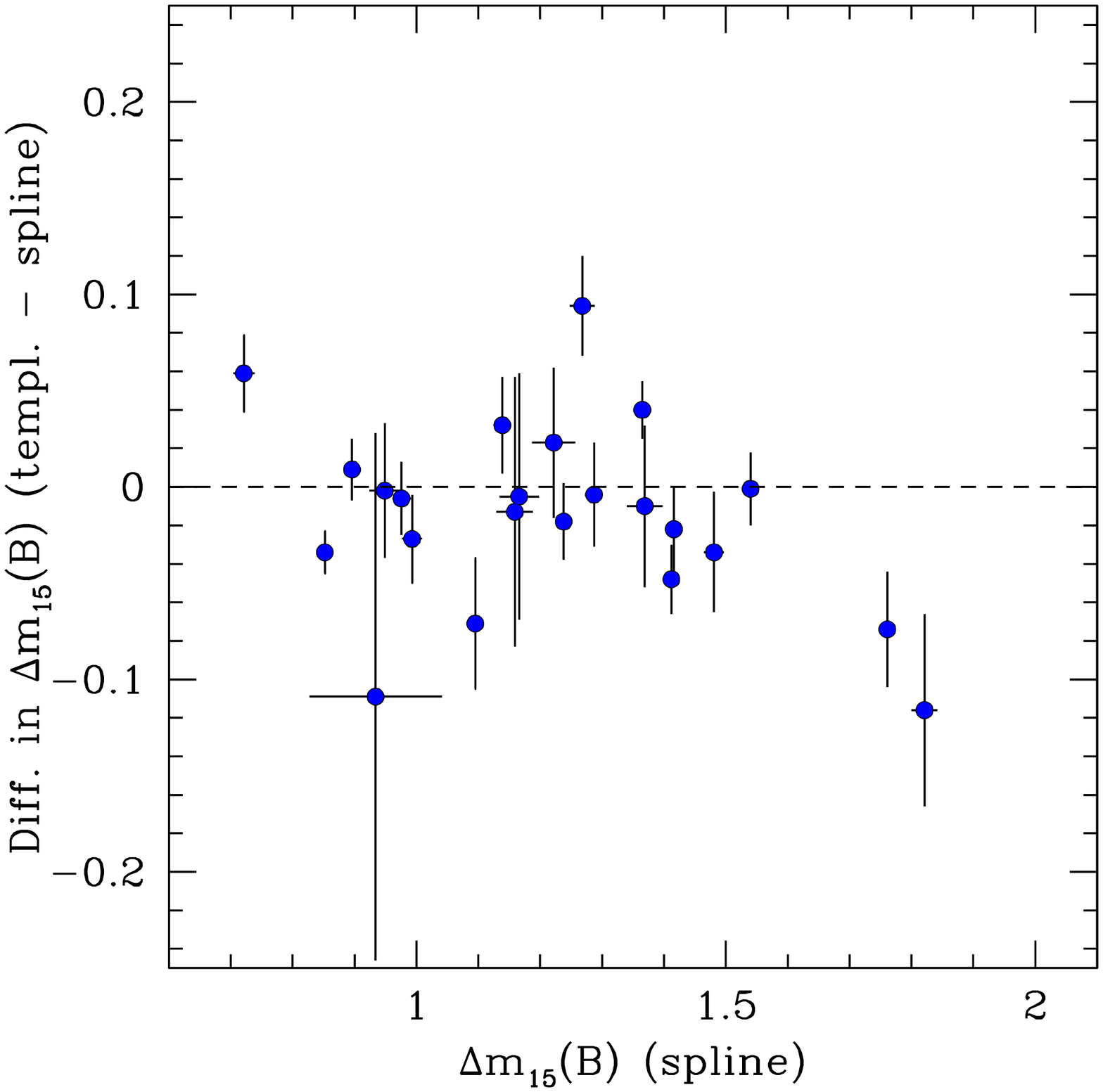}
\caption{({\em left}) Plot of the difference in the peak magnitudes
derived from template versus spline
fits as a function of filter for the subset of best-observed SNe.
({\em right}) Difference in $\dm$ derived from template versus spline
fits for the same sample.  
\label{fig:spline_templ}}
\end{figure}

In order to test the accuracy of the method, template fits were performed for 
the subset of well-observed SNe and the resulting parameters were compared
with those obtained directly from spline fits (\S~\ref{sec:direc}). These
comparisons are shown in Figure~\ref{fig:spline_templ} for the peak
magnitudes in $ugriBVYJH$, and for the decline-rate parameter $\dm$.  
The average difference in peak magnitudes is $\pm0.05$~mag or less for all 
of the optical filters except the $i$ band, with no obvious systematic 
differences except, perhaps, in $u$.
The dispersion in the $iYJH$ bands is $\sim \pm0.1$~mag, with some evidence
for systematic differences in $YJH$.  A similar situation holds for the
differences in the time of maximum as measured with spline and template
fits, where good consistency ($\pm0.6$~days or less) is found in $ugrBV$,
but the $iYJH$ bands yield worse agreement.   The relatively poor precision 
of the  $YJH$ template fits is at least partly due to the small sample used to
derive the templates.  However, the large dispersion in $i$ is
the product of variations in the morphology of the secondary 
maximum (see \S~\ref{sec:secmax}).  Figure~\ref{fig:spline_templ}  shows that,
except at the fastest decline rates, there is
reasonable consistency between the decline rate as measured by spline 
and template fits.

The uncertainties associated with extrapolating the templates to obtain 
peak magnitudes and decline rates for those SNe caught after maximum light are
more difficult to quantify.  Clearly, the later the observations begin, the
larger are the possible extrapolation errors.  One way to approach this
issue is to take the spline fits of the light curves of the SNe in the sample 
that were well-observed before and after maximum, resample these to simulate
a SN whose observations did not begin until after maximum, and then fit these
data using the $ugriBVYJH$ templates to recover the time of $B$ maximum, $\dm$, 
and the maximum-light magnitudes.  Calculations of this type will be presented
in more detail in a future paper \citep{burns09}, but for the representative
case of a SN for which coverage began $\sim$1 week after maximum and continued
until $\sim$70 days after maximum, we find a typical random uncertainty of 
$\sim$0.1~mag in the peak magnitude, averaged over all filters, and a much 
smaller systematic difference of $\sim$0.03~mag.  Experience shows that for 
SNe with photometry beginning later than this, template fits are not reliable.  
Fortunately, only one event in the \citet{contreras09} sample, SN~2004dt, 
falls into this category; hence, it is excluded from the analysis in the
present paper.

The lack of a sufficiently large set of SNe with good coverage around
maximum light in the $K_s$ band prevents the production of template light 
curves as was done for all other bands. Instead, we use the
polynomial template light curve introduced by \citet{krisciunas04a}. 
A {\em stretch}\, factor \citep{perlmutter97} based on the decline rate of 
each SN is applied 
to the time axis of the polynomial ---along with the redshift-dependent 
time dilation--- which is then fit to the data in
order to obtain a peak magnitude. This kind of
fit may be utilized for SNe with data in the range of validity of the
template, which is between $-12$ and $+10$ days
relative to $t^{\mathrm{max}}(B)$ for a SN of stretch factor $S = 1$. 
The specific formula employed to convert
between $\dm$ and  $S$ is that given by \citet{jha06}:

\begin{equation} 
\label{eq:dm15stretch}
S\,=\,\frac{3.06\,-\,\dm}{2.04}.
\end{equation}

\subsection{The Secondary Maximum}
\label{sec:secmax}

\citet{hoeflich95} and \citet{pinto00} were the first to attempt a theoretical
understanding of the NIR secondary maximum in SNe~Ia.  More recently,
\citet{kasen06} modeled the NIR light curves of SNe~Ia, arguing that the
timing and strength of the secondary maximum is a consequence of the 
ionization evolution of the iron-peak elements in the ejecta.  A specific 
prediction is that more luminous SNe should have a later and more prominent
secondary maximum. In order to test this, we examined
the delay and relative strength of the secondary maximum in the $i$-band
light curves of the CSP SNe with the best coverage. Using the
nomenclature of \citet{kasen06}, the following parameters were measured:
\begin{enumerate}
\item The phase of the peak of the secondary maximum with respect to 
$B$-band maximum, $t_2-t_B$;
\item the difference in magnitudes between the secondary maximum
and the local minimum between the primary and the
secondary maxima, $m_2-m_0$; and
\item  the difference in magnitudes between the primary and secondary
maxima, $m_1-m_2$.
\end{enumerate}

\begin{figure}
\epsscale{0.75}
\plotone{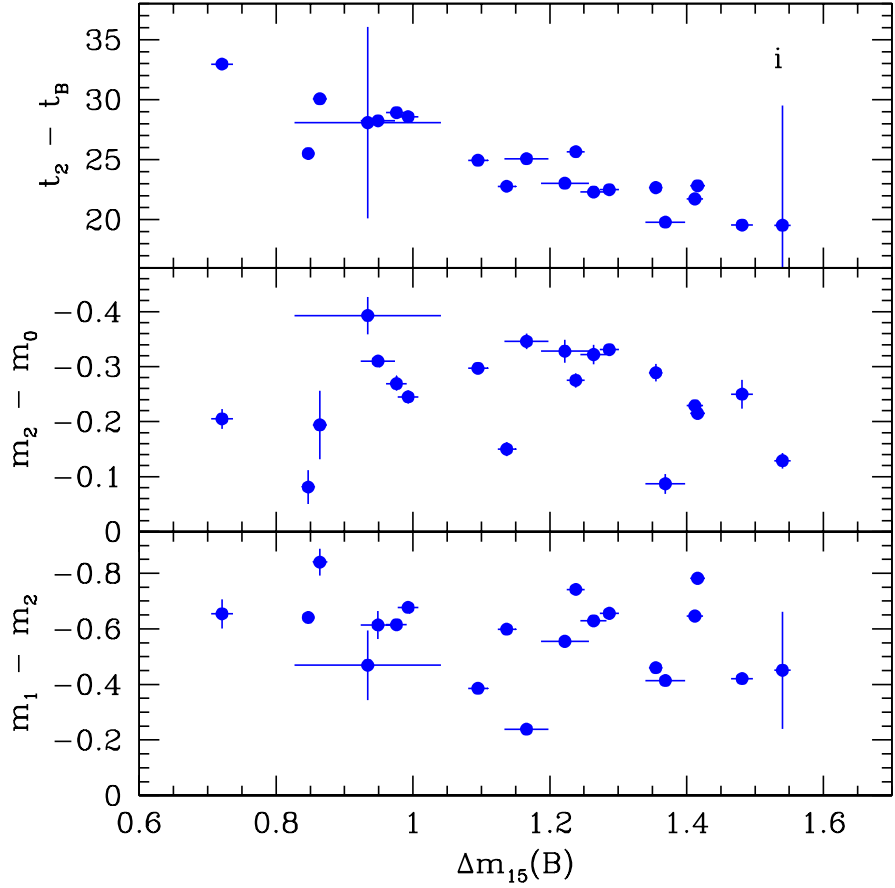}
\caption{Timing and strength of the secondary maximum in $i$ as a function of 
  the decline-rate parameter $\dm$ for the sample of best-observed
  SNe with pre-maximum data in each band. The top panel shows the 
  phase of the secondary maximum with respect to the $B$-band maximum. 
  The middle panel displays the difference in magnitudes between the 
  secondary maximum and the local minimum between the primary and
  secondary maxima. The bottom panel shows the difference in magnitude 
  between the primary and secondary maxima.
  \label{fig:max2}}
\end{figure}

Figure~\ref{fig:max2} shows the results plotted as a function of $\dm$. 
If the decline-rate parameter $\dm$ (or peak luminosity) is associated 
with the amount of $^{56}$Ni mass synthesized in the ejecta 
\citep{arnett82,arnett85,nugent95,stritzinger06}, 
then Figure~\ref{fig:max2} can be compared directly with Figure~11 of
\citet{kasen06}.
Indeed, the upper panel of Figure~\ref{fig:max2}
shows a strong correlation between $\dm$ and the {\em timing} of the
secondary maximum, confirming previous results
by \citet{hamuy96b}
and \citet{elias-rosa08}.  However, the middle and lower panels of
Figure~\ref{fig:max2} show that there is little evidence for a 
dependence between $\dm$ and the {\em strength} of the secondary 
maximum, as measured with respect to either the primary maximum or 
the local minimum between the two maxima. \citet{kasen06} points out
that both the timing and strength of the secondary maximum can also
be affected by the outward mixing of $^{56}$Ni in the ejecta, the amount
of stable iron group elements produced in the explosion, the progenitor
metallicity, and the abundance of calcium in the ejecta.  Of these,
mixing appears to have the greatest
effect on the {\em strength} of the secondary maximum, and
therefore may be responsible for the lack of obvious correlations in
the middle and bottom panels of Figure~\ref{fig:max2}.

\begin{figure}
\epsscale{0.9}
\plotone{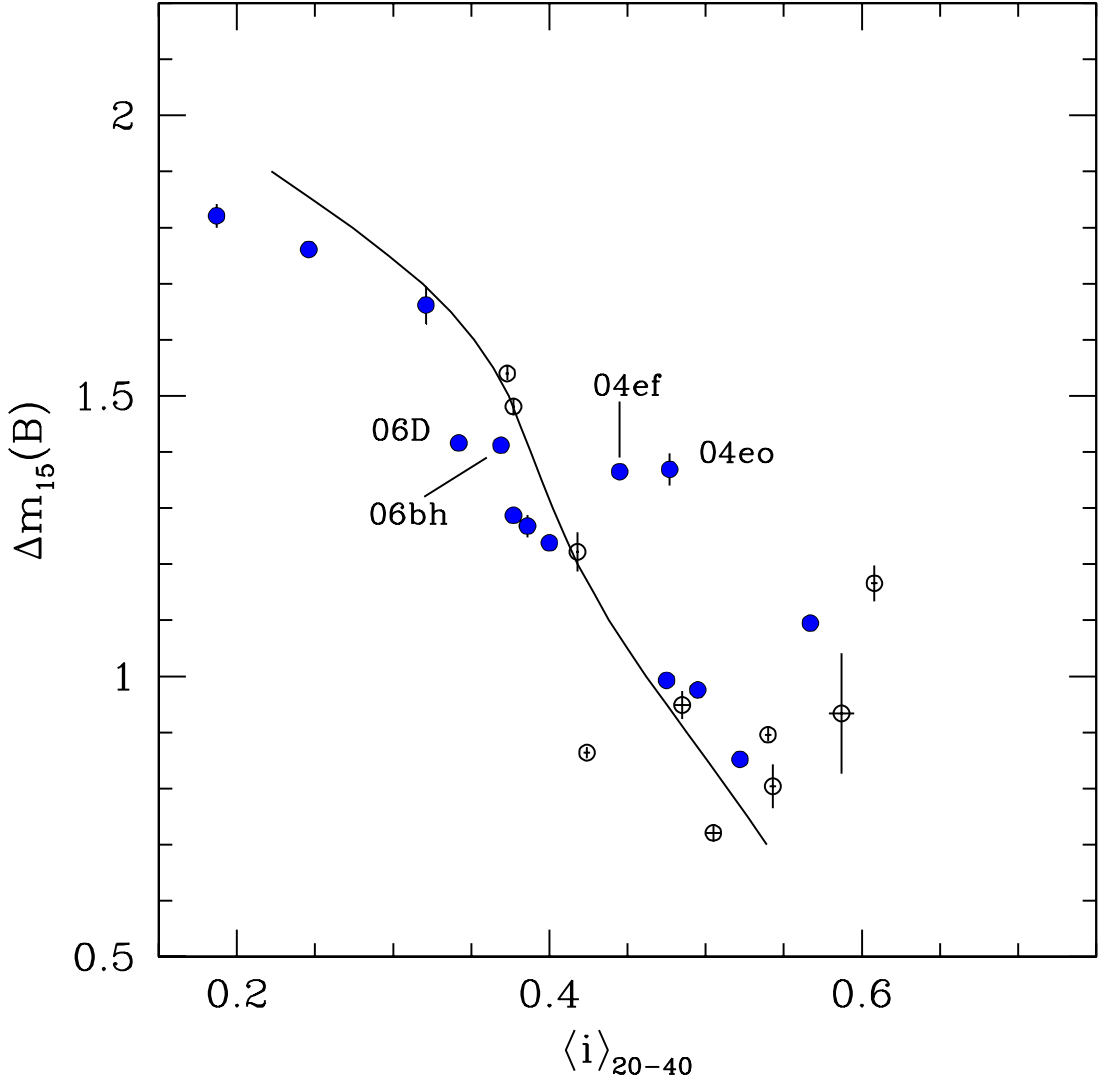}
\caption{The mean $i$-band flux 20 to 40 days after the time of $B$-band 
  maximum plotted versus $\dm$ for the best-observed subsample of CSP SNe.
  Solid symbols correspond to SNe with peak magnitudes measured from
  spline fits; open symbols are SNe with peak magnitudes derived from
  template fits.
  The solid line corresponds to the relation for the SNOOPy
  templates. Four SNe with similar $\dm$ but discrepant average
  $i$-band fluxes at the second maximum are labeled.
  \label{fig:iflux20_40}}
\end{figure}

\begin{figure}
\epsscale{1.0}
\plottwo{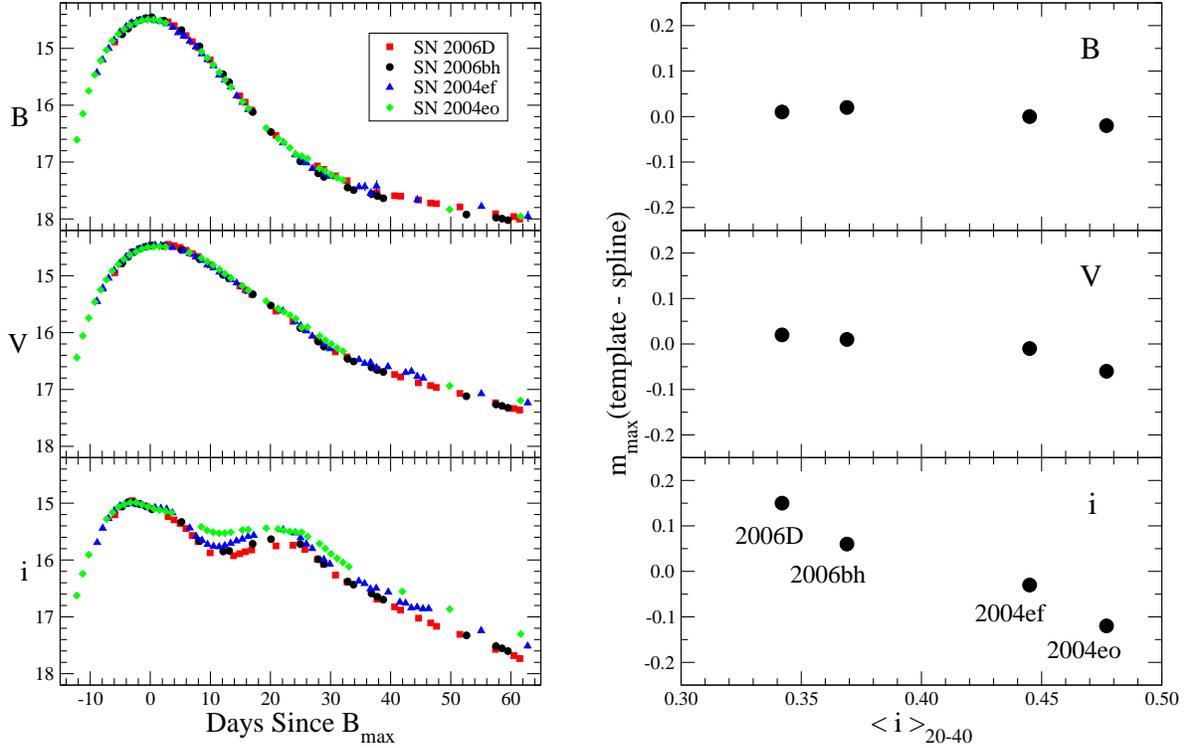}{f8b.eps}
\caption{({\em left}) Comparison of the $B$, $V$, and $i$ light curves of 
  SNe~2006D, 2006bh, 2004eo, and 2004ef.  The data for SNe 2006D, 
  2004eo, and 2004ef have been shifted to coincide at maximum with the
  observations 
  of SN~2006bh.  All four SNe have very similar decline rates in the range
  $\dm = $1.37--1.42 mag.  Note the close similarity of the $B$ and $V$ light 
  curves, but the real differences in the strength and morphology of the 
  secondary maxima in the $i$-band. ({\em right}) The difference between 
  the maximum-light magnitudes in the $BVi$ bands as measured from 
  template and spline fits for SNe~2006D, 2006bh, 2004eo, and 2004ef, 
  plotted as a function of the  mean $i$-band flux 20 to 40 days after the 
  time of $B$-band maximum.
  \label{fig:bvi_lcurves}}
\end{figure}

\citet{krisciunas01} suggested a different way of measuring the strength 
of the $I$-band second maximum which consists of converting the
light-curve observations to fluxes normalized to the flux at the primary 
maximum, fitting a high-order polynomial to these, and then determining
the mean flux (based on an integration of the polynomial) from 20 to 40~days 
after the time of $B$ maximum.  They found a good correlation between 
this parameter, $\langle I \rangle_{20-40}$, and the decline rate.  Similar
measurements were carried out for the $i$-band light curves of the 
best-observed subsample of CSP SNe, and plotted vs. $\dm$ in 
Figure~\ref{fig:iflux20_40}.  Illustrated as a solid line 
are the same measurements for the family of SNOOPy templates.  
There is a clear trend in this diagram in the sense that luminous, 
slowly declining SNe generally display stronger secondary maxima.
However, there is also a significant and real dispersion which is 
illustrated by the four labelled SNe --- 2006D, 2006bh, 2004ef, and 
2004eo.  The left half of Figure~\ref{fig:bvi_lcurves} shows that all four 
have very 
similar $B$ and $V$ light curves covering a narrow range of decline 
rates ($\dm =$ 1.37--1.42).  Nevertheless, there are significant 
differences in the strength and morphology of the $i$-band secondary 
maximum which, again, may reflect varying amounts of mixing of the 
$^{56}$Ni into the ejecta \citep{kasen06}.  

These variations in the strength of the secondary maximum
represent a significant problem for template fitting in the
$iYJHK$ bands, where the strength of the secondary maximum
is assumed to be a smoothly varying function of $\dm$.  
This is illustrated in the right half of Figure~\ref{fig:bvi_lcurves}, where
the difference in the maximum-light magnitudes in $BVi$ as derived from 
template and spline fits is plotted for the same four SNe (2006D, 
2006bh, 2004ef, and 2004eo) with nearly identical decline rates.
In the $B$ band, the template fits give results that are in excellent 
agreement ($\pm0.02$~mag) with the direct measurements from the 
spline fits.  The agreement in $V$ is nearly as good, but in the $i$ 
band, the template fits give errors as large as $\sim \pm0.15$~mag.
Figure~\ref{fig:bvi_lcurves} shows that these errors are a function
of the strength of the secondary maximum: template fits to SNe 
with a weak secondary maximum will yield maximum-light 
magnitudes that are too faint, whereas fits to SNe with a strong
secondary maximum will give maximum-light magnitudes that
are too bright.  This effect most likely explains the relatively 
large dispersion in the difference in the peak $i$ magnitudes
measured from template and spline fits observed in
Figure~\ref{fig:spline_templ}.  Until an observational
parameter can be found which makes these differences predictable, 
they will remain a significant impediment to the fitting of template 
light curves in the $iYJHK$ bands. Quantities such as
$\langle I \rangle_{20-40}$ or $m_2-m_0$ may prove to be such
a parameter, but a larger sample of SNe is needed to test
their utility.

\subsection{Light-Curve Parameters}
\label{sec:lcpar}

The results of the light-curve fits for the SNe in our sample are
summarized in Table~\ref{tab:lcpar}, where the times of maximum
($t^{\mathrm{max}}$) and apparent peak magnitudes ($m^{\mathrm{max}}$)
are listed for all bands.  Also given are decline-rate parameters
($\Delta m_t$) for each filter expressed as the amount in magnitudes that
the light curve declines between maximum light and an epoch $t$ days after 
maximum.  Finally, the number of data points and range of epochs covered 
with respect to the time of maximum, as well as the fitting method (spline 
functions, \S~\ref{sec:direc}, or templates, \S~\ref{sec:tmplfit}), are 
specified for each band.

Based on the quality of the data and fits, we define a subsample of
SNe with the best-measured light-curve parameters, mostly via direct fits 
but also through template fitting. Twenty-nine SNe belong to this group 
of ``best-observed'' events.  These objects are identified in 
Table~\ref{tab:sne}; in Table~\ref{tab:lcpar}, their names are set in
bold face. Special use will be made of this
subsample when studying the relation between peak luminosity and
decline rates in \S~\ref{sec:absmag}. 

\section{HOST-GALAXY REDDENING}
\label{sec:redd}

The $E(B-V)_{\mathrm{Gal}}$ values listed in Table~\ref{tab:sne} derived by 
\citet{schlegel98} from dust maps of the Galaxy may be used to correct for 
the interstellar reddening produced in our Galaxy. Beyond the Milky Way, 
the situation is considerably more complicated as there are at least three 
separate potential sources of reddening due to (a) dust in the intergalactic 
medium, (b) dust in the interstellar medium of the host galaxy, and (c) dust 
associated with the circumstellar material of the SN progenitor.  We have no 
{\it a priori} knowledge of the dust properties in these different environments 
and the reddening laws could, in principle, be different for all three.  
Nevertheless, by comparing the colors of SNe with similar decline rates and
taking advantage of the wide wavelength coverage of our observations, one
can hope to discern, in general terms, the combined effect of these different 
sources of reddening, which we lump together and refer to as the ``host-galaxy 
reddening.''

\subsection{Low-Reddening Sample}
\label{sec:lowr}

In order to use color information to determine reddening of the SN light 
produced in the host galaxy, we follow the procedure of \citet{phillips99}
of selecting a subsample of SNe which are suspected to have suffered little or
no dust extinction. This subsample will be used to derive intrinsic colors 
for SNe~Ia as a function of decline rate which, in turn, will serve as a 
reference for computing color excesses for the whole SN sample. 

The criteria employed to select the low-reddening subsample of SNe are (a)
SNe which occurred in E/S0-type galaxies, or SNe located away
  from the arms or nuclei of spiral galaxies, and
(b) absence of detectable interstellar \ion{Na}{1} D lines in
  early-time spectra.

Ten SNe from the sample fulfill both criteria: SNe~2004eo, 2005M,
2005al, 2005am, 2005el, 2005hc, 2005iq, 2005ke,
2005ki, and 2006bh.  In what follows, these objects are used to derive
intrinsic color relations as a function of the decline-rate parameter $\dm$.
We use colors corrected for Galactic reddening since such
  correction is, in most cases, small (see Table~\ref{tab:sne}) and
  known to higher accuracy than the host-galaxy
  reddening. We thus do not perform any cut based on Galactic
  reddening. We assume a value of $R_V = 3.1$, which may not be
  exactly correct for each SN. However, since the maximum Galactic
  reddening for this sample of ten SNe is $E(B-V)=0.14$~mag,
  uncertainties in this correction will be small.

\subsection{The $B-V$ Tail}
\label{sec:bvtail}

\begin{figure}
\includegraphics[angle=-90,width=16cm]{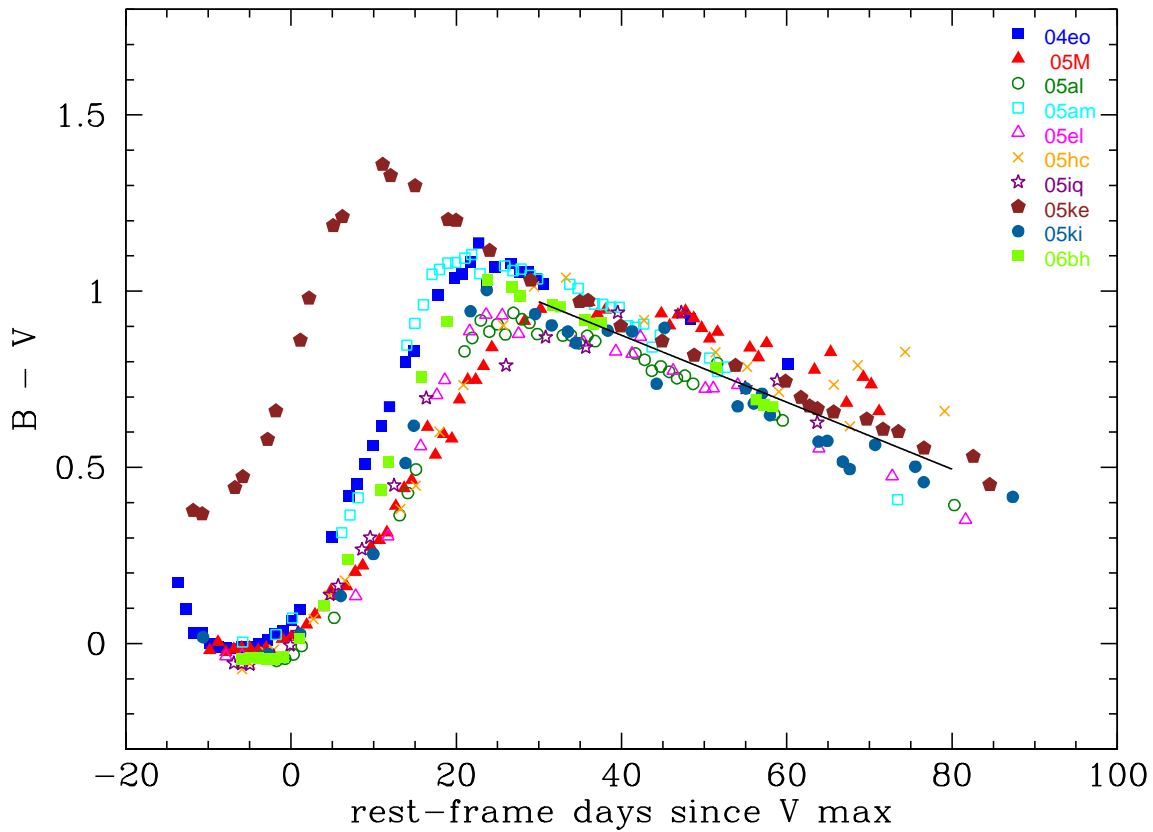}
\caption{Late-time $B-V$ color evolution of SNe assumed to have low or no
  host-galaxy reddening. Observed colors were corrected for Galactic
  reddening. The solid line shows a linear fit to the data in the range $30
  < t_V < 80$ days (see text).  The rms scatter about this fit is 0.077~mag.
  \label{fig:bvtail}}    
\end{figure}

The subsample of low-reddening SNe can be used to study the behavior of the 
$B-V$ color approximately one month after maximum light. It was first found
by \citet{lira95} that unreddened SNe~Ia follow a very similar linear regime 
in $B-V$ between
one and three months after maximum. As shown in Figure~\ref{fig:bvtail}, we 
confirm this result and derive the following intrinsic color law for the 
``tail'' of the $B-V$ color evolution: 

\begin{equation}
\label{eq:lira}
(B-V)_0 = 0.732 (0.006) - 0.0095 (0.0005) \, (t_V - 55).
\end{equation}

\noindent For consistency with \citet{phillips99}, rest-frame days since 
maximum $V$ light, $t_V$, is used for the time axis.  Equation~(\ref{eq:lira})
is valid over the range $30 < t_V < 80$ days and yields a
dispersion of $0.077$~mag.  The SNe employed in the fit cover a
broad range of decline rates, $ 0.85 < \dm < 1.76$~mag.
This relation is similar to the one of \citet{phillips99}, who found,
for a sample of six SNe and a different photometric system, 
$(B-V)_0 = 0.784 - 0.0118 \, (t_V- 55)$ with a dispersion of $0.06$
mag. Figure~\ref{fig:bvtail} shows that the residuals for
  individual SNe are strongly correlated, which implies 
that some of the 10 SNe in this subsample actually do suffer 
measurable host-galaxy reddening, or that there is an intrinsic dispersion in 
the Lira relationship, or both. We also note that the three SNe in our
sample which occurred in early-type galaxies (E/S0), namely SNe 2005al,
2005el, and 2005ki, show remarkably similar $B-V$ color evolution
--- and also in other
optical colors --- during the tail, with an average deviation of
$-0.05 \pm 0.003$~mag from equation~(\ref{eq:lira}), and a dispersion
of $\sim 0.035$~mag.

Equation~(\ref{eq:lira}) can be used to estimate color excesses,
$E(B-V)_{\mathrm{tail}}$, for the SNe in our sample with observations
in the tail.\footnote[11]{We remind the reader that equation~(\ref{eq:lira}) 
and all other relationships given in this paper which involve the CSP 
observed magnitudes are valid {\em} only for photometry in the natural 
system of the Swope+CSP bands.}  The results are listed in column (2) of 
Table~\ref{tab:cex}.  The uncertainties in these color excesses
  reflect only the photometric errors in the data points.  In 
considering the color excess for any particular  
SN, the observed dispersion of $0.077$~mag in the Lira law should also be 
added in quadrature to these errors.  However, in the analysis presented
in \S~\ref{sec:ld} of the precision to which SNe~Ia may be used as 
standardizable candles, this dispersion is included in the $\sigma_{\rm SN}$ 
values that are yielded by the fits, and the errors used for the color
excesses are those given in Table~\ref{tab:cex}.  Note that since
equation~(\ref{eq:lira}) was derived by fitting to the average color evolution 
of the low-reddening sample, it is guaranteed to generate some small negative 
color excesses.  \citet{phillips99} dealt with this by applying a Bayesian 
prior consisting of a one-sided Gaussian distribution of $A_B$ values with a 
maximum at zero and $\sigma = 0.3$~mag.  More recently, \citet{jha07} used
a prior consisting of a Gaussian with $\sigma_{(B-V)} = 0.068$ mag, representing
the intrinsic dispersion in color, convolved
with an exponential reddening distribution with a scale length 
$\tau_{E(B-V)} = 0.138$~mag
which they derived from fits to nearby
SN~Ia light curves by assuming a particular intrinsic color distribution.
However, until it is possible to confidently separate dust reddening from 
intrinsic color variation, we choose not to apply a prior to the 
color-excess measurements.

For each SN, we tested the agreement of the $B-V$ tail 
slope with that of equation~(\ref{eq:lira}). Good consistency was 
generally found, with the notable exception of SN~2006X, the most highly
reddened SN in the sample. As mentioned in \S~\ref{sec:direc}, this SN
showed a peculiar behavior in the $B$ band at the epochs considered here,
probably due to the appearance of a light echo \citep{wangx08b}. 

\begin{figure}
\epsscale{0.9}
\plotone{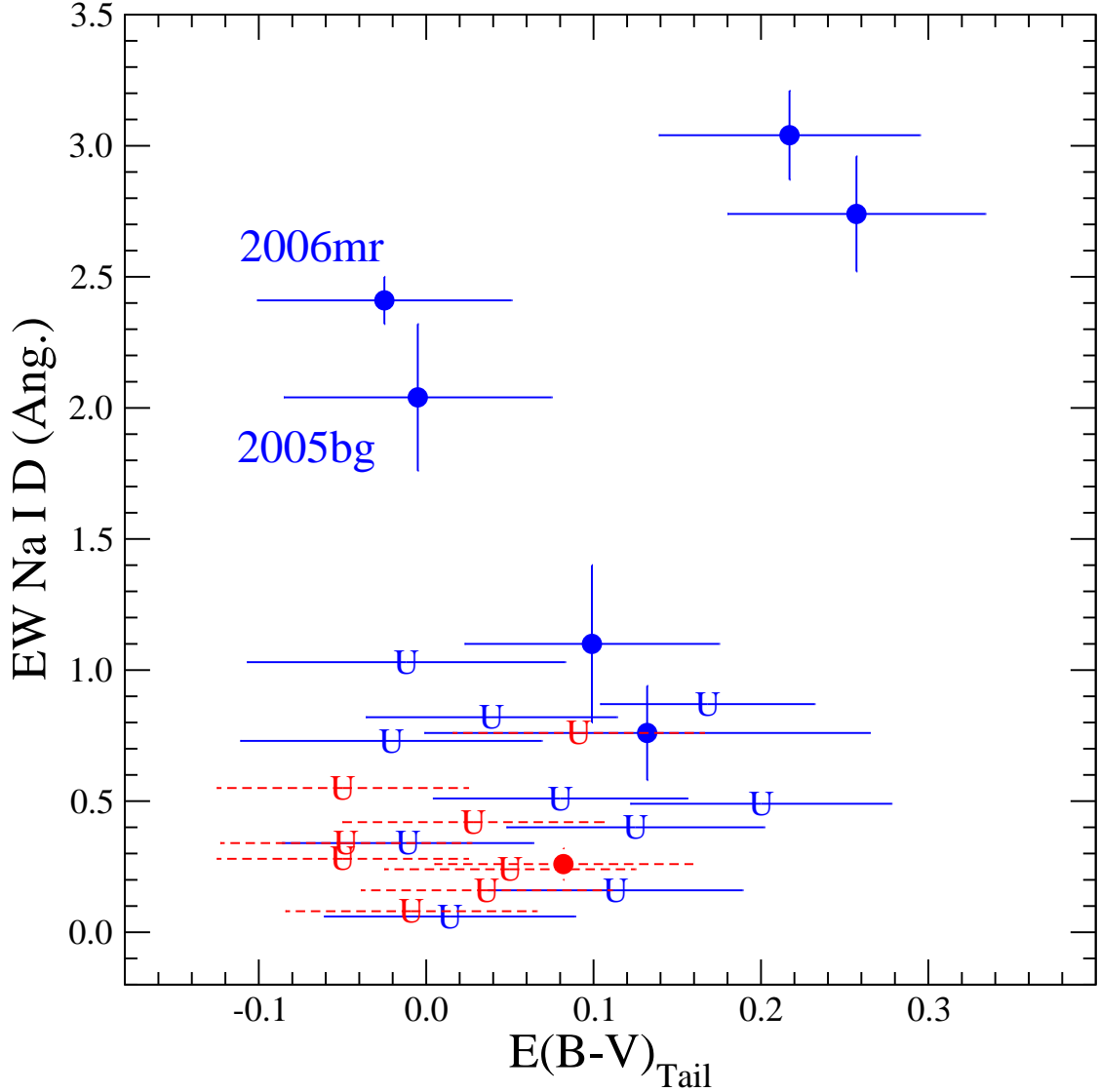}
\caption{Equivalent-width measurements (or upper limits) of the Na~I~D lines 
 plotted versus the
 color excess in $B-V$ as measured from the \citet{lira95} law.  The rms dispersion
 of 0.077~mag in the Lira-law fit (equation~\ref{eq:lira}) has been added in
 quadrature to the errors in $E(B-V)_{\rm tail}$.  Upper limits to the
 equivalent width are indicated as ``U.''  The red symbols with dashed error
 bars correspond to the low-reddening sample, while the blue symbols with 
 solid error bars show all the other SNe in our
 sample with measurements of $E(B-V)_{\rm tail}$ and the Na~I~D equivalent
 width (or upper limit).
  \label{fig:ewnai}}    
\end{figure}

Figure~\ref{fig:ewnai} displays a plot of $E(B-V)_{\mathrm{tail}}$ versus the 
equivalent width of the absorption in the Na~I~D lines produced in the host 
galaxies as measured from spectra we obtained of the SNe.  Generally speaking, 
these spectra are of low wavelength resolution,  and the Na~I~D lines are 
only visible when the equivalent width reaches $\sim$1~\AA~or more.  The red
symbols with dashed error bars in Figure~\ref{fig:ewnai} indicate the 
low-reddening sample.  Four of the eight SNe with the largest color excesses 
show detectable Na~I~D absorption, and the two with the largest color excesses, 
2004gu and 2005lu, correspond to the two with the highest Na~I~D equivalent 
widths.  However, two SNe with slightly negative values of 
$E(B-V)_{\mathrm{tail}}$, SNe
2005bg and 2006mr, also show strong Na~I~D absorption.  In the case of 
SN~2005bg, the estimate of $E(B-V)_{\mathrm{tail}}$ is based on a single 
$B-V$ color measurement in the tail, and therefore may be unreliable.  
SN~2006mr was a very rapidly declining event which occurred in the dusty 
central regions of Fornax~A (NGC~1316).  \citet{garnavich04} studied a 
small sample of fast-declining SNe~Ia and found that they followed the Lira 
relationship fairly well, but SN~2006mr suggests that there may be exceptions.
In any case, the correlation between color excess and the Na~I~D equivalent
width is known to be poor \citep{blondin09}.

Figure~\ref{fig:ewnai} shows that five additional SNe (2004ey, 2005ag, 2006D, 
2006ax, and 2006gt) have color excesses $E(B-V)_{\mathrm{tail}} < 0.08$~mag 
{\em and} undetectable Na~I~D absorption. In the following analysis, these 
five SNe are added to the original subsample of ten assumed to have suffered 
little or no host-galaxy reddening. 

\subsection{Pseudocolors at Maximum Light}
\label{sec:maxpseu}

Since the CSP data --- and generally most SN~Ia observations --- have
better coverage around maximum light than during the declining tail, it is
important to develop methods to determine color excesses from observations 
around the time of maximum. Using the peak magnitudes (corrected for Galactic 
reddening and the K-correction) derived in \S~\ref{sec:lcs}, 
pseudocolors\footnote[12]{The term {\em pseudocolor} is employed to stress 
the fact that these quantities do not represent the actual color of the SNe 
at any time but are just the difference between the magnitudes at maximum 
light of two bands, which occur at {\em different\,} times.} can be computed 
as the differences between the maximum-light magnitudes in two given bands.
The dependence of these pseudocolors on the decline
rate, parameterized by $\dm$, can then be investigated for a subsample of
SNe with low reddening. 

\begin{figure}
\epsscale{1.0}
\plotone{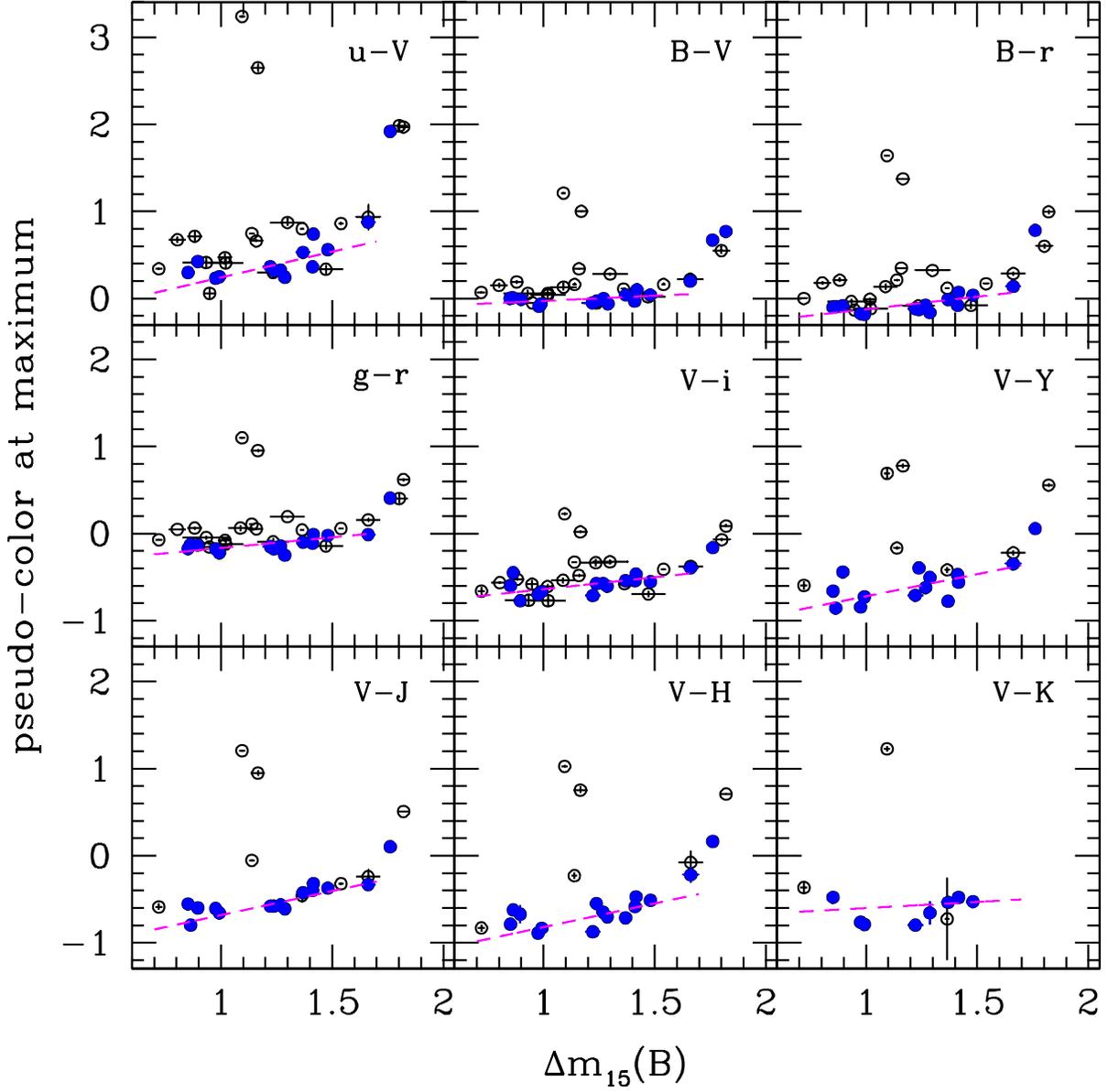}
\caption{Pseudocolors at maximum light corrected for Galactic reddening, as a
  function of $\dm$, for several filter combinations. Filled circles correspond 
  to the low-reddening subsample of SNe (see \S~\ref{sec:lowr}). The dashed
  lines are linear fits to the data of the low-reddening SNe in the range
  of  $0.8 < \dm < 1.7$ mag.\label{fig:maxpseu}}
\end{figure}

Figure~\ref{fig:maxpseu} shows the pseudocolors as a function of $\dm$
for several band combinations which will be used in the analysis of
\S~\ref{sec:absmag}. The solid points in the figure correspond to the 15
SNe with low reddening.  As expected, in all colors these 
SNe lie at the lowest values, following a roughly linear
trend with decline rate over the range $0.8 < \dm < 1.7$ mag.
At the fastest decline rates ($\dm>1.7$ mag), however,
the pseudocolors at maximum 
light are significantly redder than these linear trends would predict. 
Table~\ref{tab:cmax} summarizes straight-line fits to the colors of the 
low-reddening SNe with decline rates $0.8 < \dm < 1.7$ mag; these fits are 
plotted as dashed lines in Figure~\ref{fig:maxpseu}. 
Note that the slopes of the relations are largest when the 
$uYJH$ bands are involved.  In general, there is a correlation between 
the dispersion in color and the difference in the effective wavelengths,
$\Delta\lambda$, of the two filters that form the color.  Within this 
trend, the colors involving the $u$ filter have a relatively higher 
dispersion for their $\Delta\lambda$ values, whereas the colors 
involving the $J$ filter have a lower dispersion.

In the case of the $(B^{\mathrm{max}}-V^{\mathrm{max}})$ pseudocolor, the 
following linear law for the 14 SNe with low reddening and $0.8 < \dm < 1.7$  
mag is found: 

\begin{equation}
\label{eq:bvmax}
(B^{\mathrm{max}}-V^{\mathrm{max}})_0 = -0.016 (0.014) + 0.12 (0.05)
\, [\dm - 1.1].
\end{equation}

\noindent
The root-mean square (rms) of the fit is $0.06$~mag. 
This relation is slightly {\em redder\,} 
than that given by \citet{phillips99} and \citet{altavilla04}. We use this
intrinsic-color law to derive 
color excesses $E(B-V)_{\mathrm{max}}$ for all SNe in the range $0.7 <
\dm < 1.7$ mag, computed as 
$(B^{\mathrm{max}}-V^{\mathrm{max}})-(B^{\mathrm{max}}-V^{\mathrm{max}})_0$.  
The resulting values are listed in column 3 of Table~\ref{tab:cex}. 
The uncertainties in these color excesses are calculated from the 
photometric errors in the peak magnitudes.

\begin{figure}
\epsscale{0.8}
\plotone{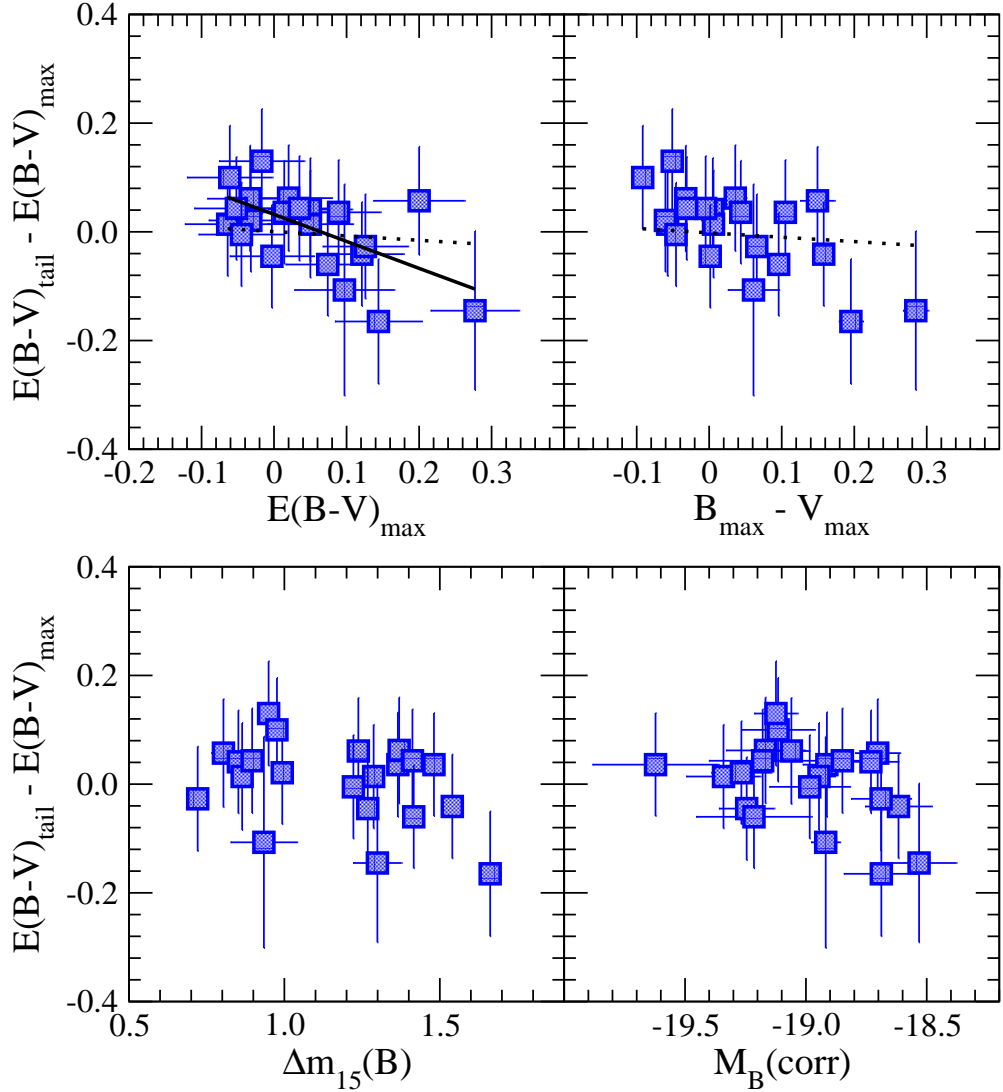}
\caption{Comparison of color-excess estimates for $B-V$ from
  measurements at maximum light [$E(B-V)_{\mathrm{max}}$] 
  and at late epochs [$E(B-V)_{\mathrm{tail}}$]. The difference
  between these two quantities is plotted vs. 
  $E(B-V)_{\mathrm{max}}$, $B_{\rm max} - V_{\rm max}$, $\dm$, and the
  absolute magnitude in $B$ corrected for decline rate (using Fit~1 of
  Table~\ref{tab:ld}; see \S~\ref{sec:ld}).  The dotted lines
  in the top two panels indicate the weak correlations that would
  be expected due to the fact that the color excesses given in
  this paper are ``observed'' [as opposed to ``true'' color excesses
  \citep[see][]{phillips99}], and are measured at very different 
  epochs in the color evolution of the SN.\label{fig:ebvcmp}}
\end{figure}

The top-left panel of Figure~\ref{fig:ebvcmp} shows a comparison of the
$E(B-V)_{\mathrm{max}}$ measurements with the $E(B-V)_{\mathrm{tail}}$
values from \S~\ref{sec:bvtail}.  The sample of 21 objects consists of the
best-observed SNe for which both color excesses could be measured.  In 
addition, the highly reddened SN~2006X has been excluded since it clearly 
deviates from the behavior of the rest of the SNe in the $B-V$ tail, as 
mentioned in \S~\ref{sec:bvtail}. One can see that there is a small but 
systematic difference between these two color-excess measurements which, 
unexpectedly, appears to be correlated with the magnitude of the color 
excess.  A formal fit to the data (plotted as a solid line in 
Figure~\ref{fig:ebvcmp}) gives 
$E(B-V)_{\mathrm{tail}}-E(B-V)_{\mathrm{max}} = 0.032 - 0.496\, 
E(B-V)_{\mathrm{max}}$, with a dispersion of $0.06$~mag.  The top-right
panel of Figure~\ref{fig:ebvcmp} shows that this difference also
seems to correlate with the color of the SN.  Since $E(B-V)_{\mathrm{tail}}$ 
and $E(B-V)_{\mathrm{max}}$ are ``observed'' color excesses, as opposed to 
``true'' color excesses \citep[see][]{phillips99}, and are measured at 
different epochs in the color evolution of the SN, one would expect there 
to be a small systematic difference between the two that would correlate 
with the SN color.  This effect is produced primarily because the effective 
wavelengths of the filters evolve due to the significant color evolution of 
the SN spectral energy distribution, but there is also a smaller dependence 
on the total amount of dust reddening itself.  The \citet{hsiao07} spectral 
template may be used to calculate the expected magnitude of these combined 
effects, which is plotted as a dotted line in both of the top panels 
of Figure~\ref{fig:ebvcmp}.  The observed systematic differences 
are greater than can be explained in this way, although the significance
is only at the $\sim 2 \sigma$ level.

Since the method of estimating the intrinsic colors at maximum by 
bootstrapping from the Lira relation was first utilized by 
\citet{phillips99}, we have examined the color excesses derived by these 
authors.  Interestingly, the same trends observed in the top two panels 
of Figure~\ref{fig:ebvcmp} are clearly present in those measurements.  
Not only does this appear to confirm the reality of the effect, 
but it also argues that it is not an artifact of the CSP dataset.

The lower-left panel of Figure~\ref{fig:ebvcmp} shows that
$E(B-V)_{\mathrm{tail}} - E(B-V)_{\mathrm{max}}$ does not depend significantly
on the decline-rate parameter $\dm$.  The measurements suggest that there 
may be a weak correlation with the absolute $B$ magnitude (corrected for 
decline rate), but this requires confirmation by more data.  We do not 
currently understand the underlying causes of these correlations, {\em but 
the fact that they exist implies that there is a component of one or the other 
of these color-excess measurements that is not due to interstellar dust.}

Keeping this caveat in mind, the fits corresponding to
$(V^{\mathrm{max}}-X_\lambda^{\mathrm{max}})$ for 
$X_\lambda = ugriBYJHK_s$ can be used to derive color excesses
$E(V-X_\lambda)_{\mathrm{max}}$ for all SNe with decline rates in
the range $0.8 < \dm < 1.7$ mag.  Columns 4 through 11 of
Table~\ref{tab:cex} list these color excesses.

\subsection{The Reddening Law}
\label{sec:redlaw}

The availability of photometric data spanning the $u$ through $K_s$ bands
enables the properties of the host-galaxy reddening to be examined. 
Assuming that the dust extinction in the host galaxies obeys the law 
introduced by \citet{cardelli89} and modified by \citet{odonnell94}
(hereafter referred to as the ``CCM+O law''), we can investigate
which value of the total-to-selective-absorption coefficient $R_V$ is
favored by SN~Ia data. Extinction in the Galaxy is well approximated
by the CCM+O law with an average value of $R_V \approx 3$ 
\citep[e.g.,][]{fitzpatrick07}.

\subsubsection{Optical--NIR Color Excesses}
\label{sec:cered}

\begin{figure}
\epsscale{1.0}
\plotone{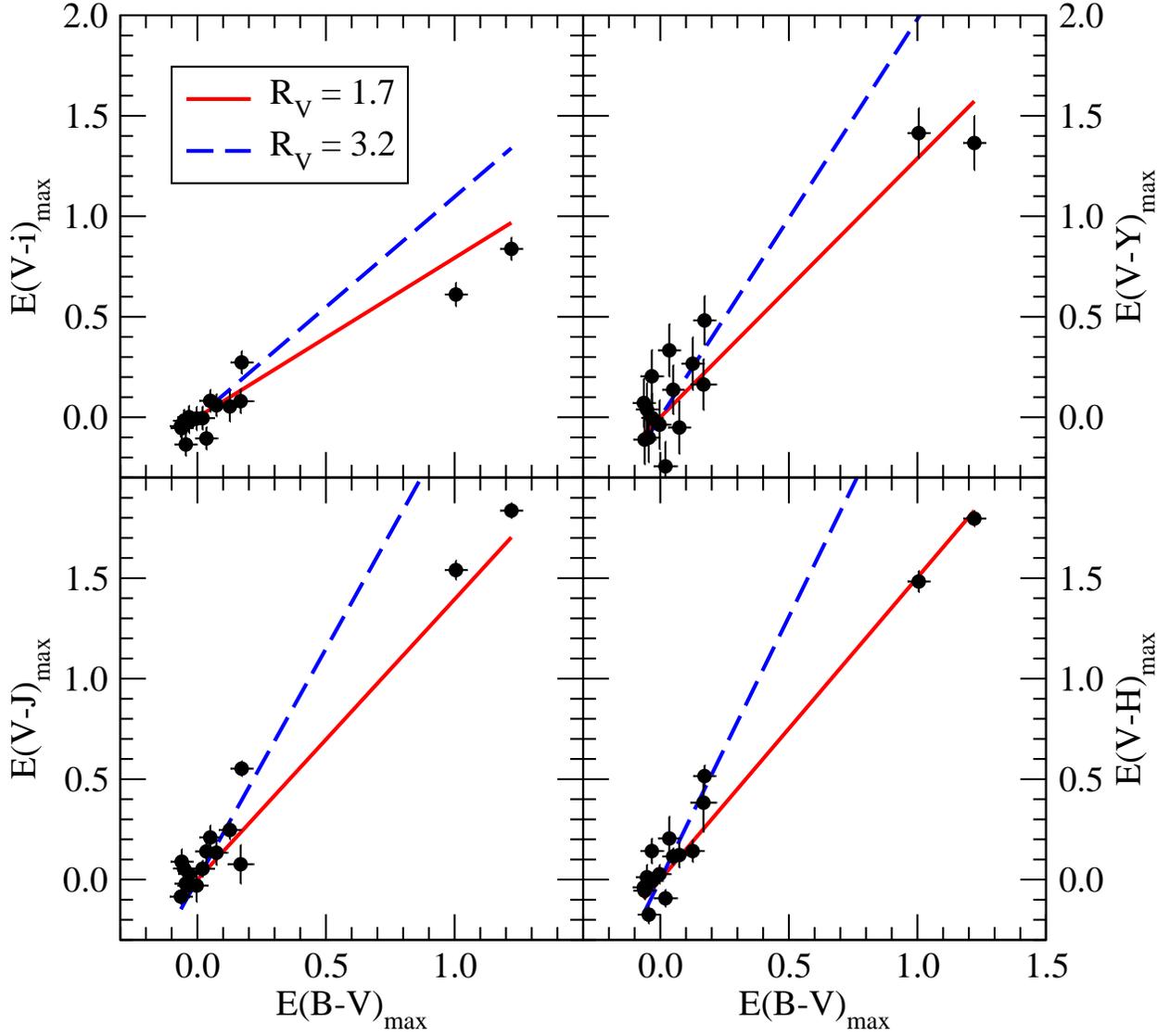}
\caption{Comparison of color-excess estimates $E(V-X)_{\mathrm{max}}$,
  for $X\equiv\,iYJH$, with $E(B-V)_{\mathrm{max}}$ for the
  best-observed SNe. The solid red lines represent the slope which
  corresponds to the average fit value of $R_V=1.7$, found using the
  whole set of points. The dashed blue lines indicate the slope
  predicted by a value of $R_V=3.2$, which is the averaged fit value
  found when excluding the two highly reddened SNe (the two points
  farthest to the right in the plots.)\label{fig:cecmp}}    
\end{figure}

An average value of $R_V$ for the subsample of best-observed SNe can be
derived by comparing the color excesses between the optical and NIR bands
presented in \S~\ref{sec:maxpseu} with the values of
$E(B-V)_{\mathrm{max}}$. The use of
optical--NIR color excesses ensures the necessary leverage in
wavelength to provide a more precise measurement of the reddening law,
even for SNe with low reddening. In Figure~\ref{fig:cecmp}, 
comparisons of $E(V-X)_{\mathrm{max}}$ with $X \equiv iYJH$ are shown. 

Fits to the slopes $E(V-X)_{\mathrm{max}} / E(B-V)_{\mathrm{max}}$ were
carried out for all four bands considered here. The slope were then
converted to values of $R_V$, using equations (A5) and (A6)
in \citet{krisciunas06} and the CCM+O law coefficients
$a_X$ and $b_X$ derived as explained in Appendix~\ref{app:rv}.
Table~\ref{tab:cecmp} summarizes the results of
these fits. The data favor a low value of $R_V$, with an average of
$R_V=1.69\pm0.05$ from the four fits. The corresponding slope for this
average is indicated by the solid red lines in Figure~\ref{fig:cecmp}.

If, however, the two highly reddened SNe, SN~2005A and SN~2006X, are excluded
from these fits, the uncertainties increase but the favored values of 
$R_V$ become larger and compatible with the standard $R_V \approx 3$. An
average of $R_V = 3.2\pm0.4$ is derived from the four fits, and
is indicated by the dashed blue lines in Figure~\ref{fig:cecmp}.

This is a very interesting result which may indicate that SNe with
low or moderate reddening are affected by standard dust extinction, 
while very highly reddened SNe suffer relatively more reddening than
extinction, as inferred from a low value of $R_V$. We will return
to this finding after analyzing the calibration of SNe~Ia as
{\em standardizable\,} candles in \S~\ref{sec:absmag}.

\subsubsection{Highly Reddened Supernovae}
\label{sec:hired}

\begin{figure}
\includegraphics[angle=-90,width=16cm]{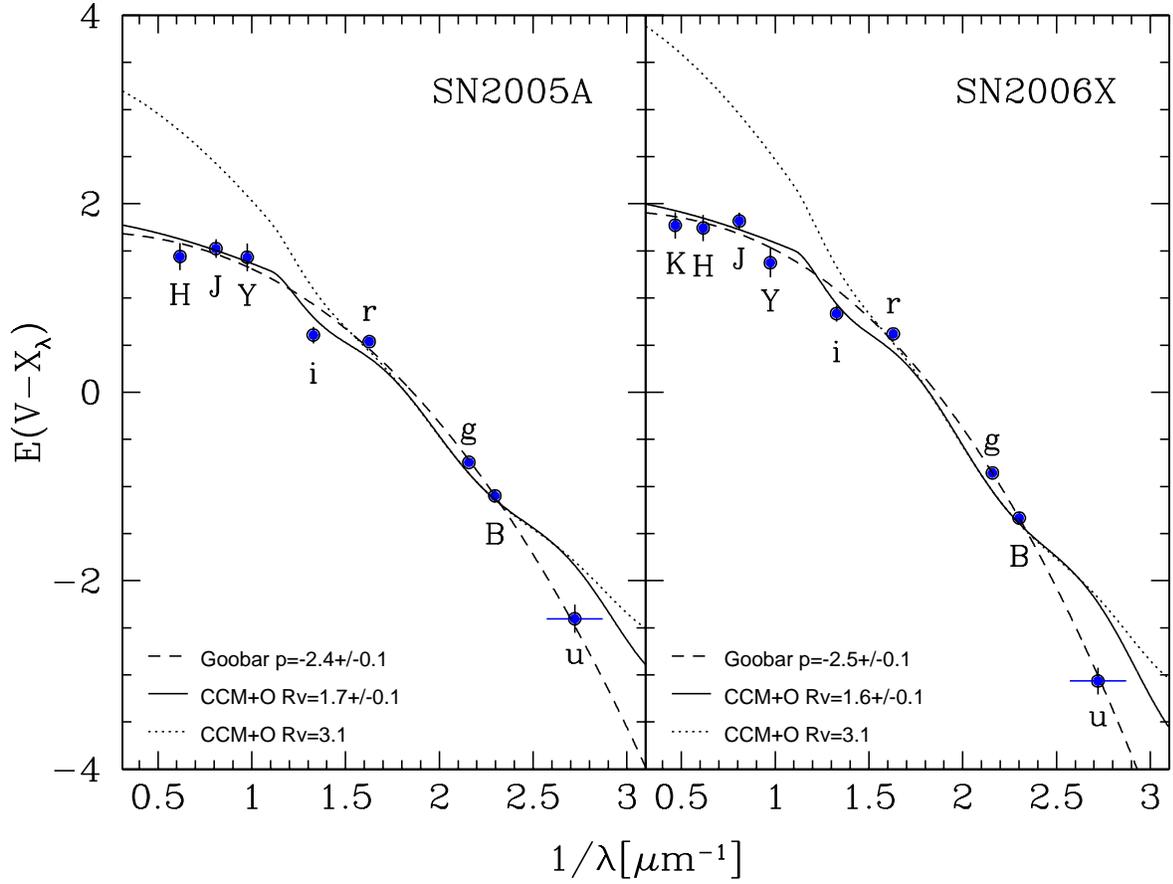}
\caption{Color excesses $E(V-X_\lambda)$, for bands
  $X_\lambda = ugriBYJHK_s$, for SN~2005A ({\em left panel}\,) and 
  SN~2006X ({\em right panel}\,). The solid lines show the best-fit CCM+O laws,
  whereas the dotted lines show the CCM+O
  model for $R_V=3.1$.  The dashed lines correspond to fits of the
  power-law model by \citet{goobar08}, valid for LMC-type dust. The
  latter provides a
  substantially better fit of the $u$-band reddening for both SNe, as
  compared to the CCM+O model.\label{fig:extlaw}} 
\end{figure}

Highly reddened objects provide the most precise information about the 
behavior of extinction as a function of wavelength. Our sample includes two 
objects, SNe 2005A and 2006X, for which $E(B-V) \gtrsim 1.0$~mag.  
Although the optical light curves of SN~2006X showed evidence of a light 
echo, this did not appear until $\sim$1~month after maximum light 
\citep{wangx08b}, and so the peak magnitudes should be uncontaminated.
Figure~\ref{fig:extlaw} shows the values of $E(V-X_\lambda)$ derived for 
both SNe (see  \S~\ref{sec:maxpseu}) plotted as a function of the effective 
wavelength of each filter $X$.  Note that at these extreme values of the 
color excesses, it is necessary to convert the {\em observed} values to 
{\em true} color excesses, and to also correct $\dm$ for reddening effects 
\citep[see][]{phillips99}.  Table~\ref{tab:obs2tru} gives the linear 
relations for converting from observed to true color excesses that are
derived using the \citet{hsiao07} template spectra.  Using the same 
spectra, the following relation for correcting the observed decline rate for 
reddening is found:

\begin{equation}
\label{eq:dm15tru}
\dm_{\rm true} = \dm_{\rm observed} + 0.065 E(B-V)_{\rm observed}.
\end{equation}

Assuming that the reddening of SNe~2005A and 2006X is produced
by dust that is described by the CCM+O reddening law, fits to the color
excesses in Figure~\ref{fig:extlaw} give values
of $R_V$ of $1.68 \pm 0.10$ and $1.55 \pm 0.07$, respectively.  These
fits are plotted as solid lines in Figure~\ref{fig:extlaw}, and are to be
compared with the dotted lines which show the best-fitting CCM+O 
reddening law where $R_V$ is fixed at its canonical Galactic value
of 3.1.  Although the CCM+O law does a good job in the optical and NIR,
it fails miserably in fitting the ($V-u$) color excesses.  Recently,
\citet{goobar08} has considered a model where multiple scattering of 
photons by circumstellar dust  steepens the effective extinction law,
converting it to a power law.  When this model is fit to the color
excesses of SNe~2005A and 2006X, excellent agreement at all
wavelengths is obtained with Large Magellanic Cloud (LMC) dust and a power-law 
index $p \approx -2.4$.  These fits are plotted as dashed lines in 
Figure~\ref{fig:extlaw}.  Thus, the data would appear to favor
a model where the reddening of these two objects is produced by
dust that is local to the SN.  In this context, it is interesting to note 
that short-term variations in the interstellar Na~I~D absorption lines in 
the spectrum of SN~2006X were observed by \citet{patat07} and 
interpreted as evidence for the existence of significant  
circumstellar material.

SN~2006X has also been studied in detail by
\citet{wangx08a}, who found a best fit to the data with $R_V=1.48\pm0.06$, 
in excellent agreement with our results.  However, these authors concluded
that the data, including their $U$-band photometry, were matched quite
well by the CCM reddening law.  The \citet{wangx08a} color-excess 
measurements for SN~2006X
were derived through comparison with a few individual SNe as well as via
the method we employ here of looking at the maximum-light colors as a 
function of the decline rate.  In general, there is good agreement between
their measurements and our own {\em except} for the critical $U$ band,
where \citet{wangx08a} find $E(U-V)_{\rm true} = 2.65 \pm 0.14$ but we derive
$E(u-V)_{\rm true} = 3.11 \pm 0.14$.  
Some of this discrepancy may be due to the 
fact that the sample of SNe~Ia with well-observed light curves in the $U$ or 
$u$ bands is still fairly small, although in looking at 
Figure~\ref{fig:maxpseu}, it is difficult to see how the value of $E(u-V)$ 
could be {\em overestimated} by 0.4~mag.  Although the $u$ filter 
response function is the least well-determined of the CSP bands, 
the color excesses for SNe~2005A and 2006X were derived with respect to 
other SNe~Ia observed with exactly the same filter, and in the same 
photometric system.  Thus, the color excess is well determined.  The largest
uncertainty is the effective wavelength that is associated with the
CSP $u$ band, which we estimate to be $<200$~\AA.  This maximum possible
error is indicated by the horizontal error bars on the $u$ points in
Figure~\ref{fig:extlaw}.  The conclusion is that the uncertainty in the
$u$-band response function cannot explain the discrepancy.

\begin{figure}
\includegraphics[angle=-90,width=16cm]{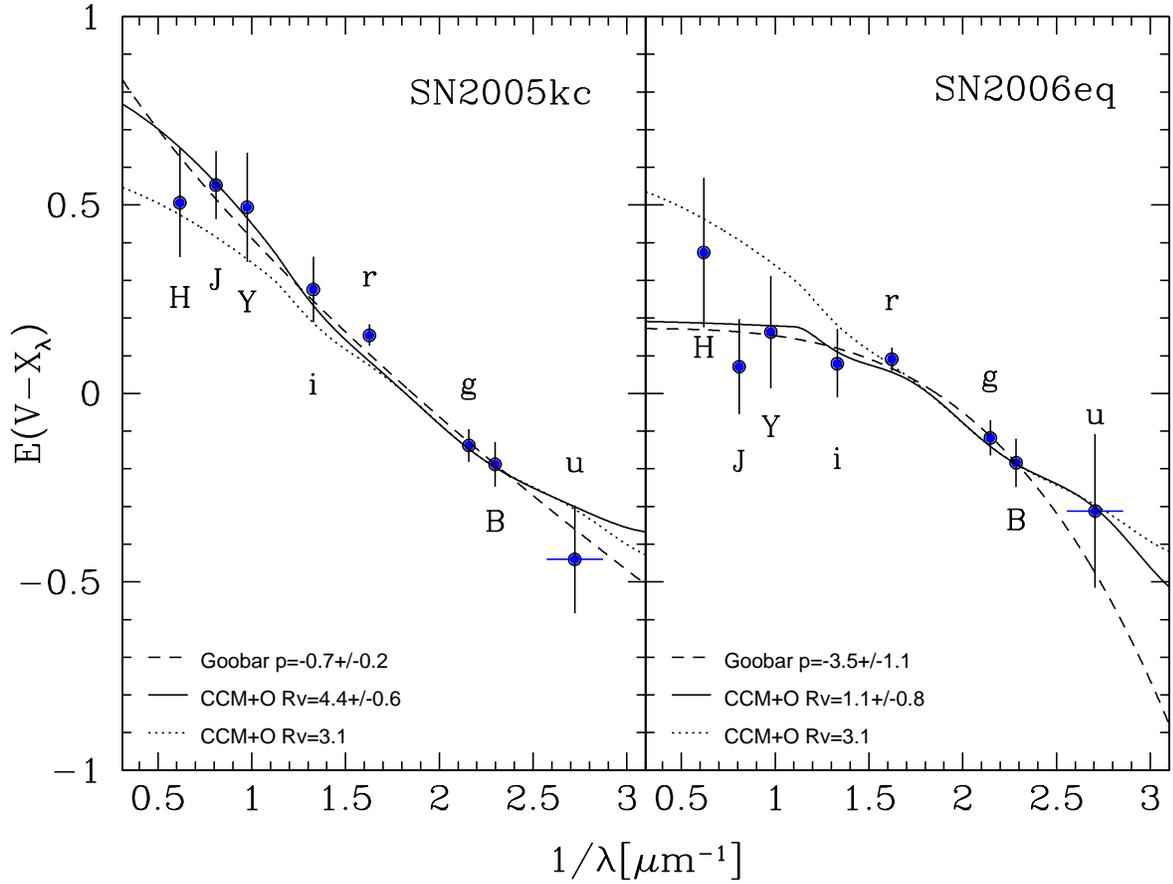}
\caption{Color excesses $E(V-X_\lambda)$, for bands
  $X_\lambda = ugriBYJHK_s$, for SN~2005kc ({\em left panel}\,) and 
  SN~2006eq ({\em right panel}\,). See the Figure~\ref{fig:extlaw} caption  
  for further details.\label{fig:extlaw2}}
\end{figure}

Only two other objects in our sample, SNe 2005kc and 2006eq, have
significant host-galaxy reddenings {\em and} observations in
the $YJH$ bands.  Fits to the color excesses of these
two SNe with both the CCM+O and \citet{goobar08} reddening
laws are displayed in Figure~\ref{fig:extlaw2}.  The observations
of SN~2005kc are consistent with either the CCM+O law with 
$R_V = 4.4 \pm 0.6$, or a Goobar power law with $p = -0.7 \pm 0.2$.
In the case of SN~2006eq, the measurements may be fit with
either a CCM+O law with 
$R_V = 1.1 \pm 0.8$, or a Goobar power law with $p = -3.5 \pm 1.1$.
Thus, the results for SN~2006eq are similar to those obtained for SNe~2005A
and 2006X, but the reddening law for SN~2005kc appears to be quite different,
perhaps even consistent with normal Galactic reddening.  These results
confirm those of Figure~\ref{fig:cecmp}, where SNe~2005kc and 2006eq
are the two points with highest $E(B-V)$ values among the
  moderately reddened SNe.  These two SNe illustrate the power of
optical--NIR color  
excesses in differentiating the reddening law.  More well-observed, 
moderately reddened events are clearly needed to advance further in the 
study of the reddening laws of SNe~Ia, and to look for possible 
correlations with other properties of the SN or its environment.

\section{TYPE IA SUPERNOVAE AS STANDARDIZABLE CANDLES}
\label{sec:absmag}

In this section, the homogeneous sample of CSP light curves is used to 
re-evaluate the precision to which SNe~Ia may be used as {\em standardizable 
candles}.  Two different approaches for calibrating the absolute magnitudes 
of SNe~Ia are considered. We start with the two-parameter method introduced by 
\citet{tripp98} in which the correlation between absolute peak magnitude, 
decline rate, and color is modeled for different choices of magnitudes and 
colors.  We then move on to use the color excesses determined in
\S~\ref{sec:cered} to examine the correlation of reddening-corrected, 
absolute peak magnitudes versus decline rate, following the approach of
\citet{phillips99}.  The former method makes no assumptions about the 
intrinsic colors of SNe~Ia; it simply considers decline rates and colors as 
two parameters that serve to improve the precision of SNe~Ia as distance
indicators. This formalism was originally employed by \citet{tripp99} to derive
cosmological parameters from SN~Ia data, and is part of the SALT
\citep{guy05} method used by 
\citet{astier06}. The color term in this type of analysis makes no distinction
between differences in intrinsic colors or in the amount of reddening 
suffered by the SNe.  On the other hand, the second approach explicitly 
assumes that one can derive color excesses by determining the intrinsic 
colors of SNe~Ia, and apply these as extinction corrections to the
observed peak magnitudes.

To calculate luminosity distances for the SNe, the following approximation 
is employed which is valid to a precision of better than 0.25\% at redshifts 
$z < 0.1$ in a two-component (matter $+$ dark energy) model of the Universe:

\begin{equation}
\label{eq:mu}
d_L(z_{\mathrm{CMB}};H_0,\om,\ol)\;=\;\frac{(1+z_{\mathrm{helio}})}{(1+z_{\mathrm{CMB}})}\,\frac{c}{H_0}\, \left[z_{\mathrm{CMB}}+\frac{1}{2}\,\left(\ol\,-\frac{\om}{2}+1\right)\,z^2_{\mathrm{CMB}}\right],
\end{equation}

\noindent where $z_{\mathrm{helio}}$ is the heliocentric redshift of
the host galaxy, $z_{\mathrm{CMB}}$ is the redshift
referred to the CMB rest frame, and the cosmological parameters were set to 
the standard values $H_0=72$ km s$^{-1}$ Mpc$^{-1}$, $\om=0.28$, and $\ol=0.72$ 
\citep{spergel07}. The factor
$(1+z_{\mathrm{helio}})/(1+z_{\mathrm{CMB}})$ accounts for
  the fact that the photon redshift is observed with respect to the
  heliocentric reference system.
The heliocentric redshifts for the sample SNe are given 
in Table~\ref{tab:sne}.  Also given in this table are the redshifts in the 
CMB frame calculated from the heliocentric redshifts using the velocity 
vector determined by \citet{fixsen96} from the {\em COBE\,} data (and which 
are available via the NASA/IPAC Extragalactic Database, NED).  The 
uncertainty in the redshifts due to peculiar velocities is assumed to be 
$\sigma_z=0.001$ (300~km~s$^{-1}$ in velocity).  With the exception
of the three SNe listed in Table~\ref{tab:dist} with $z<0.01$ that were 
hosted by galaxies with direct distance measurements based either on Cepheids 
or the surface brightness fluctuation (SBF) method, we restrict the sample 
to those SNe with redshifts $z>0.01$, where peculiar recession velocities 
become at least one order of magnitude smaller than the cosmological velocities.

In the analyses presented here, the observed peak magnitudes of
Table~\ref{tab:lcpar} are corrected for Galactic reddening using the
values of $E(B-V)_{\mathrm{Gal}}$ given in Table~\ref{tab:sne}, and
adopting the CCM+O extinction law with $R_V^{\mathrm{Gal}}=3.1$.

\subsection{Calibration Using Decline Rates and Colors}
\label{sec:tb}

The two-parameter model of \citet{tripp98} assumes that the distance
modulus of a SN has the following dependence on decline rate and color:

\begin{equation}
\label{eq:modtb}
\tilde{\mu}_X\;=\;m_X\;-\;M_X(0)\;-\;b_X\,[\dm-1.1]\;-\;\beta_X^{YZ}\,(m_Y-m_Z),
\end{equation}

\noindent where $m_X$ is the observed peak magnitude in a given band
$X$, and $(m_Y-m_Z)$ is a pseudocolor at maximum brightness from any choice of
bands $Y$ and $Z$. The fit parameters 
are the peak absolute magnitude of SNe~Ia with $\dm=1.1$ and zero
color, $M_X(0)$; the
slope of the luminosity vs. decline-rate relation, $b_X$; and the slope of
the luminosity-color relation $\beta_X^{YZ}$. Using measurements
of peak magnitudes, pseudocolors, and decline rates \{${m_X}_i$;
$\dm_i$; $(m_Y-m_Z)_i$\} ($i=1,\dots,N$) for a sample of
$N$ SNe, the best-fit parameters are solved for via $\chi^2$
minimization, as explained in Appendix~\ref{app:fits}.  This is 
equivalent to minimizing the dispersion in the Hubble diagram.  Note
that a constant term has been added to the measurement
uncertainties that appear in the denominator of the $\chi^2$ in
equation~(\ref{eq:chisq}) to account for possible intrinsic dispersion of
the SN~Ia data about the model given by equation~(\ref{eq:modtb}).

The \citet{tripp98} model has traditionally been applied to fits of $B$
versus $\dm$ and $B-V$.  Such an analysis is presented here, along 
with a fit of $J$ versus $\dm$ and $V-J$.

\begin{figure}
\epsscale{0.7}
\plotone{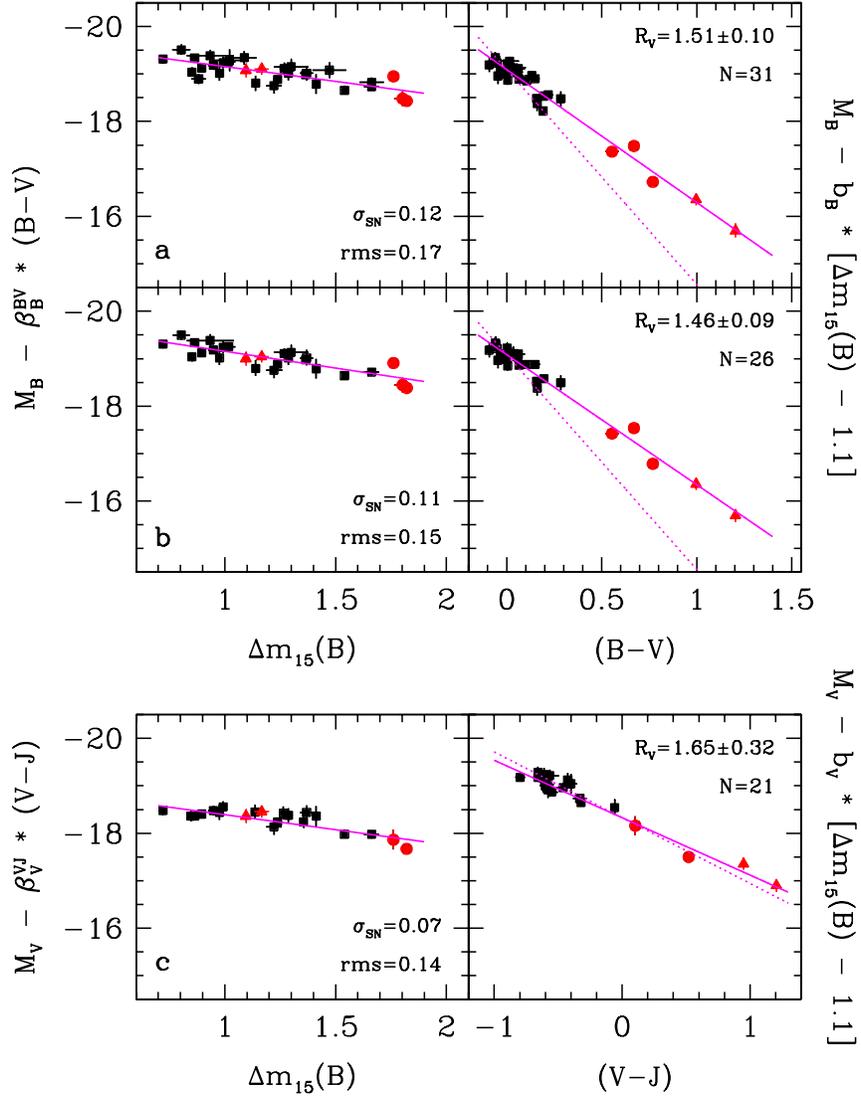}
\caption{Sample fits of absolute peak magnitude versus decline rate
  ($\dm$) and color for SNe with reliable distances. One fit per row is 
  displayed.  The left panels show the relations between color-corrected 
  absolute peak magnitude and decline rate.  The right panels show the 
  absolute peak magnitude vs. color relations, corrected for decline rate.
  Red circles denote the fast-declining SNe; red triangles mark the 
  very red SNe 2005A and 2006X. The resulting fits are 
  shown with solid lines.  The dotted lines in these panels show the 
  relations predicted by reddening with the CCM+O law and $R_V=3.1$. 
  The number of SNe used in the fit, the rms scatter, the
  intrinsic dispersions ($\sigma_{\rm SN}$), and the predicted $R_V$ are
  labeled. {\em Rows\,}: (a) $B$ vs. $\dm$ and $B-V$ for the whole 
  sample; (b) $B$ vs. $\dm$ and $B-V$ for the subsample of 
  best-observed SNe;  (c) $J$ vs. $\dm$ and $V-J$ for the subsample of 
  best-observed SNe.
  \label{fig:tb}} 
\end{figure}

Table~\ref{tab:tb} summarizes the results of the fits. Fit~1 was
done using the entire sample of 32 SNe with reliable distances and the
combination of $B$ versus $\dm$ and  $B-V$. This fit is shown in 
panel (a) of Figure~\ref{fig:tb}, and yields a total dispersion of
$0.17$~mag and an intrinsic dispersion $\sigma_{\rm SN}=0.12$~mag. Performing
the fit with the subsample of 26 best-observed SNe (Fit~2)
reduces the scatter to $0.15$~mag and the derived intrinsic dispersion
to just $0.09$~mag without modifying substantially the fit
parameters. This is shown in panel (b) of Figure~\ref{fig:tb}.

Note that the fits described above include SNe with {\em all\,}
the available decline rates; specifically, they include the fast-declining SNe
with $\dm>1.7$ (SN~2005bl, SN~2005ke, and SN~2006mr) which are marked
with red circles in Figure~\ref{fig:tb}. These SNe follow
the behavior of the rest of the sample. This is substantiated by
the results of Fits~3 and 4, done excluding these three SNe; the
resulting fit parameters are in agreement within the 
uncertainties with those of Fits~1 and 2, and the dispersions
found are nearly the same.   This is a remarkable result since the red
colors of the fast-declining SNe~Ia ---which occur preferentially in E and
S0 galaxies --- are clearly largely intrinsic in origin, whereas the reddening
of slower-declining SNe ---which are most common in spiral galaxies ---
is presumably due at least in part to dust.

If it is assumed for the moment that the pseudocolors (after correction
for Galactic reddening) vary due to reddening by dust in the host galaxy,
the color term, $\beta$, can be converted to a total-to-selective absorption 
coefficient, $R_V$, adopting
the CCM+O reddening law and following the prescriptions of
Appendix~\ref{app:rv}. The value of $R_V$ obtained from each fit is given in
column 7 of Table~\ref{tab:tb}. As is seen, all the fits thus far considered
yield values of $R_V \approx 1.5$, which is significantly different from the
standard Galactic value of $3.1$. In the right-hand panels of
Figure~\ref{fig:tb}, the fits are labeled with the corresponding
$R_V$ values. For comparison, the reddening vector predicted by $R_V=3.1$ 
is also plotted as a dotted line. 

The Tripp (1998) model was also fit to the subset of best-observed SNe that meet the 
condition that 
$(B^{\mathrm{max}}-V^{\mathrm{max}}) < 0.4$ mag.  This excludes both the
fast-declining objects, which are intrinsically red, and SNe~2005A and 2006X, 
which suffered heavy reddening due to dust.
Fit~5 in Table~\ref{tab:tb} gives the results of such a fit.
Once again, the zero point and
slope of the luminosity vs. decline-decline rate relation ($M_B(0)$ and $b_B$) do
not change significantly in comparison with the previous fits. The
luminosity-color slope, $\beta_B^{BV}$, does change slightly (by
$\sim 1\sigma$), resulting in an even lower value of $R_V$. These
results are relevant to the study of high-redshift SNe~Ia for which
red (and therefore faint) objects [with
  $(B^{\mathrm{max}}-V^{\mathrm{max}}) > 0.4$ mag] are rarely observed
\citep[e.g.,][]{kowalski08}. 

The results of applying the Tripp (1998) model to $J$ versus $\dm$ and $V-J$
for the sample of 
best-observed SNe with $J$-band coverage (21 objects) are
given in Fit~6 of Table~\ref{tab:tb}, and are plotted in the lower part
of Figure~\ref{fig:tb}.  
Two additional fits are presented in Table~\ref{tab:tb} for the
$J$ versus $\dm$ and $V-J$ combination.  In Fit~7, the two fast-declining 
SNe with 
coverage in $J$ are excluded.  A slightly lower value of the slope 
$\beta_V^{VJ}$ is obtained, with an almost negligible intrinsic
dispersion of $\sigma_{\rm SN}=0.02$~mag.  Fit~8 employs a color cut of
$(V^{\mathrm{max}}-J^{\mathrm{max}}) < 0.0$ mag, which is equivalent to limiting
the sample to $(B^{\mathrm{max}}-V^{\mathrm{max}}) < 0.4$ mag.  Again, the
best-fit parameters are in good agreement with those of Fit~7, and the
dispersions are likewise very similar.

Note that the interpretation of $\sigma_{\rm SN}$ as the
intrinsic dispersion of the SN data about the model relies on the
correct estimation of the measurement uncertainties and their
covariances. An error in the treatment of the uncertainties would lead
to different $\sigma_{\rm SN}$. Consequently, the fit
parameters themselves could potentially change. In order to test this,  
the dependence of the fit parameters on the adopted uncertainties in the 
peak magnitudes, colors, and decline rates was examined. This was done by 
alternatively adding an arbitrary constant uncertainty on each quantity, 
up to 0.06~mag.  While the values of $\sigma_{\rm SN}$ obtained were found to
decrease with increasing uncertainties as expected, the best-fit values of 
the parameters remained well within the uncertainties.

In order to test the dependence of our results on the method used
  to derive light-curve parameters, we repeated the fits using peak
  magnitudes, decline rates and colors from template light-curve fits
  (see \S~\ref{sec:tmplfit}). We used the sample of best-observed SNe for
which we performed both spline and template fits. The results are
equivalent within the quoted errors to the ones presented above.

\subsection{Calibration Using Decline Rates and Extinction}
\label{sec:ld}

In this section, the color excesses derived in \S~\ref{sec:redd} are used
to produce fits of absolute peak magnitude versus decline 
rate and reddening.  It is explicitly assumed that the color excesses, 
$E(Y-Z)$, 
are due to dust and can be converted into an absorption in the $X$ band 
via $A_X=R_X^{YZ}\,E(Y-Z)$. This absorption is used to correct the observed
peak magnitudes according to the model

\begin{equation}
\label{eq:modld}
\tilde{\mu}_X\;=\;m_X\;-\;M_X(0)\;-\;b_X\,[\dm-1.1]\;-\;R_X^{YZ}\,E(Y-Z).
\end{equation}

\noindent Note the similarity of this model to that of
equation~(\ref{eq:modtb}) of \S~\ref{sec:tb}, especially if one
considers that the color excesses derived in \S~\ref{sec:maxpseu} are
based on linear relationships between the observed pseudocolors and
$\dm$.\footnote[13]{Indeed, as discussed in Appendix~\ref{app:fits}, the
two treatments are mathematically equivalent.}   Since
the color excesses were obtained for SNe in the
range $0.7 < \dm <1.7$ mag, the fast-declining SNe are necessarily excluded
from the analysis of this section.

We first focus on fits done with $B$ and $E(B-V)$ data.  To provide a 
reference, the sample was initially limited to the twelve SNe which were
previously identified as having suffered little or no dust extinction
(see \S~\ref{sec:bvtail}).  These low-reddening objects were used to fit the 
model of equation~(\ref{eq:modld}) with the last term on the right-hand side 
set to zero --- that is, without applying any reddening correction.  The resulting 
parameters are given in Fit~1 of Table~\ref{tab:ld}.  These twelve SNe with
low reddening yield a total dispersion of $0.19$~mag, with an intrinsic 
dispersion $\sigma_{\rm SN}$ of the same amount.

Next, the entire SN sample was fit by first fixing $R_V$ to the standard
Galactic value of 3.1.  The results are given in Fit~2 of Table~\ref{tab:ld}, 
where we solve only for $M_B(0)$ and $b_B$.  This fit implies substantially
higher peak luminosities and, most notably, a very large dispersion of
$\sim 0.5$~mag.  Thus, correcting for the observed reddening as if it
were produced by typical Galactic interstellar dust yields a dispersion
which is a factor of $\sim$2.5 greater than that obtained for the subsample of
low-reddening SNe.

\begin{figure}
\epsscale{0.7}
\plotone{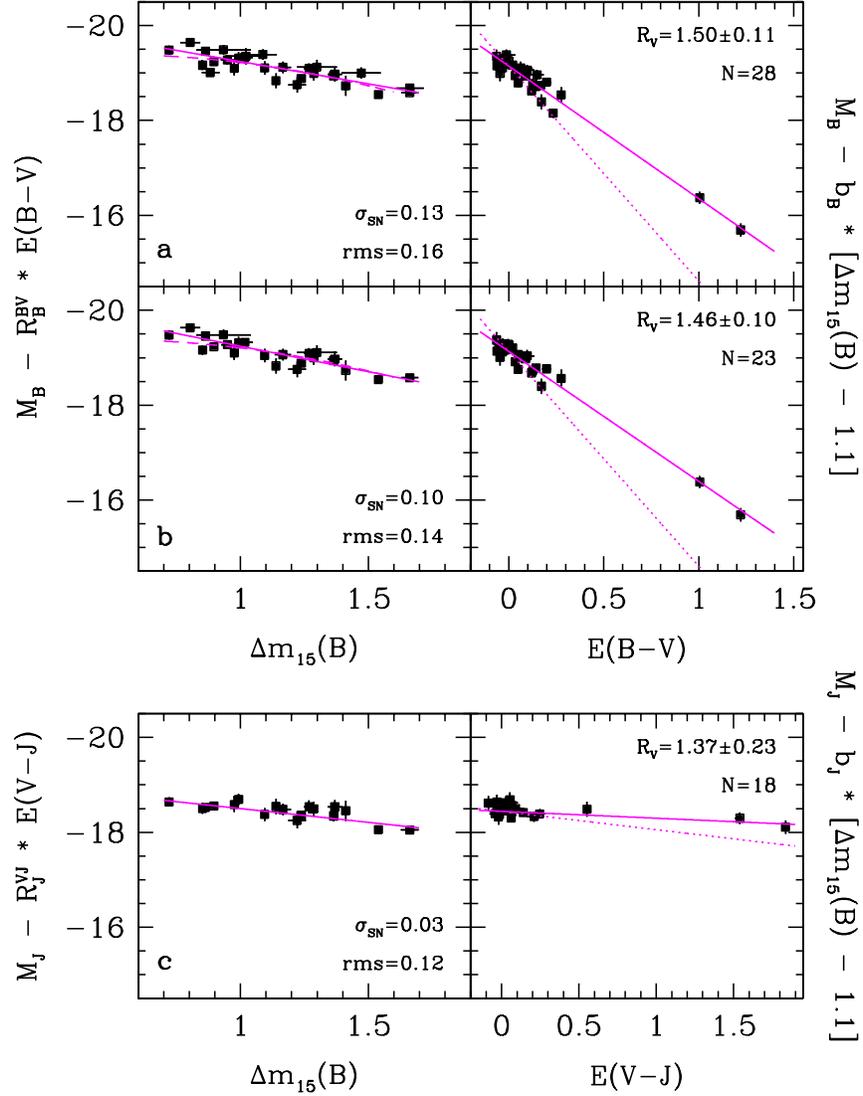}
\caption{Fits of absolute peak magnitude versus decline rate ($\dm$) and 
  reddening (measured as a color excess). One fit per row is displayed.  The 
  left panels show the relations between reddening-corrected absolute peak 
  magnitude and decline rate.  The right panels show the absolute peak 
  magnitude vs. color excess relations, corrected for decline rate.  The 
  symbols for points and lines, and the labels, are equivalent to those used 
  in Figure~\ref{fig:tb}. For comparison, we added to the left panels of the
  upper two rows a dashed line indicating the quadratic relation fits by
  \citet{phillips99}, shifted  
  to the distance scale adopted in this work which corresponds to
  $H_0=72$ km s$^{-1}$ Mpc $^{-1}$. Note that fast-declining SNe are not
  included in these plots. 
  {\em Rows\,}: (a) $B$ vs. $\dm$ and $E(B-V)$ for the whole sample; 
  (b) $B$ vs. $\dm$ and $E(B-V)$ for the subsample of best-observed SNe;  
  (c) $J$ vs. $\dm$ and $E(V-J)$ for the subsample of best-observed SNe.
  \label{fig:ld}} 
\end{figure}

Beginning with Fit~3 of Table~\ref{tab:ld}, $R_B^{BV}$ was allowed to be a 
free parameter.  Here the full sample of SNe with $0.7 < \dm <1.7$ mag was
employed, while in Fit~4 only the best-observed subsample was considered.
The resulting parameters for these two fits are indistinguishable within the 
errors.  The best-fit value of $R_B^{BV}$ in
both cases is $\sim$2.8, corresponding to $R_V \approx 1.5$.  These two fits are
shown in panels (a) and (b) of Figure~\ref{fig:ld}. 

The values of $R_B^{BV}$ derived in Fits~3 and 4 are strongly influenced by 
the two highly reddened SNe 2005A and 2006X.  Fits~5 and 6 in 
Table~\ref{tab:ld} give solutions where these two SNe are excluded from the 
sample, with Fit~5 corresponding to the whole sample and Fit~6 to the 
best-observed SNe.  As is seen, these fits still prefer a value of 
$R_V \approx 1.5$.

Bearing in mind the slight differences between photometric systems, the
slope of the luminosity vs. decline-rate relation found for Fits~1--6 is in 
reasonable agreement with previously published results
\citep{hamuy96a,altavilla04,prieto06}\footnote[14]{At maximum light, the 
$B-V$ colors of SNe~Ia are close to zero, except for the most 
heavily reddened events.  By definition, at $B-V = 0.0$ mag, magnitudes in 
the natural system are identical to those that have been color-corrected.  
Hence, comparing the CSP natural magnitudes at maximum light with those in 
systems where a color-term correction has been applied is a reasonable 
proposition.}.  As shown in the left panels of 
Figure~\ref{fig:ld}, the quadratic relationship found by \citet{phillips99} 
is also consistent with the data, although the CSP observations themselves 
are fit perfectly well by a simple linear relation. 

Next, we considered fits using the peak magnitudes in $ugriVYJHK_s$, combining
each band with a suitable color excess.  The choice of this color excess is 
somewhat arbitrary.  For the $ugr$ and $JHK_s$ bands, color excesses 
involving the filter itself were used.  However, when this is done for the 
$i$ and $Y$ filters, the resulting fits yield significantly higher 
dispersions than are obtained when using $E(B-V)$.  We attribute this to
the poor quality of the template fits in the $i$ and $Y$ bands (see 
\S~\ref{sec:secmax}), which not only introduces an error in the peak 
magnitude, but also in the color excess if the latter involves either of 
these bands.

Fits~7--15 in Table~\ref{tab:ld} give the results 
for all bands other than $B$ using the sample of best-observed SNe.
The best-fit values of $R_V$ range from $1.1$ to $1.8$ for $ugriYJ$. For
$H$ and $K_s$, somewhat larger values in the range $2.4$--$2.7$ are obtained, 
but with such large uncertainties that they are compatible with both low
and standard values of $R_V$.  A weighted average of the results for 
Fits~7--15 yields $R_V=1.38\pm0.04$, which is consistent
with the results obtained for $B$ vs. $E(B-V)$ (see Fit~4).
The scatter in the corrected magnitudes for Fits~7--15 is in the range 
$0.12$--$0.16$~mag, with the intrinsic components of the dispersion amounting
to as small as $0.04$~mag and as large as $0.13$~mag.

\subsection{SNe~Ia as Standard Candles in the Near-Infrared}
\label{sec:absnir}

The set of NIR light curves presented here constitutes a significant
contribution to the available data in the literature. The advantages of 
using NIR observations of SNe~Ia to determine distances were demonstrated
by \citet{krisciunas04} and, more recently, by \citet{wood-vasey08}.
Apart from being less affected by dust extinction and scattering --- which 
considerably
reduces the effect of uncertainties in the color excesses and reddening
law--- the peak luminosities of SNe~Ia in the NIR present a shallower 
dependence on decline rates than their optical counterparts.

Indeed, previous studies have shown no clear evidence for a correlation between
luminosity and decline rate in the $JHK_s$ bands except for the very fastest
declining SNe \citep{krisciunas04b}.  The improved quality and coverage of 
the CSP NIR light curves allow a weak dependence of the luminosity on the 
decline rate to be discerned in the $J$ band.  Fit~13 of Table~\ref{tab:ld} 
yields a slope of $b_J = 0.58 \pm 0.09$ over $0.7 < \dm <1.7$ mag, and is shown 
in panel (c) of Figure~\ref{fig:ld}.  In the $H$~band, the evidence is 
also suggestive for a weak correlation between luminosity and decline rate, 
although the significance of the measured slope is only $\sim 2\sigma$ 
(see Fit~14 of Table~\ref{tab:ld}).  In a future paper \citep{kattner09},
we will use a larger sample of CSP SNe to test the reality of these
relations.

To compare with previous studies, we compute the dispersion in absolute peak 
magnitudes {\em without} any correction for decline rate.  Table~\ref{tab:absnir} 
shows the results of weighted averages of the absolute peak magnitudes in 
$YJHK_s$. These averages correspond to the sample of best-observed SNe, 
correcting for extinction using the values of $E(V-X)$ and $R_X^{YZ}$ 
listed in Fits~12--15 of Table~\ref{tab:ld}.  For comparison, averages were
also taken limiting the sample to those SNe with low/moderate reddening 
(i.e., excluding SN~2005A and SN~2006X), and assuming two different values
of $R_X^{YZ}$: zero, which is equivalent to ignoring host-galaxy 
extinction corrections, and the values of $R_X^{YZ}$ that correspond to
$R_V = 3.1$.

The most striking implication of Table~\ref{tab:absnir} is that, when
the highly reddened SNe are excluded from the sample, the absolute 
magnitudes change only very slightly depending on the value of $R_V$.
{\em This illustrates the great advantage of working in the NIR, where
lack of knowledge of the exact value of $R_V$ has little influence,
and where dust-extinction corrections can essentially be ignored for all
but the most heavily reddened SNe.}

Table~\ref{tab:absnir} shows that the dispersions in the absolute 
magnitudes are 0.18--0.21~mag in the $J$ and $H$ bands, depending only
slightly on the assumed value of $R_V$.
These are to be compared with the
rms dispersions of 0.12~mag and 0.16~mag, respectively, obtained when
the absolute magnitudes are corrected for the decline-rate dependence
(cf. Fits~13 and 14 of Table~\ref{tab:ld}).  For reference, 
\citet{krisciunas04b} obtained rms values (uncorrected for decline rate) 
of 0.13~mag in $J$ and 0.15~mag in $H$ for a sample of
22 SNe~Ia, and \citet{wood-vasey08} found dispersions of 0.33~mag in $J$ and 
0.15~mag in $H$ for 21 SNe.

In Table~\ref{tab:absnircomp}, the average absolute magnitudes at maximum of 
the best-observed subsample of CSP SNe are compared with those obtained by 
\citet{krisciunas04b}.  The values in $J$ and $H$ are consistent at the 
$\sim 2\sigma$ level.  In this same table, a comparison is also attempted 
with the absolute magnitudes {\em at the epoch of $B$ maximum} given by 
\citet{wood-vasey08}.  For $J$ and $H$, these were derived using all the 
measurements available for the CSP SNe inside a bin of [$-2.5$,2.5] 
rest-frame days with respect to $B$ maximum, converting these to absolute 
magnitudes as explained below.  A total of 33 and 28 data points entered 
into the calculations for $J$ and $H$, respectively. For the $K_s$
band, the relative scarcity of observations forces the use of a slightly 
larger bin size of [$-4$,4] days in order to include 14 data points in the
average.  Again, the results in $J$ and $H$ are consistent with those of
\citet{wood-vasey08} at only the $\sim$1--2$\sigma$ level.  In both cases,
we suspect that the relatively poor agreement is due more to 
differences in the methods used to fit and combine the data rather
than differences in the data themselves.

\begin{figure}
  \epsscale{0.9}
  \plotone{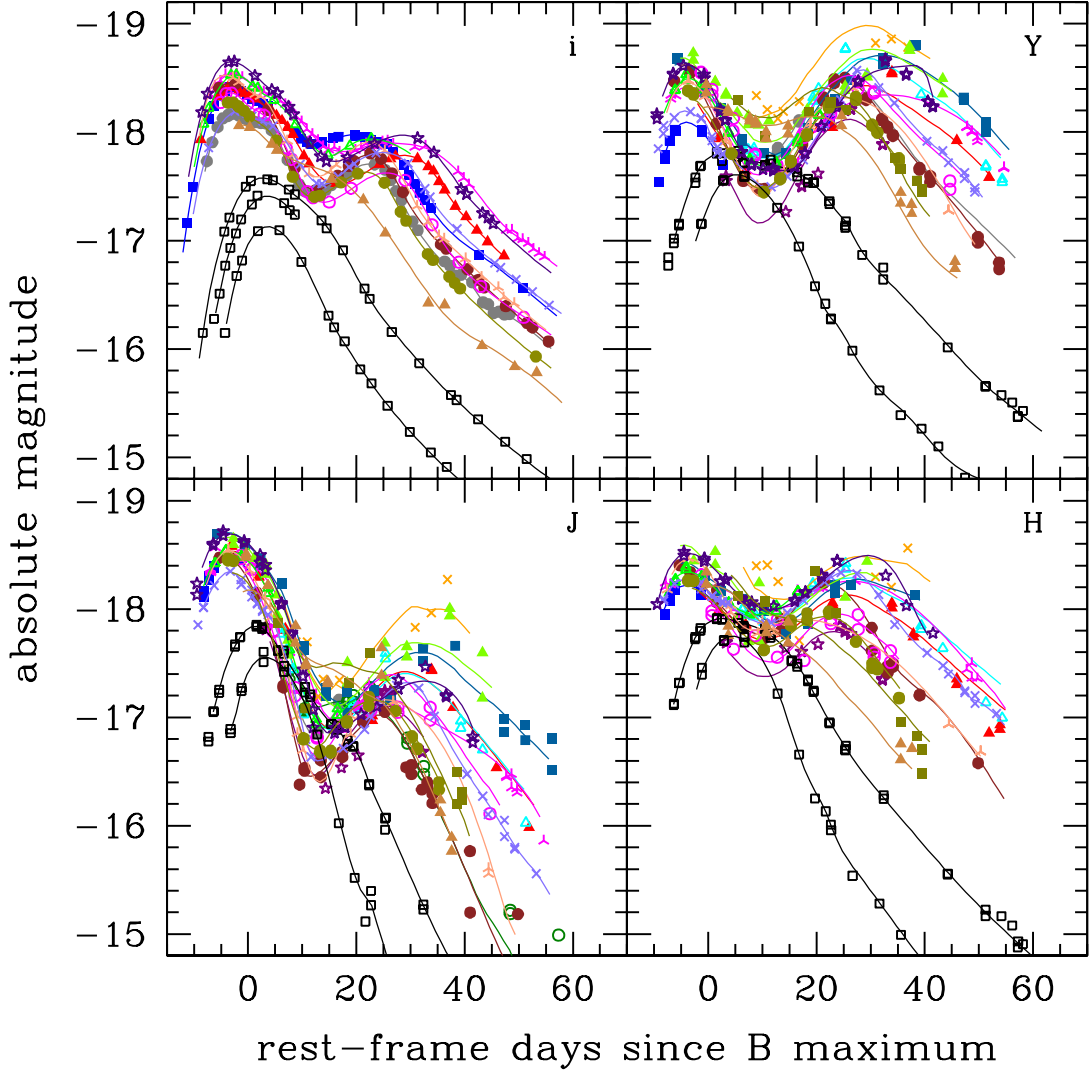}
  \caption{Absolute $iYJH$ light curves for the subsample of best-observed
    SNe. The observed magnitudes are corrected for K-corrections,
    extinction in the Galaxy (assuming $R_V=3.1$), and extinction in the
    host galaxies (using the $R_X^{YZ}$ values of Fits. 11 to 14). The times of
    observation are put in the rest frame of the SNe, using the
    heliocentric redshifts, and referred to
    the time of $B$ maximum. The distances to the SNe are determined as
    described in \S~\ref{sec:absmag}.
    The data from different SNe are shown with different
    symbols and the template light-curve fits are included for
    guidance.
    \label{fig:jointnir}}  
\end{figure}

Figure~\ref{fig:jointnir} shows the absolute $iYJH$ light curves
for the sample of best-observed CSP SNe.  The time axis is plotted as 
rest-frame days since maximum light in $B$.  The Hubble-flow distances of 
equation~(\ref{eq:mu}) along with the Cepheid and SBF distances given in 
Table~\ref{tab:dist} are used to convert the observed magnitudes to 
absolute values.  K-corrections (see \S~\ref{sec:lcs}), Galactic-extinction 
corrections with $R_V^{\mathrm{Gal}}=3.1$, and host-galaxy extinction 
corrections using the measurements of color excesses $E(V-X)$ and the 
$R_X^{YZ}$ parameters obtained from Fits~11--14 for $iYJH$ have also been 
applied.  For the fast-declining SNe~2005ke and 2006mr, zero host-galaxy
extinction is assumed. 

Ignoring for the moment the two fast-declining SNe 2005ke and 2006mr,
Figure~\ref{fig:jointnir} indicates that the smallest dispersion in absolute 
magnitude occurs around the time of the primary maximum in each filter.
By the time of the minimum between the primary and secondary maximum, the
dispersion increases in all bands.  This observation is at odds with
the suggestion by \citet{kasen06} that the minimum of the $J$-band light 
curves of SNe~Ia should show a low dispersion in luminosity. 
At epochs later than the second maximum, the data in $iYJH$ show
increasing spread due to differences in the decline rates among the
SNe (see also \S~\ref{sec:lcs} and Figure~\ref{fig:declro}).  

The two fast-declining events (SN~2005ke and SN~2006mr) do not reach maximum 
brightness until $\sim$5~days after the epoch of $B$ maximum.  
Interestingly, Figure~\ref{fig:jointnir} shows that at this epoch, their 
$J$-band luminosities are similar to those of the rest of the SNe which 
are already declining from the first maximum.

\subsection{Combined Hubble Diagram}
\label{sec:hubb}

\begin{figure}
\epsscale{1.0}
\plotone{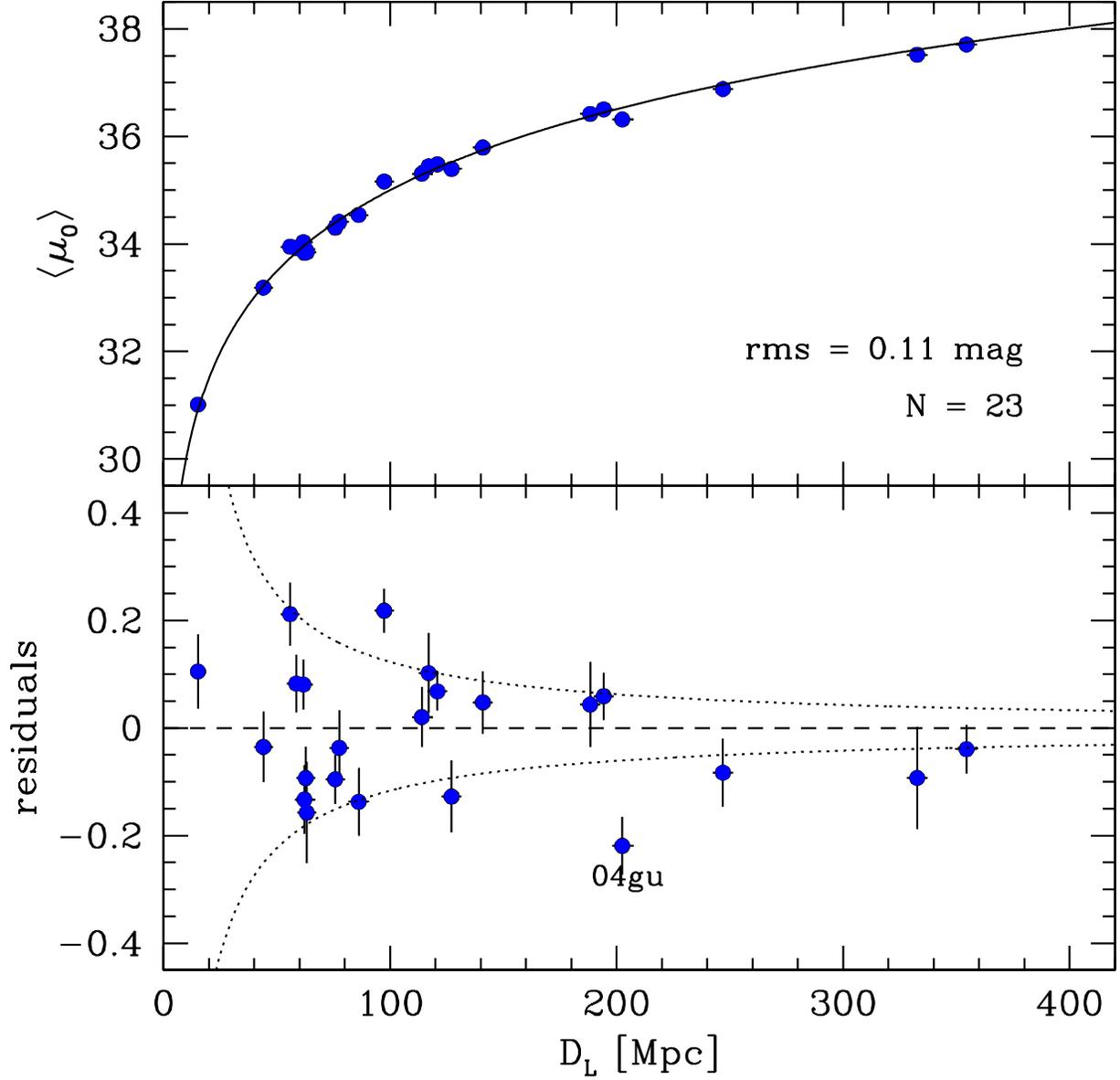}
\caption{Hubble diagram for the sample of best-observed SNe out
  to $z \approx 0.08$. 
  The top panel shows the combined distance moduli for each SN from
  the fits of absolute peak magnitude vs. decline rate and
  reddening in $ugriBVYJH$. The solid line shows the
  adopted concordance model of equation~(\ref{eq:mu}). The dotted
  lines represent the spread predicted by a dispersion of 382~km~s$^{-1}$
  in the measured redshifts due to peculiar motions of the host
  galaxies. The bottom panel shows the residuals with respect to the model fit.
  \label{fig:hubb}}    
\end{figure}

The fits presented in \S~\ref{sec:ld} can be used to place the SNe with 
$0.7 < \dm <1.7$ mag in a Hubble diagram.  Distance moduli were computed for
each SN in $ugriBVYJH$ using the parameters given in Table~\ref{tab:ld} for 
the fits to the best-observed SNe (Fits~4 and Fits~8--14).  Note that Fit~7
involving $V$ and $E(B-V)$ was not included because it is mathematically 
equivalent to that involving $B$ and $E(B-V)$ (Fit~4).  The distance
moduli for all available bands were then averaged for each SN.  
The full correlation matrix among distance moduli from all bands
  was estimated in order to weight the average. Correlation might be expected
to arise from the fact that all fits involve the same decline-rate
parameter $\dm$, and in several cases, the same bands in the 
color-excess term. These
averaged values are plotted versus the luminosity distance, $d_L$, in 
the top half of  Figure~\ref{fig:hubb}.  A combined scatter of 
$0.11$~mag, or $\sim$5\% in distance, is found for the 23 best-observed 
SNe that are included in the diagram.

In the bottom panel where the residuals from the fit are plotted, it may be
seen that the dispersion decreases with distance, implying that it is due
predominantly to the peculiar velocities of the individual host galaxies.
Plotted for reference as dotted lines is the spread predicted by the 
$1\sigma$ dispersion of 382~km~s$^{-1}$, as derived by \citet{wang06} 
from an independent sample of 56 SNe~Ia.  In general, the distance
modulus residuals are consistent with the latter value, although 
it is interesting that the four most-distant SNe all have negative 
residuals (meaning that they are overluminous).  The most discrepant
of these objects is SN~2004gu which, as discussed in C09, 
appears to have been spectroscopically similar to the peculiar SN~Ia 
2006gz, a slow-declining, overluminous SN~Ia that may have resulted 
from the merger of two white dwarfs \citep{hicken07}.  Both SNe 
displayed nearly identical decline rates ($\dm \approx 0.7$), and very
similar luminosities ($M_B \approx -19.6$ mag).  The other three SNe (2005ag,
2005ir, and 2006py) are all slow-decliners ($0.86 \leq \dm \leq 1.02$ mag) 
and, therefore, luminous events.  In principle, the luminosity
vs. decline-rate relationship should correct for this.  Nevertheless,
SN~2005ag was discovered by the LOSS, and must have been at the limit of 
detection.  Hence, Malmquist bias could explain its overluminous nature.
However, SNe~2005ir and 2006py were discovered by the much deeper 
SDSS~II survey which must have negligible bias at these redshifts.
SN~2004gu had a moderately large color excess of $E(B-V)_{max} = 0.20$,
but the other three events had color excesses very close to the median
value of $E(B-V)_{max} = 0.05$ for the full subset of best-observed SNe.  
Hence, it is unlikely that the negative residuals in Figure~\ref{fig:hubb} 
can be ascribed to inaccurate reddening corrections.  Since the points
for SNe~2005ag, 2005ir, and 2006py all lie within $1\sigma$ of the expected
velocity spread due to peculiar velocites, we consider
small-number statistics to be the most likely explanation. Clearly it will 
be interesting to re-examine the Hubble diagram residuals once the full CSP 
sample has been produced.

\begin{figure}
\epsscale{0.85}
\plotone{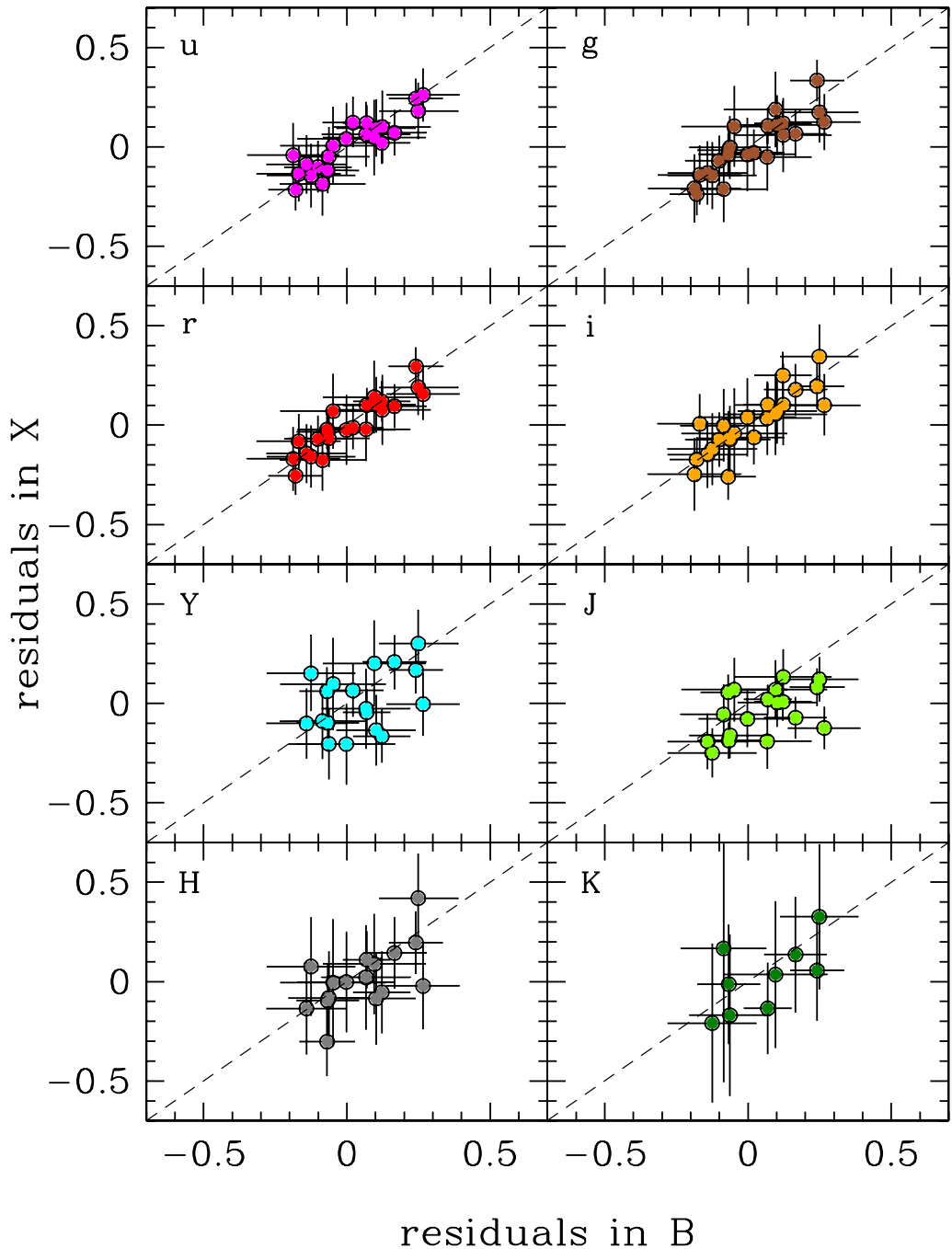}
\caption{Residuals in the distance moduli calculated in band~$X$, where
  $X = ugriYJHK$,
  plotted versus the residuals in the $B$-band distance moduli.
  Note the significant correlation between these.
  \label{fig:hubb_resid}}    
\end{figure}

Interestingly, although the Hubble diagram in Figure~\ref{fig:hubb} 
was produced by averaging the distance moduli derived in each filter,
the resulting dispersion is not much better than the dispersions
obtained in the individual filters (cf. Table~\ref{tab:ld}).  The
explanation for this is found in Figure~\ref{fig:hubb_resid}, where
the distance-modulus residuals in $ugriYJHK$ are plotted versus
the residual in $B$.  A significant correlation is observed
between the distance-modulus residuals in one band vs. those
in another, particularly in the $ugri$ bands.  
For this reason, the 
combined residuals do not reduce significantly the scatter in the 
diagram as compared with the fits to the individual bands. 

We measured
  correlation coefficients, $r$, among residuals in $Bugri$ to be
  $0.8$--$0.9$, and $0.3$--$0.6$ between any band and $YJHK$. Most
  likely the lower correlation in the NIR is attributable to the
  difficulties of   
measuring precise peak magnitudes in these bands as discussed
earlier. We compared the observed correlation coefficients with those
estimated for each SN based on the measurement uncertainties and the
form of the model in equation~(\ref{eq:modld}). We found the expected
correlation coefficients to be $r<0.05$ for all cases, except
between $B$ and $i$ bands where we estimate $r \approx 0.6$, and between
$B$ and $Y$ where we expect $r \approx0.3$. For these two cases, the
high correlation may in part be due to the use of $E(B-V)$ in the
color-excess term. For all the rest of the cases, the contribution to
the correlations due to the fit model is negligible.
 We thus interpret the dispersion 
in Figure~\ref{fig:hubb} to be mostly due to peculiar velocities. 
If this is correct, then the true precision of the SN distances is
the dispersion about the relations in Figure~\ref{fig:hubb_resid}, {\em which
for the $ugri$ bands amounts to 0.06--0.09~mag in distance modulus, 
or 3--4\% in distance}.  

\section{DISCUSSION AND CONCLUSIONS}
\label{sec:concl}

This paper presents a first analysis of the light curves of 34 SNe~Ia 
followed by the CSP and released by C09. The high photometric precision and 
dense time sampling of the observations, especially in the optical bands, 
has allowed an in-depth examination of the general properties of the light 
curves of SNe~Ia in the $ugriBVYJH$ bands.  Subsamples of well-observed 
SNe --- 26 in the optical and 9 in the NIR ---  were used to build a family 
of template light curves using a technique that allows one to interpolate 
among the template data, taking into account variations in the sampling, 
in a 3-D space parameterized by the epoch relative to maximum 
light, the magnitude relative to maximum, and the decline rate.
 
The availability of observations covering a wide range of wavelengths (from 
$u$ to $K_s$) offers the opportunity to make further progress on the 
difficult issue of correcting SN luminosities for extinction outside the 
Galaxy.  However, depending on the approach taken, two quite different
results are found.  In the first case, a subsample of 15 SNe assumed to have 
suffered little or no extinction in their host galaxies was used to measure
color excesses for the whole sample.  By plotting the ratios of color 
excesses involving optical and NIR filters and comparing the values predicted 
by the CCM+O extinction law for interstellar dust, a value of the
total-to-selective absorption coefficient of $R_V \approx 1.7$ was derived,
which is significantly lower than the Galactic average of $3.1$.  However, 
this value is largely influenced by two very red objects in the sample, SNe 2005A 
and 2006X.  When the same calculations are repeated after excluding these two 
objects, a value of $R_V=3.2\pm0.4$ is found, in agreement with the Galactic 
average. 

An alternative way of estimating $R_V$ is to express the absolute magnitudes 
of the SNe as a two-parameter function of the decline rate and color, 
and then to use the Hubble-flow distances (or Cepheid and SBF distances) of 
the SNe to derive the best fit parameters through $\chi^2$ minimization.  
This method was first employed by \citet{tripp98} and \citet{tripp99}, who 
related the absolute magnitude in $B$ to the $B-V$ color and derived values 
of $R_V \approx$ 1--1.5.  When this same technique is applied to the CSP sample of 
SNe~Ia using the peak magnitudes in $B$ and $J$ bands and the $B-V$ and 
$V-J$ colors, a value of $R_V \sim$1--2 is obtained, regardless of whether 
the heavily reddened events, SNe 2005A and 2006X, are included or excluded.  
An equivalent analysis can also be carried out relating absolute magnitude to 
the decline rate and color excess.  Performing such fits for all available 
bands ($ugriBVYJHK_s$) again leads to systematically low values of the CCM+O-law 
parameter ($R_V \approx 1.5$) regardless of whether the two highly reddened 
SNe are included in the sample.

These conflicting results on the value of $R_V$ reflect the variety of results 
found in the literature on this subject.  Most early attempts to derive the 
average properties of the host-galaxy reddening indicated that $R_V$ was 
unusually low \citep[see][and references therein]{branch92}.  However, the 
available light curves at the time were largely photographic and there was 
considerable additional uncertainty regarding the nature of the intrinsic
colors of SNe~Ia at maximum light.  The first modern study based on CCD data
was made by \citet{riess96}, who used the ratios of color excesses measured 
in $B-V$, $V-R$, and $V-I$ for a sample of 20 SNe~Ia to derive a value of 
$R_V = 2.55\pm0.30$.  This analysis differed from previous studies in that 
the intrinsic color variation as a function of decline rate was taken fully 
into account in deriving the color excesses.  Similar values of $R_V$ 
have been found by \citet{phillips99}, \citet{altavilla04}, \citet{reindl05}, 
and \citet{wangx06}.  However, beginning with
\citet{tripp98}, studies based on the technique of minimizing the dispersion 
in the Hubble diagram of SNe~Ia have nearly uniformly led to values
of $R_V$ in the range of $\sim$1--2 \citep[e.g.,][]{tripp99, astier06, 
conley07}.  If there is a pattern to these results, including our own
in this paper, it would seem that attempts to measure $R_V$ via comparison
of colors or color excesses tend to give larger values than the procedure
of minimizing the scatter in the Hubble diagram with $R_V$ treated as a free
parameter.  Nevertheless, there are exceptions to this rule such as the
recent paper by \citet{nobili08}, who studied the color evolution in $U-B$, 
$B-V$, $V-R$, and $R-I$ of 69 SNe~Ia with moderate reddening and 
obtained a value of $R_V = 1.01\pm0.25$.

Common sense suggests that at least some portion of the observed reddening
of SNe~Ia in spiral galaxies like the Milky Way must be due to dust in the 
interstellar medium of these galaxies.  Although only a few studies have been 
carried out of the nature of the reddening law in external galaxies, these are
generally consistent with $R_V \approx 3$, albeit with a large dispersion
\citep{goudfrooij94,patil07,ostman08}.  As shown in \S~\ref{sec:cered}, the 
ratios of the measured color excesses involving optical and NIR bands with 
respect to $E(B-V)$ are consistent with values of $R_V \approx 3$.  As discussed 
by \citet{krisciunas07}, inclusion of the NIR bands in this type of analysis 
improves considerably the accuracy with which $R_V$ can be measured.  Hence, 
it is tempting to conclude that we are measuring reddening produced by dust 
with very similar properties to that found in the interstellar medium of the 
disk of the Milky Way.  It will be interesting to see if this finding holds 
up as many more SNe~Ia with reliable optical$-$NIR color excesses are added
to the CSP sample.

If we are, indeed, observing host-galaxy reddening due to ``normal'' 
Galactic-type dust, then the fact that low values of $R_V$ ($\sim$1--2) 
are obtained when it is treated as a free parameter in minimizing the 
dispersion in the Hubble diagram is puzzling.  The implication is that there 
is an intrinsic dispersion in the colors of SNe~Ia which is correlated
with luminosity, but is independent of the decline rate.  
Our finding in \S~\ref{sec:cered} that there is a small but systematic 
difference between color-excess measurements, $E(B-V)$, made at 
late epochs using the Lira law and those derived from the maximum
light magnitudes, and that this difference correlates with color
and perhaps also with
absolute magnitude, may well be related to this.

Very red SNe potentially allow the dust reddening law to be studied on
an individual basis.  Including the results for SN~2005A given in this paper,
there are now six SNe~Ia for which a determination of $R_V$ has been made
from optical and NIR photometry.  The results for these objects are
summarized in Table~\ref{tab:redsne}.  A weighted mean of the six 
determinations gives $R_V = 1.6$, with a surprisingly small rms of 0.3.
The relatively narrow range of decline rates is also noteworthy, averaging
$\dm = 1.2$ with a dispersion of only 0.1.  Except for SN~2001el, the color 
excesses for these SNe are all very large.  \citet{jha07} found that the 
distribution of $E(B-V)$ for SNe~Ia is well approximated by an exponential 
function with $\tau = 0.138 \pm 0.023$~mag.  Hence, values of $E(B-V) > 1$ mag
should be extremely rare, although it must be kept in mind that the sample of 
SNe analyzed by Jha et al. was culled from several sources with different 
selection criteria.  Assuming that the Jha et al. distribution is correct, 
less than one SN for every thousand observed would be expected to have such a 
large amount of reddening --- yet according to the Asiago Supernova 
Catalog\footnote[15]{http://web.oapd.inaf.it/supern/cat/ .}, the total number of 
nearby ($z < 0.02$) SNe~Ia discovered from 1985 through 2008 was only 358.  The
fact that these heavily reddened SNe are much more common than expected,
combined with the remarkable similarity of decline rates and $R_V$ values for 
the five objects in Table~\ref{tab:redsne} with $E(B-V) > 1$ mag, leads us to 
speculate that these events actually represent a physically distinct 
subclass of SNe~Ia.  Hence, while 
extremely interesting in themselves, these objects like SNe 2005A and 2006X
are not representative of the class of SNe~Ia employed in cosmological
studies.

\cite{wang05} has suggested that the presence of circumstellar dust
distributed in a shell around the SN can produce reddening compatible
with the observed low values of  $R_V$.  This hypothesis is supported
by our observations of SNe~2005A and 2006X.  Specifically, we find that 
the red colors of these two SNe are better matched by  a model 
where multiple scattering of photons by circumstellar dust steepens the 
effective extinction law \citep{goobar08}, than by the standard CCM+O 
Galactic reddening law with a low value of $R_V$.

Our results on the absolute peak magnitude calibrations for this new sample
of CSP SNe confirm those obtained in previous studies, while extending the
calibration to the optical $ugri$ and the NIR $YJHK_s$ bands.  Fits to
the subsample of best-observed SNe employing the \citet{tripp98} model, which 
assumes the absolute magnitudes to be a two-parameter function of the decline 
rate and color, give rms dispersions of 0.12--0.15~mag.  Remarkably, the
same fits are found to apply equally well to either highly-reddened objects 
such as SNe 2005A and 2006X, or fast-declining, intrinsically-red events like 
2005ke and 2006mr.  The simplicity of this method is appealing.
We have also carried out fits which assume the absolute peak magnitudes 
to be a function of the decline rate and the measured color excesses.  These
give quite similar results to those obtained with the Tripp model.  Fits to
a large number of filters and colors give dispersions in the range of
0.12--0.20~mag for SNe with $0.7 < \dm <1.7$ mag.  

The quality of our NIR light 
curves allows, for the first time, detection of a weak dependence of the 
$J$-band luminosity on decline rate.  Combining the calibrations 
from all bands, we created a single Hubble diagram for the 23 best-observed
SNe.  Although the resulting scatter of $0.12$~mag (6\% in distance) is
excellent, it does not represent a significant improvement over the precision
obtained with individual filters.  This is attributed to the fact that,
for any particular SN, there is a significant correlation in the error in
the distance moduli obtained using different filters.  The source of these 
correlated errors appears to be the peculiar velocities of the SN host
galaxies.  If true, this implies that the derived distances to the SNe are 
actually
precise to 3--4\%, making SNe~Ia competitive with even Cepheid variables
as extragalactic distance indicators.

Finally, the set of NIR light curves presented here confirms the advantages 
of working in this wavelength region for cosmological studies.  Fit~8 of 
Table~\ref{tab:tb} and Fit~13 of Table~\ref{tab:ld} show that using either 
the $V-J$ color or color excess to correct the $J$ absolute 
magnitudes of the best-observed SNe (excluding the fast decliners) yields a 
dispersion of only 0.12~mag.  Moreover, this result is insensitive to the 
exact form of the reddening law since the extinction corrections in the NIR 
are small.  Note that the values derived for $\sigma_{rm SN}$ of 0.02--0.04~mag
suggest that the intrinsic dispersion in the $J$ band may be
quite small.  This is obviously a preliminary result which must be treated
with caution --- however, we will soon have a sample of $\sim$80
SNe~Ia to test this. 
If confirmed, it implies that if the observational uncertainties can be 
reduced, there is much to be gained by extending the rest-wavelength coverage 
of future dark-energy experiments to include the $J$ band. 

\acknowledgments 

This material is based upon work supported by the National Science 
Foundation (NSF) under grant AST--0306969. 
We also acknowledge support from {\it Hubble Space Telescope} grant 
GO-09860.07-A from the Space Telescope Science Institute, which is operated by
the Association of Universities for Research in Astronomy, Inc., under
NASA contract NAS 5-26555. M.H. acknowledges support provided by NASA 
through Hubble Fellowship grant HST-HF-01139.01-A, by Fondecyt through 
grant 1060808, from Centro de Astrof\'\i sica FONDAP 15010003, and by the 
Center of Excellence in Astrophysics and Associated Technologies (PFB 06).
G.F., M.H., and F.S. acknowledge support from the Millennium Center for
Supernova Science through grant P06-045-F funded by ``Programa Bicentenario
de Ciencia y Tecnolog\'ia de CONICYT'' and ``Programa Iniciativa Cient\'ifica
Milenio de MIDEPLAN.''  N.B.S. acknowledges the support of the 
Mitchell/Heep/Munnerlyn Chair in Astronomy at Texas A\&M University, and 
support though the Dean of the College of Sciences.
A.V.F.'s supernova research has been funded by NSF
grants AST-0607485 and AST-0908886, as well as by the TABASGO Foundation.
We thank James Hughes for supporting our network of computers, and the 
technical staff of Las Campanas Observatory for its help during many observing 
nights.  This research has made use of the NASA/IPAC Extragalactic Database 
(NED) which is operated by the Jet Propulsion Laboratory, California Institute 
of Technology, under contract with the National Aeronautics and Space 
Administration.

\appendix
\section{Chi-Squared Fitting}
\label{app:fits}
The problem of calibrating the low-redshift CSP SNe in a particular band $X$
amounts to fitting a surface in a 3-dimensional
parameter space defined by (1) a decline-rate parameter, $\dm$; (2) a color
parameter $c$, which could represent a pseudocolor at maximum, a color
excess, or some other color indicator; and (3) the luminosity, $M_X$.  For our
analysis, we choose to fit this surface with a simple plane:
\[
   M_X\left(\dm,c\right)  = M_X(0) + b_X\,[\dm - 1.1]\;+\;\beta_X\,c .
\]
The $M_X$ values themselves are derived from the observed magnitudes $m_X$ by 
a distance modulus $\mu$ which, for the closest objects, can be derived from
independent data (e.g., Cepheids, SBF).  For the more distant objects, we
use Hubble's law, assuming $H_0 = 72$ km~s$^{-1}$ Mpc$^{-1}$.  
Our model
for the observed magnitudes is therefore (cf. equations~(\ref{eq:modtb}) and
(\ref{eq:modld}))

\begin{equation}
\label{eq:model}
\bar{m}_X\;=\;M_X(0)\;+\;b_X\,[\dm-1.1]\;+\;\beta_X\, c\;+\mu .
\end{equation}

In the case where $c$ is interpreted as a color excess, we use the 
intrinsic color relation derived in \S~\ref{sec:maxpseu} to compute
\begin{equation}
\label{eq:ce}
c\,\equiv\,E(Y-Z)\,=\,(m_Y-m_Z)\;-\;A_0\;-\;B_0\,[\dm\,-\,1.1].
\end{equation}
Substituting this in equation (\ref{eq:model}) gives
\begin{equation}
\label{eq:model2}
\bar{m}_X\;=\;(M_X(0) - \beta_X A_0)\;+\;(b_X-\beta_X B_0)\,[\dm-1.1]\;+\;\beta_X\, c\;+\mu .
\end{equation}
In the case where $c$ is implemented as a pseudocolor (\S~\ref{sec:tb}),
the same model can be used, simply setting $A_0$ and $B_0$ to zero.  In this way, we
see that the two treatments of $c$ are mathematically equivalent.  We also see
very clearly that any uncertainties in the values of $A_0$ and $B_0$ are to be
treated as systematic errors and should not be included in the statistical error
budget for the color excesses.  Rather, they should be included in the
systematic error budget for the $M_X(0)$ and $b_X$.

Given a set of $N$ SNe with measured
values \{$\mu_i$; ${m_X}_i$;  $\dm_i$; $c_i$\} for $i=1,\dots,N$, we
construct the $\chi^2$ function 
\begin{equation}
\label{eq:chisq}
\chi^2\left(M_X(0),b_X,\beta_X,\sigma_{SN}\right)\,=\,\sum_{i=1}^N{\frac{({m_X}_i\;-\;{\bar{m}_X})^2}{\sigma_i^2\;+\;\sigma_{SN}^2}},
\end{equation}
where $\sigma_i^2$ are the measurement variances,
and $\sigma_{SN}$ is an additional constant term to account for any possible
intrinsic dispersion of the SN data which has not been taken into account
by the model \citep[see][]{Tremaine2002}.  The distance moduli, $\mu_i$, are computed using
equation~(\ref{eq:mu}).

\subsection{Variances}

In computing the denominator of the $\chi^2$ function, care must be taken in 
estimating not only the variances of the data, but also the covariances between
the different observables.  We outline below how these variances and covariances
are estimated.  For each supernova, the total variance is (we omit the
subscript $i$ for clarity)
\begin{eqnarray}
\label{eq:sigmai}
\sigma^2 & = & \sigma^2(\mu)\;+\;\sigma^2(m_X)\;+\;b_X^2\,\sigma^2(\dm)\;+\;\beta_X^2\,\sigma^2(c)\nonumber\\
           & - & 2\,b_X\,\mathrm{cov}[m_X,\dm]\;-\;2\,\beta_X\,\mathrm{cov}[m_X,c]\nonumber\\
           & + & 2\,b_X\,\beta_X\,\mathrm{cov}[\dm,c] ,
\end{eqnarray}
where $\sigma^2(q)$ denotes the variance in quantity $q$, $\mathrm{cov}[p,q]$ denotes the
covariance between quantity $p$ and $q$, and we have assumed $\mathrm{cov}[\mu,q] = 0$. 
Furthermore, we assume the peak magnitudes in two different bands are uncorrelated 
($\mathrm{cov}[m_X,m_Y] = \delta_{XY}\sigma^2(m_X)$).  The three covariance terms in
equation (\ref{eq:sigmai}) are computed as follows.
\begin{itemize}
\item For $\mathrm{cov}[m_X,\dm]$, we recall that $\dm \equiv m_B(t=15) - m_B(t=0)$ and so
\begin{eqnarray}
\label{eq:cmdm}
\mathrm{cov}[m_X,\dm] & = & \mathrm{cov}[m_X,m_B(t=15)]\;-\;\mathrm{cov}[m_X,m_B(t=0)]\nonumber\\
                & = & -\delta_{XB}\,\sigma^2(m_X),
\end{eqnarray}
where we have assumed $\mathrm{cov}[m_X,m_B(t=15)]=0$. Note that
we have neglected any possible correlated error introduced by the subtraction 
of the host-galaxy image in the $B$ band.  As discussed by
\citet{contreras09}, such errors are less than the quoted 
statistical errors of the photometry.

\item For $\mathrm{cov}[m_X,c]$ we have, from equation~(\ref{eq:ce}), 
\begin{eqnarray}
\label{eq:cmce}
\mathrm{cov}[m_X,c] & = & \mathrm{cov}[m_X,(m_Y-m_Z)]\;-\;B_0\,\mathrm{cov}[m_X,\dm]\nonumber\\
& = & (\delta_{XY}\;-\;\delta_{XZ}\;+\;B_0\,\delta_{XB})\,{\sigma^2(m_X)}.
\end{eqnarray}

\item For $\mathrm{cov}[\dm,c]$, we also have from equation~(\ref{eq:ce})
\begin{eqnarray}
\label{eq:cdmce}
\sigma_{(\dm,c)} & = & \mathrm{cov}[\dm,(m_Y-m_Z)]\;-\;\mathrm{cov}[\dm,B_0\,\dm]\nonumber\\
 & = & (\delta_{ZB} - \delta_{YB})\,\sigma^2{m_B} - B_0\,\sigma^2(\dm) .
\end{eqnarray}
\end{itemize}

Substituting equations~(\ref{eq:cmdm}), (\ref{eq:cmce}), and
(\ref{eq:cdmce}) into equation~(\ref{eq:sigmai}) for $\sigma^2$
we obtain the final expression for the measurement variances,
\begin{eqnarray}
\sigma^2 & = &
\sigma^2(\mu)\;+\;\left[1\;+\;2\,\delta_{XB}\,b_X\;-\;2\,\beta_X\,(\delta_{XY}\,-\,\delta_{XZ}\,+\,B_0\,\delta_{XB})\right]\,\sigma^2(m_X)\;+\nonumber\\
 & + & \left[b_X^2\;-\;2\,b_X\,\beta_X\,B_0\right]\,\sigma^2(\dm)\;+\;\beta_X^2\,\sigma^2_{c} + 
 2b_X\beta_X(\delta_{ZB} - \delta_{YB})\sigma^2(m_B),
\end{eqnarray} 
where we recall that $B_0=0$ for the case when $c$ represents
a pseudocolor.

\subsection{Solution}

The $\chi^2$ minimization is performed by analytically 
marginalizing equation~(\ref{eq:chisq}) over $M_X(0)$ 
\citep[see][]{goliath01} at each point in a grid of values of
\{$b_X;\beta_X$\}, followed by a search for the minimum value of $\chi^2$.
This process is repeated for various values of
$\sigma_{\rm SN}$ ---usually between $0$ and $0.3$ mag--- in order to find
the value ${\hat{\sigma}}_{\rm SN}$ that produces a minimum reduced $\chi^2$ 
of $\chi^2_{\mathrm{min}}/(N-3)=1$. 

Once ${\hat{\sigma}}_{\rm SN}$  is found, it is used for computing the
probability density $\rho(\chi^2_{\mathrm{marg}})$ at each point in
the grid of \{$b_X;\beta_X$\} values. This probability is given by

\begin{equation} 
\rho(b_X,\beta_X)\,=\,\exp\left\{\frac{-\chi^2_{\mathrm{marg}}}{2}\right\}.
\end{equation}
Finally, the expectation value of each parameter,
$\hat{\theta}$, in the grid of ${\vec{\theta}} \equiv\,
\{M_X(0);b_X;\beta_X\}$ is computed as the moment 
\begin{equation}
\hat{\theta}\,=\,\frac{\int\theta\,\rho({\vec{\theta}})\,d{\vec{\theta}}}{\int\rho({\vec{\theta}})\,d{\vec{\theta}}},
\end{equation}
and the variances in the fit parameters, $\sigma_{\hat{\theta}}$,
are estimated as the second moment
\begin{equation}
\sigma_{\hat{\theta}}^2\,=\,\frac{\int(\theta-\hat{\theta})^2\,\rho({\vec{\theta}})\,d{\vec{\theta}}}{\int\rho({\vec{\theta}})\,d{\vec{\theta}}}.
\end{equation}

\section{Deriving $R_V$}
\label{app:rv}

Here we explain how we compute the coefficients $a_X$ and
$b_X$ for each band $ugriBVYJHK_s$ that allow us to derive $R_V$ from
the slopes $E(V-X)_{\mathrm{max}}/E(B-V)_{\mathrm{max}}$ of
\S~\ref{sec:cered}, and from the fit parameters $R_X^{YZ}$ of
\S~\ref{sec:ld}.

The reddening law is assumed to be a function of wavelength of the
form given by \citet{cardelli89}: 

\begin{equation}
\label{eq:ccm}
\frac{A_\lambda}{A_V}\,=\,a_\lambda\;+\;\frac{b_\lambda}{R_V} .
\end{equation}

\noindent We simulate different amounts of extinction
$A_V$ on the typical spectrum of a SN~Ia at maximum light by 
multiplying equation~(\ref{eq:ccm}) into the spectrum. The spectrum 
we use is the template SN~Ia spectrum of epoch 20 from 
\citet{hsiao07,hsiao09} which corresponds to maximum light in $B$. 
We perform the multiplication for various values of $R_V$, computing 
synthetic photometry from the resulting spectra using the transmission 
functions of the CSP filters described in C09. The {\em observed} 
absorptions, $A_X$ and $A_V$, are then given by the difference 
between the synthetic magnitudes obtained from the reddened template 
spectrum and those from the original unreddened one, for
bands $X$ and $V$. We finally fit a linear relation between $A_X$ and $R_V$ 
as in equation~(\ref{eq:ccm}), and thus derive $a_X$ and $b_X$ for all
bands $ugriBVYJHK_s$.

We have performed these calculations for the original \citet{cardelli89} 
(CCM) reddening law, and also with the modifications introduced by
\citet{odonnell94} (CCM+O). Columns 2--5 of Table~\ref{tab:axbx} list
the values of $a_X$ and $b_X$ obtained for each case. As will be
explained below, these coefficients serve to convert an {\em observed} 
reddening measurement, such as those in \S~\ref{sec:cered} and 
\S~\ref{sec:ld}, to a value of $R_V$. We note that this would be a 
{\em true\,} value of $R_V$ --- that is, the parameter defining the reddening 
law of equation~(\ref{eq:ccm}).

If we instead consider the {\em observed\,} values of $R_V$ derived
from synthetic photometry as $R_V^{\mathrm{obs}} \equiv
A_V^{\mathrm{obs}}/E(B-V)^{\mathrm{obs}}$, we obtain different
coefficients $a_X$ and $b_X$ from the fits of $A_X$ versus
$R_V^{\mathrm{obs}}$. These coefficients are listed in columns 6 and 7
of Table~\ref{tab:axbx} for the case of the CCM+O law. 

The formulae to convert observed slopes
$E(V-X)_{\mathrm{max}}/E(B-V)_{\mathrm{max}}$ to $R_V$ are given in
Appendix A of \citet{krisciunas06} [see their equations~(A5) and (A6)]. We
briefly explain here how to derive $R_V$ from fit parameters
$R_X^{YZ}$ of \S~\ref{sec:ld}.

>From the definition of $R_X\,=\,A_X/E(B-V)$, $R_X^{YZ}$ can be 
expressed as

\begin{equation}
R_X^{YZ}\,=\,\frac{R_X}{R_Y\;-\;R_Z}.
\end{equation}

Substituting equation~(\ref{eq:ccm}) into this expression, and 
recalling that $A_\lambda / A_V$ is equivalent to $R_\lambda / R_V$,
we derive 

\begin{equation}
\label{eq:Rv}
R_V\,=\,-\frac{R_X^{YZ}\,(b_Y\,-\,b_Z)\;-\;b_X}{R_X^{YZ}\,(a_Y\,-\,a_Z)\;-\;a_X}.
\end{equation}

\noindent By error propagation of the uncertainty in $R_X^{YZ}$, the
uncertainty in $R_V$ becomes 

\begin{equation}
\label{eq:eRv}
{\sigma}_{R_V}\,=\,\left|\frac{(b_Y\,-\,b_Z)\,[R_X^{YZ}\,(a_Y\,-\,a_Z)\;-\;a_X]\;+\;(a_Y\,-a_Z)\,[b_X\,-\,R_X^{YZ}\,(b_Y\,-\,b_Z)]}{[R_X^{YZ}\,(a_Y\,-\,a_Z)\;-\;a_X]^2}\right|\,{\sigma}_{R_X^{YZ}}.
\end{equation}

\clearpage
\begin{deluxetable}{lcccccc} 
\tabletypesize{\small} 
\tablecolumns{7} 
\tablewidth{0pt} 
\tablecaption{\label{tab:sne}} 
\tablehead{ 
\colhead{} & \colhead{} & \colhead{} & \colhead{$E(B-V)$} &  \colhead{$\dm$} & \colhead{$\langle \mu \rangle$} & \colhead{Best} \\ 
\colhead{SN} & \colhead{$z_{\mathrm{helio}}$} & \colhead{$z_{\mathrm{CMB}}$} & \colhead{Gal (mag)} & \colhead{(mag)} & \colhead{(mag)} & \colhead{Obs.?} \\ 
\colhead{(1)} & \colhead{(2)} & \colhead{(3)} & \colhead{(4)} & \colhead{(5)} & \colhead{(6)} & \colhead{(7)} } 
\startdata
2004ef & 0.0310 & 0.0298 & 0.056 & $1.36$($01$) & $35.46$($03$) & yes \\ 
2004eo & 0.0157 & 0.0147 & 0.108 & $1.37$($03$) & $33.92$($05$) & yes \\ 
2004ey & 0.0158 & 0.0146 & 0.139 & $0.98$($02$) & $34.06$($05$) & yes \\ 
2004gc & 0.0321 & 0.0307\tablenotemark{a} & 0.206 & $1.09$($06$) & $35.34$($04$) & no \\ 
2004gs & 0.0267 & 0.0275 & 0.031 & $1.54$($01$) & $35.51$($03$) & yes \\ 
2004gu & 0.0459 & 0.0469 & 0.026 & $0.80$($04$) & $36.41$($10$) & yes \\ 
2005A & 0.0191 & 0.0184 & 0.030 & $1.17$($03$) & $34.46$($06$) & yes \\ 
2005M & 0.0220 & 0.0230 & 0.031 & $0.85$($01$) & $35.13$($03$) & yes \\
\enddata 
\tablenotetext{a}{Value of $z_{\mathrm{CMB}}$ corresponding to the systemic recession velocity of the group ARP~327}.
\tablecomments{Columns: (1) SN name; (2) Heliocentric redshift from the NASA/IPAC Extragalactic Database (NED); (3) Redshift in the frame of the 3K CMB (NED); (4) Galactic reddening (NED); (5) Observed $\dm$; (6) Average distance modulus from fits of \S~\ref{sec:ld}; (7) Does the SN belong to the ``best-observed'' sample?.}

\end{deluxetable}

\clearpage
\begin{deluxetable}{lcccccc} 
\tabletypesize{\scriptsize} 
\tablecolumns{7} 
\tablewidth{0pt} 
\tablecaption{Light-curve parameters\label{tab:lcpar}} 
\tablehead{ 
\colhead{} & \colhead{$t^{\mathrm{max}}$} & \colhead{$m^{\mathrm{max}}$} & \colhead{$\Delta m_t$} & \colhead{Epoch} & \colhead{} & \colhead{Fit} \\ 
\colhead{Band} & \colhead{$-2,453,000$} & \colhead{(mag)} & \colhead{(mag)} & \colhead{Range} & \colhead{Npts.} & \colhead{Method} \\ 
\colhead{(1)} & \colhead{(2)} & \colhead{(3)} & \colhead{(4)} & \colhead{(5)} & \colhead{(6)} & \colhead{(7)} } 
\startdata
\multicolumn{7}{c}{{\bf SN~2004ef}}\\ 
$B$ & 264.44$\pm$0.05 & 17.059$\pm$0.005 & 0.721$\pm$0.014 & $-8.5$~~~$+63.2$ & 44 & Spline \\ 
$V$ & 266.07$\pm$0.03 & 16.900$\pm$0.002 & 1.120$\pm$0.008 & $-10.1$~~~$+61.6$ & 48 & Spline \\ 
$u$ & 263.01$\pm$0.14 & 17.791$\pm$0.012 & 1.596$\pm$0.068 & $-6.2$~~~$+56.8$ & 41 & Spline \\ 
$g$ & 264.82$\pm$0.06 & 16.964$\pm$0.003 & 1.077$\pm$0.009 & $-8.9$~~~$+62.8$ & 50 & Spline \\ 
$r$ & 266.34$\pm$0.07 & 16.862$\pm$0.003 & 1.625$\pm$0.009 & $-10.3$~~~$+61.3$ & 47 & Spline \\ 
$i$ & 261.21$\pm$0.08 & 17.415$\pm$0.005 & 1.199$\pm$0.013 & $-5.3$~~~$+66.3$ & 47 & Spline \\ 
$Y$ & 261.25$\pm$0.03 & 17.209$\pm$0.034 & \nodata  & $+8.8$~~~$+69.5$ & 7 & Templ. \\
\enddata 
\tablecomments{SNe with names in bold face belong to the ``best-observed'' group (see text). \\ Columns: (1) Filter name; (2) Date of maximum; (3) Apparent peak magnitude (K-corrected); (4) Decline rate measured as the difference in magnitudes at an epoch $t$ and maximum light in each band ($t=10$ d for $B$; $t=15$ d for $ug$; $t=20$ d for $V$; $t=30$ d for $r$; $t=35$ d for $iY$; $t=40$ for $JH$); (5) Epochs covered by data in rest-frame days with respect to the time of maximum; (6) Number of data points; (7) Fitting method (spline or template; see text)}

\end{deluxetable}

\clearpage
\begin{deluxetable}{lcccc} 
\tabletypesize{\small} 
\tablecolumns{5} 
\tablewidth{0pt} 
\tablecaption{Pseudocolors versus $\dm$ relation parameters\label{tab:cmax}} 
\tablehead{ 
\colhead{Pseudocolor} & \colhead{$a$} & \colhead{$b$} & \colhead{rms} & \colhead{$N_{\rm SNe}$} \\ 
\colhead{(1)} & \colhead{(2)} & \colhead{(3)} & \colhead{(4)} & \colhead{(5)} } 
\startdata
$u^{\mathrm{max}}-V^{\mathrm{max}}$ & \phs$0.30$$\pm$$0.04$ & \phs$0.59$$\pm$$0.25$ & 0.14 & 13 \\ 
$B^{\mathrm{max}}-V^{\mathrm{max}}$ & $-0.02$$\pm$$0.01$ & \phs$0.12$$\pm$$0.05$ & 0.06 & 14 \\ 
$g^{\mathrm{max}}-r^{\mathrm{max}}$ & $-0.14$$\pm$$0.02$ & \phs$0.24$$\pm$$0.06$ & 0.06 & 14 \\ 
$B^{\mathrm{max}}-r^{\mathrm{max}}$ & $-0.09$$\pm$$0.02$ & \phs$0.29$$\pm$$0.07$ & 0.07 & 14 \\ 
$V^{\mathrm{max}}-i^{\mathrm{max}}$ & $-0.61$$\pm$$0.02$ & \phs$0.27$$\pm$$0.08$ & 0.09 & 14 \\ 
$V^{\mathrm{max}}-Y^{\mathrm{max}}$ & $-0.67$$\pm$$0.03$ & \phs$0.51$$\pm$$0.17$ & 0.14 & 13 \\ 
$V^{\mathrm{max}}-J^{\mathrm{max}}$ & $-0.63$$\pm$$0.02$ & \phs$0.55$$\pm$$0.10$ & 0.09 & 14 \\ 
$V^{\mathrm{max}}-H^{\mathrm{max}}$ & $-0.77$$\pm$$0.03$ & \phs$0.54$$\pm$$0.11$ & 0.14 & 14 \\ 
$V^{\mathrm{max}}-K^{\mathrm{max}}$ & $-0.59$$\pm$$0.05$ & \phs$0.14$$\pm$$0.18$ & 0.14 & 8 \\ 
\enddata 
\tablecomments{Fits of the type: Pseudocolor $=a+b\,[\dm-1.1]$ valid over the range $0.85 < \dm < 1.76$ mag. \\ Columns: (1) Pseudocolor name; (2) Fit intercept; (3) Fit slope; (4) rms of fit in magnitudes; (5) Number of SNe used in fit.}

\end{deluxetable}

\clearpage
\begin{deluxetable}{lcccccccccc} 
\tabletypesize{\tiny} 
\tablecolumns{11} 
\tablewidth{-5pt} 
\rotate 
\tablecaption{Host-galaxy color excesses from maximum light colors\label{tab:cex}} 
\tablehead{ 
\colhead{SN} & \colhead{$E(B-V)$} & \colhead{$E(B-V)$} & \colhead{$E(u-V)$} & \colhead{$E(g-r)$} & \colhead{$E(B-r)$} & \colhead{$E(V-i)$} & \colhead{$E(V-Y)$} & \colhead{$E(V-J)$} & \colhead{$E(V-H)$} & \colhead{$E(V-K)$} \\ 
\colhead{} & \colhead{Tail} & \colhead{Max} & \colhead{Max} & \colhead{Max} & \colhead{Max} & \colhead{Max} & \colhead{Max} & \colhead{Max} & \colhead{Max} & \colhead{Max} \\ 
\colhead{(1)} & \colhead{(2)} & \colhead{(3)} & \colhead{(4)} & \colhead{(5)} & \colhead{(6)} & \colhead{(7)} & \colhead{(8)} & \colhead{(9)} & \colhead{(10)} & \colhead{(11)} } 
\startdata
2004ef & \phs$0.125$($017$) & \phs$0.089$($006$) & \phs$0.343$($014$) & \phs$0.117$($005$) & \phs$0.137$($007$) & $-0.037$($006$) & \phs$0.117$($034$) & \phs$0.024$($080$) & \nodata & $-0.176$($473$) \\ 
2004eo & \phs$0.082$($018$) & \phs$0.020$($016$) & \phs$0.070$($026$) & $-0.023$($018$) & \phs$0.006$($015$) & $-0.004$($018$) & $-0.244$($021$) & \phs$0.054$($021$) & $-0.093$($022$) & \phs$0.013$($017$) \\ 
2004ey & \phs$0.039$($005$) & $-0.061$($008$) & \phs$0.006$($015$) & $-0.004$($006$) & $-0.043$($008$) & $-0.054$($008$) & $-0.111$($020$) & \phs$0.089$($051$) & $-0.055$($024$) & $-0.159$($046$) \\ 
2004gc & \phs$0.200$($021$) & \phs$0.149$($037$) & \nodata & \phs$0.206$($038$) & \phs$0.235$($041$) & \phs$0.078$($038$) & \nodata & \nodata & \nodata & \nodata \\ 
2004gs & \phs$0.080$($010$) & \phs$0.121$($007$) & \phs$0.300$($020$) & \phs$0.092$($007$) & \phs$0.138$($008$) & \phs$0.080$($009$) & \nodata & \phs$0.063$($019$) & \nodata & \nodata \\ 
2004gu & \phs$0.257$($017$) & \phs$0.200$($024$) & \phs$0.549$($045$) & \phs$0.263$($013$) & \phs$0.358$($025$) & \phs$0.128$($019$) & \nodata & \nodata & \nodata & \nodata \\ 
2005A & \nodata & \phs$1.005$($011$) & \phs$2.309$($047$) & \phs$1.078$($013$) & \phs$1.451$($014$) & \phs$0.611$($018$) & \phs$1.414$($031$) & \phs$1.540$($031$) & \phs$1.483$($039$) & \nodata \\ 
2005M & \phs$0.091$($007$) & \phs$0.050$($005$) & \phs$0.144$($019$) & \phs$0.028$($008$) & \phs$0.067$($008$) & \phs$0.082$($006$) & \phs$0.137$($015$) & \phs$0.210$($048$) & \phs$0.115$($020$) & \phs$0.144$($036$) \\
\enddata 
\tablecomments{Columns: (1) SN name; (2) Host-galaxy reddening measured from the $(B-V)$ tail (\S~\ref{sec:bvtail}); (3)-(11) Host-galaxy reddening measured from pseudocolor at maximum light (\S~\ref{sec:maxpseu}).}

\end{deluxetable}

\clearpage
\begin{deluxetable}{lccccc} 
\tabletypesize{\small} 
\tablecolumns{6} 
\tablewidth{0pt} 
\tablecaption{$R_V$ values from fits to $E(V-X)_{\mathrm{max}}$ versus $E(B-V)_{\mathrm{max}}$ \label{tab:cecmp}} 
\tablehead{ 
\colhead{} & \colhead{$E(V-i)$} & \colhead{$E(V-Y)$} & \colhead{$E(V-J)$} & \colhead{$E(V-H)$} & \colhead{Weighted} \\
\colhead{Sample} & \colhead{Max} & \colhead{Max} & \colhead{Max} & \colhead{Max} & \colhead{Average} 
}
\startdata
Whole &  $1.1\pm0.2$ & $1.6\pm0.2$ & $1.8\pm0.1$ & $1.7\pm0.1$ & $1.7\pm0.1$ \\
Excl. 2005A and 2006X & $3.8\pm1.5$ & $3.5\pm1.4$ & $3.1\pm0.7$ & $3.1\pm0.7$ & $3.2\pm0.4$ \\
\enddata 
\end{deluxetable}

\clearpage
\begin{deluxetable}{lc} 
\tabletypesize{\small} 
\tablecolumns{2} 
\tablewidth{0pt} 
\tablecaption{Conversion of observed to true color excesses\label{tab:obs2tru}} 
\tablehead{ 
\colhead{Color Excess} & \colhead{Correction Factor $C_X$} }
\startdata
$E(V-u)$   & 1.0576 \\ 
$E(V-B)$   & 1.1023 \\ 
$E(V-g)$   & 1.1855 \\ 
$E(V-r)$   & 1.2028 \\ 
$E(V-i)$   & 1.0218 \\ 
$E(V-Y)$   & 1.0370 \\ 
$E(V-J)$   & 1.0142 \\ 
$E(V-H)$   & 0.9951 \\ 
$E(V-K_s)$ & 0.9818 \\ 
\enddata 
\tablecomments{$E(V-X)_{\rm true} = C_X * E(V-X)_{\rm observed}$}

\end{deluxetable}

\clearpage
\begin{deluxetable}{lccll} 
\tabletypesize{\small} 
\tablecolumns{5} 
\tablewidth{0pt} 
\tablecaption{Adopted distances for three SNe with $z<0.01$\label{tab:dist}} 
\tablehead{ 
\colhead{} & \colhead{Host} & \colhead{$\mu$} & \colhead{Distance} & \colhead{} \\
\colhead{SN} & \colhead{Galaxy} & \colhead{(mag)} & \colhead{Method} & \colhead{Reference}
}
\startdata
2005ke & NGC~1371 & $31.84 \pm 0.08$ & SBF &  \citet{tonry01}\tablenotemark{a}\\
2006X  & NGC~4321 & $30.91 \pm 0.14$ & Cepheids &  \citet{freedman01} \\
2006mr & NGC~1316 & $31.59 \pm 0.08$ & SBF &  \citet{cantiello07} \\
\enddata 
\tablenotetext{a}{NGC 1371 is a member of the Eridanus group/cluster.  
\citet{tonry01} give a distance modulus of $32.00\pm0.08$ mag for this group 
based on SBF measurements of 7 members.  
Subtracting 0.16 mag \citep{jensen03} to put this 
on the \citet{freedman01} scale gives $31.84\pm0.08$ mag.}
\end{deluxetable}

\clearpage
\begin{deluxetable}{lcccccccccl} 
\tabletypesize{\scriptsize} 
\tablecolumns{11} 
\tablewidth{0pt} 
\rotate 
\tablecaption{Fits of peak magnitudes versus $\dm$ and pseudocolor\label{tab:tb}} 
\tablehead{ 
\colhead{Fit} & \colhead{Filter} & \colhead{Pseudocolor} & \colhead{} & \colhead{} & \colhead{} & \colhead{$R_V$} & \colhead{$\sigma_{SN}$} & \colhead{RMS} & \colhead{} & \colhead{} \\ 
\colhead{No.} & \colhead{$X$} & \colhead{$(Y-Z)$} & \colhead{$M_X(0)$} & \colhead{$b_X$} & \colhead{$\beta_X^{YZ}$} & \colhead{(CCM+O)} & \colhead{[mag]} & \colhead{[mag]} & \colhead{$N_{SNe}$} & \colhead{Sample} \\ 
\colhead{(1)} & \colhead{(2)} & \colhead{(3)} & \colhead{(4)} & \colhead{(5)} & \colhead{(6)} & \colhead{(7)} & \colhead{(8)} & \colhead{(9)} & \colhead{(10)} & \colhead{(11)} 
}
\startdata
$1$ & $B$ & $(B-V)$ & $-19.10$$\pm$$0.02$ & \phs$0.64$$\pm$$0.10$ & \phs$2.81$$\pm$$0.11$ & \phs$1.52$$\pm$$0.10$ & $0.14$ & $0.17$ & 31 & whole \\ 
$2$ & $B$ & $(B-V)$ & $-19.09$$\pm$$0.02$ & \phs$0.74$$\pm$$0.10$ & \phs$2.74$$\pm$$0.11$ & \phs$1.46$$\pm$$0.09$ & $0.12$ & $0.15$ & 26 & best-observed \\ 
$3$ & $B$ & $(B-V)$ & $-19.10$$\pm$$0.01$ & \phs$0.61$$\pm$$0.11$ & \phs$2.79$$\pm$$0.12$ & \phs$1.50$$\pm$$0.10$ & $0.12$ & $0.16$ & 28 & excluding~fast-decliners \\ 
$4$ & $B$ & $(B-V)$ & $-19.09$$\pm$$0.01$ & \phs$0.74$$\pm$$0.11$ & \phs$2.74$$\pm$$0.11$ & \phs$1.46$$\pm$$0.10$ & $0.09$ & $0.14$ & 23 & best-observed~excl.~fast-decliners \\ 
$5$ & $B$ & $(B-V)$ & $-19.07$$\pm$$0.02$ & \phs$0.81$$\pm$$0.12$ & \phs$2.24$$\pm$$0.37$ & \phs$1.03$$\pm$$0.32$ & $0.09$ & $0.14$ & 21 & best-observed~and~$(B-V)<0.4$ \\ 
$6$ & $J$ & $(V-J)$ & $-18.33$$\pm$$0.03$ & \phs$0.65$$\pm$$0.10$ & \phs$0.20$$\pm$$0.07$ & \phs$1.60$$\pm$$0.32$ & $0.08$ & $0.14$ & 21 & best-observed \\ 
$7$ & $J$ & $(V-J)$ & $-18.36$$\pm$$0.03$ & \phs$0.52$$\pm$$0.09$ & \phs$0.13$$\pm$$0.06$ & \phs$1.32$$\pm$$0.22$ & $0.04$ & $0.12$ & 19 & best-observed~excl.~fast-decliners \\ 
$8$ & $J$ & $(V-J)$ & $-18.44$$\pm$$0.14$ & \phs$0.57$$\pm$$0.14$ & \phs$0.01$$\pm$$0.24$ & \phs$0.96$$\pm$$0.59$ & $0.05$ & $0.12$ & 17 & best-observed~and~$(V-J)<0.0$ \\
\enddata 
\tablecomments{Fits of the type: $\mu_X=m_X^{\mathrm{max}}-M_X(0)-b_X\,[\dm-1.1]-\beta_X^{YZ}\, (Y-Z).$ \\ Columns: (1) Fit identifier; (2) Filter corresponding to $m_X^{\mathrm{max}}$; (3) Color; (4) Absolute magnitude for $\dm=1.1$ and zero $(Y-Z)$ color; (5) Luminosity-decline rate slope; (6) Luminosity-color slope; (7) Corresponding parameter $R_V$ of the CCM+O reddening law; (8) Resulting intrinsic dispersion of SN data; (8) RMS of fit in magnitudes; (10) Number of SNe used in fit; (11) Sample of SNe used in fit (see text).}

\end{deluxetable}

\clearpage
\begin{deluxetable}{lcccccccccl} 
\tabletypesize{\scriptsize} 
\tablecolumns{11} 
\tablewidth{0pt} 
\rotate 
\tablecaption{Fits of peak magnitudes versus $\dm$ and reddening\label{tab:ld}} 
\tablehead{ 
\colhead{Fit} & \colhead{Filter} & \colhead{Color excess} & \colhead{} & \colhead{} & \colhead{} & \colhead{$R_V$} & \colhead{$\sigma_{SN}$} & \colhead{RMS} & \colhead{} & \colhead{} \\ 
\colhead{No.} & \colhead{$X$} & \colhead{$E(Y-Z)$} & \colhead{$M_X(0)$} & \colhead{$b_X$} & \colhead{$R_X^{YZ}$} & \colhead{(CCM+O)} & \colhead{[mag]} & \colhead{[mag]} & \colhead{$N_{SNe}$} & \colhead{Sample} \\ 
\colhead{(1)} & \colhead{(2)} & \colhead{(3)} & \colhead{(4)} & \colhead{(5)} & \colhead{(6)} & \colhead{(7)} & \colhead{(8)} & \colhead{(9)} & \colhead{(10)} & \colhead{(11)} 
}
\startdata
$1$ & $B$ & $E(B-V)$ & $-19.07$$\pm$$0.01$ & \phs$1.03$$\pm$$0.25$ & \nodata & \nodata & $0.19$ & $0.19$ & 12 & low~reddening \\ 
$2$ & $B$ & $E(B-V)$ & $-19.39$$\pm$$0.02$ & \phs$0.97$$\pm$$0.41$ & \phs$4.53$ & \phs$3.10$ & $0.54$ & $0.53$ & 28 & whole \\ 
$3$ & $B$ & $E(B-V)$ & $-19.15$$\pm$$0.02$ & \phs$0.94$$\pm$$0.11$ & \phs$2.79$$\pm$$0.12$ & \phs$1.50$$\pm$$0.11$ & $0.13$ & $0.16$ & 28 & whole \\ 
$4$ & $B$ & $E(B-V)$ & $-19.14$$\pm$$0.01$ & \phs$1.07$$\pm$$0.11$ & \phs$2.74$$\pm$$0.11$ & \phs$1.46$$\pm$$0.10$ & $0.10$ & $0.14$ & 23 & best-observed \\ 
$5$ & $B$ & $E(B-V)$ & $-19.14$$\pm$$0.03$ & \phs$0.94$$\pm$$0.12$ & \phs$2.75$$\pm$$0.38$ & \phs$1.46$$\pm$$0.33$ & $0.13$ & $0.16$ & 26 & excluding~05A~and~06X \\ 
$6$ & $B$ & $E(B-V)$ & $-19.11$$\pm$$0.02$ & \phs$1.08$$\pm$$0.11$ & \phs$2.22$$\pm$$0.37$ & \phs$1.01$$\pm$$0.31$ & $0.09$ & $0.14$ & 21 & best-observed~excl.~05A~and~06X \\ 
$7$ & $V$ & $E(B-V)$ & $-19.12$$\pm$$0.01$ & \phs$0.95$$\pm$$0.11$ & \phs$1.74$$\pm$$0.11$ & \phs$1.46$$\pm$$0.10$ & $0.09$ & $0.14$ & 23 & best-observed \\ 
$8$ & $u$ & $E(u-V)$ & $-18.83$$\pm$$0.02$ & \phs$1.59$$\pm$$0.12$ & \phs$1.74$$\pm$$0.05$ & \phs$1.10$$\pm$$0.07$ & $0.10$ & $0.13$ & 22 & best-observed \\
\enddata 
\tablecomments{Fits of the type: $\mu_X=m_X^{\mathrm{max}}-M_X(0)-b_X \, [\dm-1.1]-R_X^{YZ}\, E(Y-Z).$ \\ Columns: (1) Fit identifier; (2) Filter corresponding to $m_X^{\mathrm{max}}$; (3) Color excess; (4) Absolute magnitude for $\dm=1.1$ and no reddening; (5) Luminosity-decline rate slope; (6) Total-to-selective absorption coefficient $R_X^{YZ}\equiv A_X/E(Y-Z)$; (7) Corresponding parameter $R_V$ of the CCM+O reddening law; (8) Resulting intrinsic dispersion of SN data; (9) RMS of fit in magnitudes; (10) Number of SNe used in fit; (11) Sample of SNe used in fit (see text).}

\end{deluxetable}

\clearpage
\begin{deluxetable}{lcccccl} 
\tabletypesize{\small} 
\tablecolumns{7} 
\tablewidth{0pt} 
\tablecaption{Precision of the SNe~Ia as standard candles in the NIR\label{tab:absnir}} 
\tablehead{ 
\colhead{Band} & \colhead{$M_X^{\mathrm{max}}$} & \colhead{Adopted} & \colhead{$R_V$} & \colhead{RMS} & \colhead{} & \colhead{} \\
\colhead{$X$} & \colhead{(avg.)} & \colhead{$R_X^{YZ}$} & \colhead{(CCM+O)} & \colhead{[mag]} & \colhead{$N_{SNe}$} & \colhead{Sample} 
}
\startdata
$Y$ & $-18.44\pm0.07$ & $R_Y^{VY}=0.32$ & 1.7 & 0.24 & 17 & best-observed \\
$Y$ & $-18.47\pm0.08$ & $R_Y^{VY}=0.60$ & 3.1 & 0.26 & 15 & excluding~05A~and~06X \\
$Y$ & $-18.43\pm0.08$ & $R_Y^{VY}=0.0$ & 0.0 & 0.19 & 15 & excluding~05A~and~06X \\
\\
$J$ & $-18.43\pm0.07$ & $R_J^{VJ}=0.13$ & 1.3 & 0.18 & 18 & best-observed \\
$J$ & $-18.44\pm0.07$ & $R_J^{VJ}=0.39$ & 3.1 & 0.20 & 16 & excluding~05A~and~06X \\
$J$ & $-18.42\pm0.07$ & $R_J^{VJ}=0.0$ & 0.0 & 0.18 & 16 & excluding~05A~and~06X \\
\\
$H$ & $-18.42\pm0.08$ & $R_H^{VH}=0.19$ & 2.4 & 0.19 & 16 & best-observed \\
$H$ & $-18.43\pm0.08$ & $R_H^{VH}=0.22$ & 3.1 & 0.21 & 14 & excluding~05A~and~06X \\
$H$ & $-18.40\pm0.08$ & $R_H^{VH}=0.0$ & 0.0 & 0.19 & 14 & excluding~05A~and~06X \\
\\
$K_s$ & $-18.47\pm0.12$ & $R_K^{VK}=0.12$ & 2.7 & 0.27 & 9 & best-observed \\
$K_s$ & $-18.49\pm0.13$ & $R_K^{VK}=0.13$ & 3.1 & 0.29 & 8 & excluding~06X \\
$K_s$ & $-18.49\pm0.13$ & $R_K^{VK}=0.0$ & 0.0 & 0.27 &  8 & excluding~06X \\
\enddata 
\end{deluxetable}

\clearpage
\begin{deluxetable}{ccccc} 
\tabletypesize{\small} 
\tablecolumns{5} 
\tablewidth{0pt} 
\tablecaption{Comparison of Absolute Magnitudes in $JHK_s$\label{tab:absnircomp}} 
\tablehead{ 
 & \multicolumn{2}{c}{Maximum} & \multicolumn{2}{c}{$t(B_{max})$} \\
\colhead{Band} & \colhead{\citet{krisciunas04b}} & \colhead{This Paper} & \colhead{\citet{wood-vasey08}} & \colhead{This Paper} 
}
\startdata
$J$ & $-18.61\pm0.03$ & $-18.43\pm0.07$ & $-18.29\pm0.09$ & $-18.42\pm0.02$ \\
$H$ & $-18.28\pm0.03$ & $-18.42\pm0.08$ & $-18.08\pm0.09$ & $-18.23\pm0.03$ \\
$K_s$ & $-18.44\pm0.03$ & $-18.47\pm0.12$ & $-18.30\pm0.11$ & $-18.30\pm0.05$ \\
\enddata 
\end{deluxetable}

\clearpage
\begin{deluxetable}{lcccl} 
\tabletypesize{\small} 
\tablecolumns{5} 
\tablewidth{0pt} 
\tablecaption{Measurements of $R_V$ for Individual SNe~Ia\label{tab:redsne}} 
\tablehead{ 
\colhead{SN} & \colhead{$\dm$} & \colhead{$E(B-V)$} & \colhead{$R_V$} & \colhead{Reference}
}
\startdata
1999cl & $1.29\pm0.08$ & $1.24\pm0.07$ & $1.55\pm0.08$ & \citet{krisciunas06} \\
2001el & $1.15\pm0.04$ & $0.21\pm0.05$ & $2.15\pm0.23$ & \citet{krisciunas07} \\
2002cv & $1.46\pm0.17$ & $5.45\pm0.28$ & $1.59\pm0.07$ & \citet{elias-rosa08} \\
2003cg & $1.25\pm0.05$ & $1.33\pm0.11$ & $1.80\pm0.19$ & \citet{elias-rosa06} \\
2005A  & $1.34\pm0.03$ & $1.11\pm0.07$ & $1.68\pm0.10$ & This paper \\
2006X  & $1.31\pm0.05$ & $1.42\pm0.04$ & $1.48\pm0.06$ & \citet{wangx08a} \\
       & $1.29\pm0.05$ & $1.34\pm0.07$ & $1.55\pm0.07$ & This paper \\
\enddata 
\tablecomments{The values $\dm$ given here have been corrected for reddening
using equation~(6) of \citet{phillips99}.}
\end{deluxetable}

\clearpage
\begin{deluxetable}{lcccccc}
\tabletypesize{\small} 
\tablecolumns{7} 
\tablewidth{0pt} 
\tablecaption{Reddening-law coefficients for a SN~Ia at maximum\label{tab:axbx}}
\tablehead{
 & \multicolumn{2}{c}{CCM} & \multicolumn{2}{c}{CCM+O} & \multicolumn{2}{c}{CCM+O (Obs.)} \\ 
 \colhead{Filter} & \colhead{$a_X$} & \colhead{$b_X$} & \colhead{$a_X$} & \colhead{$b_X$} & \colhead{$a_X$} & \colhead{$b_X$}
 }
\startdata
 $u$ & 0.990 &  1.605 &  1.006  &  1.480 &  0.959  &  1.849 \\ 
 $B$ & 1.016 &  0.860 &  1.026  &  0.799 &  1  &  1 \\ 
 $g$ & 1.017 &  0.540 &  1.020  &  0.525 &  1.003  &  0.657 \\ 
 $V$ & 1 &  0 &  1  &  0 &  1  &  0 \\ 
 $r$ & 0.939 & -0.245 &  0.942  & -0.284 &  0.952  & -0.356 \\ 
 $i$ & 0.807 & -0.499 &  0.797  & -0.449 &  0.813  & -0.564 \\ 
 $Y$ & 0.539 & -0.502 &  0.538  & -0.503 &  0.557  & -0.636 \\ 
 $J$ & 0.398 & -0.370 &  0.398  & -0.371 &  0.412  & -0.469 \\ 
 $H$ & 0.259 & -0.243 &  0.259  & -0.244 &  0.268  & -0.308 \\ 
 $K$ & 0.166 & -0.153 &  0.165  & -0.153 &  0.171  & -0.193 \\ 
\enddata
\end{deluxetable}

\clearpage

\clearpage

\clearpage

\clearpage

\clearpage

\clearpage

\clearpage

\clearpage

\clearpage

\clearpage

\clearpage

\clearpage

\clearpage

\clearpage

\clearpage

\clearpage

\clearpage

\clearpage

\clearpage

\clearpage

\end{document}